\documentclass[preprint,number,sort&compress]{elsarticle}

\usepackage{graphicx}
\usepackage{amssymb}
\usepackage{dcolumn}
\usepackage{url}

\newcommand{\degreesC}{\ensuremath{\,^{\circ}\mathrm{C}}}
\newcommand{\rphi}{$r$-\textit{$\phi$}}
\newcommand{\rz}{$r$-$z$}

\journal{Nuclear Instrumentation Methods A}

\begin{document}
\begin{frontmatter}

%
%
\title{
Operational experience, improvements, and performance of the
       CDF Run II silicon vertex detector}
\tnotetext[]{FERMILAB-PUB-13-015-E}

%
%
%
%

\author[HELSINKI]{T.~Aaltonen}
\author[JHU,FNAL]{S.~Behari\corref{cor1}}
\ead{behari@fnal.gov}
\cortext[cor1]{Corresponding author}
\author[UCSB,CHICAGO]{A.~Boveia}
\author[UCSB,AMHERST]{B.~Brau}
\author[PURDUE]{G.~Bolla}
\author[PURDUE]{D.~Bortoletto}
\author[CIEMAT]{C.~Calancha}
\author[FNAL,SLAC]{S.~Carron}
\author[FNAL]{S.~Cihangir}
\author[PARIS]{M.~Corbo}
\author[BRANDEIS]{D.~Clark}
\author[FNAL,BNL]{B.~Di~Ruzza}
\author[FNAL,TAMU]{R.~Eusebi}
\author[CIEMAT]{J.~P.~Fernandez}
\author[FNAL]{J.~C.~Freeman}
\author[PISA,GENEVA]{J.~E.~Garcia}
\author[LBNL]{M.~Garcia-Sciveres}
\author[FNAL] {D.~Glenzinski} 
\author[CIEMAT]{O.~Gonz\'alez}
\author[HARVARD,BARCELONA]{S.~Grinstein} 
\author[PITTSBURGH,TORONTO]{M.~Hartz}
\author[JHU,WISCONSIN]{M.~Herndon}
\author[UCSB,OSU]{C.~Hill}
\author[FNAL]{A.~Hocker}
\author[UROCH,YALE,KARLSRUHE]{U.~Husemann}
\author[UCSB]{J.~Incandela}
\author[UCSB,OXFORD]{C.~Issever}
\author[FNAL]{S.~Jindariani}
\author[FNAL]{T.~R.~Junk}
\author[FNAL]{K.~Knoepfel}
\author[FNAL]{J.~D.~Lewis}
\author[SINICA,NTU] {R.~S.~Lu}
\author[CIEMAT]{R.~Mart\'{\i}nez-Ballar\'{\i}n}
\author[JHU,WMC]{M.~Mathis}
\author[WAYNE]{M.~Mattson}
\author[FNAL,PURDUE]{P.~Merkel}
\author[HARVARD] {L.~Miller}
\author[SINICA] {A.~Mitra}
\author[FNAL]{M.~N.~Mondragon}
\author[FNAL,MGHHARVMED]{R.~Moore}
\author[JHU]{J.~R.~Mumford}
\author[YALE,MIT]{S.~Nahn}
\author[LBNL,SANTACRUZ]{J.~Nielsen}
\author[SLAC]{T.~K.~Nelson}
\author[FNAL]{V.~Pavlicek}
\author[JHU,HARVMED]{J.~Pursley}
\author[CIEMAT]{I.~Redondo}
\author[FNAL]{R.~Roser}
\author[FNAL]{K.~Schultz}
\author[FNAL]{J.~Slaughter}
\author[FNAL]{J.~Spalding}
\author[FNAL]{M.~Stancari}
\author[YALE,DESY]{M.~Stanitzki}
\author[UCSB]{D.~Stuart}
\author[UFL,FNAL]{A.~Sukhanov}
\author[FNAL]{R.~Tesarek}
\author[FNAL]{K.~Treptow}
\author[UCLA,ETH]{R.~Wallny}
\author[FNAL]{P.~Wilson}
\author[RUTG,RUTHER,CERN]{S.~Worm}

\address[SINICA]{Academia Sinica, Taipei, Taiwan 11529, Republic of China}
\address[BARCELONA]{Institut de Fisica d’Altes Energies, ICREA, Universitat Autonoma de Barcelona, E-08193, Bellaterra (Barcelona), Spain}
\address[BRANDEIS]{Brandeis University, Waltham, MA 02453, United States}
\address[BNL]{Physics Department, Brookhaven National Laboratory,
Upton, NY 11973, United States}
\address[CERN]{CERN, CH-1211 Geneva, Switzerland}
\address[CHICAGO]{Enrico Fermi Institute, University of Chicago, Chicago, IL 60637, United States}
\address[UCLA]{University of California, Los Angeles, CA 90095, United States}
\address[UCSB]{University of California, Santa Barbara, CA 93106, United States}
\address[SANTACRUZ]{Santa Cruz Institute for Particle Physics, University of California, Santa Cruz, CA 95064, United States}
\address[DESY]{DESY, Notkestr. 85, D-22603 Hamburg and Platanenallee 6, D-15738 Zeuthen, Germany}
\address[ETH]{ETH Institute for Particle Physics, Schafmattstrasse 20, 8093 Zurich, Switzerland}
\address[FNAL]{Fermi National Accelerator Laboratory, Batavia, IL 60510, United States}
\address[UFL]{University of Florida, Gainesville, FL  32611, United States}
\address[GENEVA]{University of Geneva, CH-1211 Geneva 4, Switzerland}
\address[HARVARD]{Harvard University, Cambridge, MA 02138, United States}
\address[HELSINKI]{University of Helsinki and Helsinki Institute of Physics, FIN-00014, Helsinki, Finland}
\address[JHU]{The Johns Hopkins University, Baltimore, MD 21218, United States}
\address[KARLSRUHE]{Institut f\"ur Experimentelle Kernphysik, Karlsruhe Institute of Technology, D-76131 Karlsruhe, Germany}
\address[LBNL]{Ernest Orlando Lawrence Berkley National Laboratory,
Berkley, CA 94720, United States}
\address[NTU]{National Taiwan University (NTU), Taipei, Taiwan, Republic of China}
\address[CIEMAT]{Centro de Investigaciones Energeticas Medioambientales y Tecnologicas, E-28040 Madrid, Spain}
\address[AMHERST]{Department of Physics, University of Massachusetts, Amherst, MA, United States}
\address[MGHHARVMED]{Massachusetts General Hospital, Harvard Medical
School, Boston, MA 02114, United States}
\address[MIT]{Massachusetts Institute of Technology, Cambridge, MA 02139, United States}
\address[HARVMED]{Brigham and Women's Hospital, Harvard Medical School, Boston, MA 02115, United States}
\address[OSU]{The Ohio State University, Columbus, OH 43210, United States}
\address[OXFORD]{University of Oxford, Oxford OX1 3RH, United Kingdom}
\address[PARIS]{LPNHE, Universite Pierre et Marie Curie/IN2P3-CNRS, UMR7585, Paris, F-75252 France}
\address[PISA]{Istituto Nazionale di Fisica Nucleare Pisa, Universities of Pisa,Siena and Scuola Normale Superiore, I-56127 Pisa, Italy}
\address[PITTSBURGH]{University of Pittsburgh, Pittsburgh, PA 15260, United States}
\address[PURDUE]{Purdue University, West Lafayette, IN 47907, United States}
\address[UROCH]{University of Rochester, Rochester, NY 14627, United States}
\address[RUTG]{Rutgers University, Piscataway, NJ 08855, United States}
\address[RUTHER]{Rutherford Appleton Laboratory, Science and Technology Facilities Council, Harwell Science and Innovation Campus, Didcot OX11 0QX, United Kingdom}
\address[SLAC]{SLAC National Accelerator Laboratory, Menlo Park, CA 94025, United States}
\address[TAMU]{Texas A\&M University, College Station, TX 77843, United States}
\address[TORONTO]{University of Toronto, Toronto, Ontario M5S 1A7, Canada}
\address[WAYNE]{Wayne State University, Detroit, MI 48202, United States}
\address[WISCONSIN]{University of Wisconsin, Madison, WI 53706, United States}
\address[WMC]{College of William \& Mary, Williamsburg, VA 23187, United States}
\address[YALE]{Yale University, New Haven, CT 06520, United States}

%
%
%
%
\begin{abstract}

The Collider Detector at Fermilab (CDF) pursues a broad physics program at
Fermilab's Tevatron collider. Between Run II commissioning in 
early 2001 and the end of operations in September 2011, the Tevatron 
delivered 12 fb$^{-1}$ of integrated luminosity of $p{\bar{p}}$ collisions
at $\sqrt{s}=1.96$~TeV.  The physics at CDF
includes precise measurements of the masses of the top quark and $W$ boson,
measurement of CP violation and $B_s$ mixing, and searches for Higgs 
bosons and new physics signatures, all of which require heavy flavor tagging 
with large charged particle tracking acceptance.  To realize these goals, 
in 2001 CDF installed eight layers of silicon microstrip detectors around its 
interaction region. These detectors were designed for 2--5 years of operation, 
radiation doses up to 2 Mrad (0.02 Gy), 
and were expected to be replaced in 2004.
The sensors were not replaced, and the Tevatron run was extended for
several years beyond its design, exposing the sensors and electronics
to much higher radiation doses than anticipated.  
In this paper we describe the operational challenges encountered over the past 
10 years of running the 
CDF silicon detectors, the preventive measures undertaken, and the 
improvements made along the way to ensure their optimal performance for 
collecting high quality physics data. In addition, we describe the quantities 
and methods used to monitor radiation damage in the sensors for optimal 
performance and summarize
the detector performance quantities important to CDF's physics program, 
including vertex resolution, heavy flavor tagging, and silicon vertex trigger 
performance.
\end{abstract}


%
%
\end{frontmatter}

%
%
\newpage

%
%

%
%
\section{Introduction}
\label{sec:intro}

The Tevatron collider at the Fermi National Accelerator Laboratory
(FNAL) collided proton and antiproton beams at a center-of-mass energy
of 1.96~TeV. The collisions occurred at two interaction points where
multipurpose detectors Collider Detector at Fermilab (CDF II) and D0,
were positioned.

\begin{figure}[h]
\centering
\includegraphics[width=0.65\textwidth]{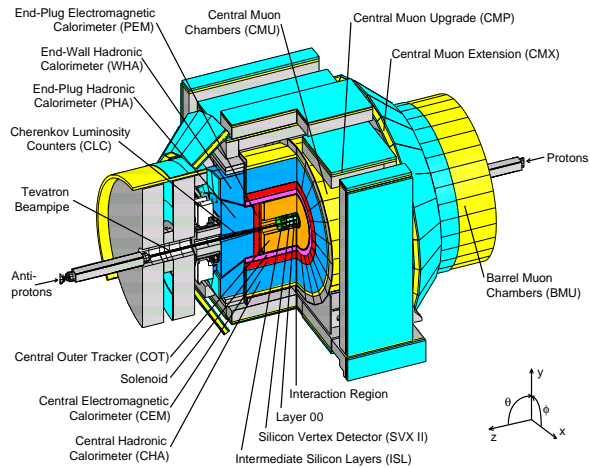}
\caption{Isometric view of the entire CDF II detector.}
\label{fig:tracking_syst1}
\end{figure}

\begin{figure}[h]
\centering
\includegraphics[width=0.65\textwidth]{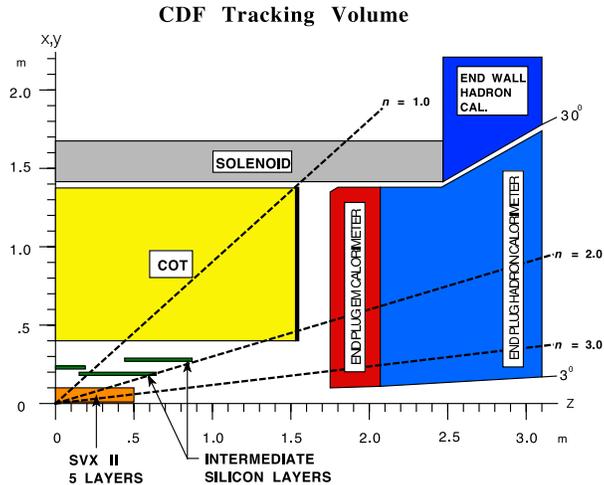}
\caption{Schematic layout of the CDF II tracking system.}
\label{fig:tracking_syst2}
\end{figure}

The CDF II detector~\cite{cdftdr} was a general purpose detector with
cylindrical geometry. The innermost part of the detector consisted of
charged-particle tracking detectors, shown in
Figs.~\ref{fig:tracking_syst1} and \ref{fig:tracking_syst2}, which 
were located inside a superconducting solenoidal magnet which provided 
a highly uniform 1.4~T magnetic field oriented parallel to the beam 
axis.  Calorimeters and muon systems outside the solenoid provided 
lepton identification and momentum measurement as well as jet energy 
measurements. The tracking detectors and calorimeters were used for 
jet reconstruction, where the former provided identification of jets 
from heavy (charm and bottom) quarks.

The inner component of the tracking system was a series of silicon
microstrip detectors that constituted the CDF silicon
detector. Beyond the silicon detector lay the Central Outer Tracker
(COT), an open-cell drift chamber. Together with the additional
constraints coming from the position of the primary vertex, the COT
and Silicon Detector provided resolution on the track momentum
transverse to the beam direction, $p_T$, of ${\sigma}(p_T)/p_T = 0.15
\% \cdot p_T / (\mathrm{GeV}/c)$.


CDF II used a cylindrical coordinate system with the $z$ axis oriented
along the proton beam direction and azimuthal angle $\phi$ measured
around the beam axis. The polar angle $\theta$ was measured with
respect to the positive $z$ (proton-beam) direction and was used to
define the pseudorapidity $\eta \equiv -{\ln}({\tan}({\theta}/2))$.

The physics program of Run II at the Tevatron includes precision
measurements of the mass of the top quark and $W$ boson; bottom and
charm physics, including the determination of the $B_s$ and $D^0$
mixing parameters; studies of the strong interaction (jet
multiplicities, diffractive physics, etc.); and searches for objects
and phenomena as varied as the Higgs boson, 
supersymmetric particles, hidden space-time dimensions, and quark
substructure~\cite{RunIIPhys,RunIIPhys2, RunIIPhys3, RunIIPhys4}.  All
these measurements benefit from a high-resolution tracking detector
and many rely heavily on the efficient identification of heavy quarks
by detection of displaced secondary vertices, and are enhanced by the 
capability to trigger on tracks originating away from the beam.

The CDF silicon detector was designed to withstand radiation
doses up to 2\,MRad (0.02\,Gy), the dose expected during the first 2--5
years of CDF operations, with replacement of inner layers planned in
2004~\cite{Akimoto2006459}. However, the upgrade project was canceled
in 2003, and Run II was later extended into late 2011, with total
delivered integrated luminosity of 12~$\mathrm{fb}^{-1}$. Several
preventive measures were taken to keep the original silicon detector
operational and maintain its performance. The most important of these
was the decrease in the operating temperature of the detector, which
reduced the impact of chronic radiation exposure
(Section~\ref{sec:cool}).  Steps were also taken to minimize thermal
cycles, damage from resonances of wire bonds
(Section~\ref{DAQ:Wirebond}), and instabilities and sudden
loss of the Tevatron beams (Section~\ref{sec:beam}).

Issues arising from radiation damage of the sensors, aging
infrastructure, and electronics were addressed continuously in addition
to the basic challenges posed by the inaccessibility of the detector
volume and large number of readout channels (approximately 722,000).
The operational challenges, improvements to, and the performance of the
CDF silicon detector are presented in this paper. Similar experiences 
with their silicon detectors have been reported by the D0
collaboration~\cite{Jung2012ii,Weber2008zz,Desai2009zza}.

This paper is organized as follows:  
Section~\ref{sec:detdesc} provides a general description of the detector, 
Section~\ref{sec:daqtrig} gives an overview of the data acquisition, the trigger, and the interface between them,
Section~\ref{sec:ps} describes the power supplies and the operational experience with them and response to their failures,
Section~\ref{sec:cool} details the design, history, and response to failures in the cooling system,
Section~\ref{sec:beam} gives a review of particle beam incidents, and response to them.
Section~\ref{sec:calib} details the readout calibration,
Section~\ref{sec:mon} is dedicated to the routine monitoring and operations support systems,
Section~\ref{sec:age} describes the response of the CDF silicon detector to accumulated radiation doses, 
Section~\ref{sec:perf} details the performance of the silicon detector and the displaced vertex trigger,
and Section~\ref{sec:sum} gives a summary.
As well as new results, this paper compiles final results on material dispersed in several conference proceedings
produced over the years by the members of operations team~\cite{Merkel20031,l00_tdr2,Hou2003166,Bolla2004277,Miller2004281,Hill20041}.


%
%
\section{Detector description}
\label{sec:detdesc}

\begin{figure}[h]
\centering
\includegraphics[width=0.47\textwidth]{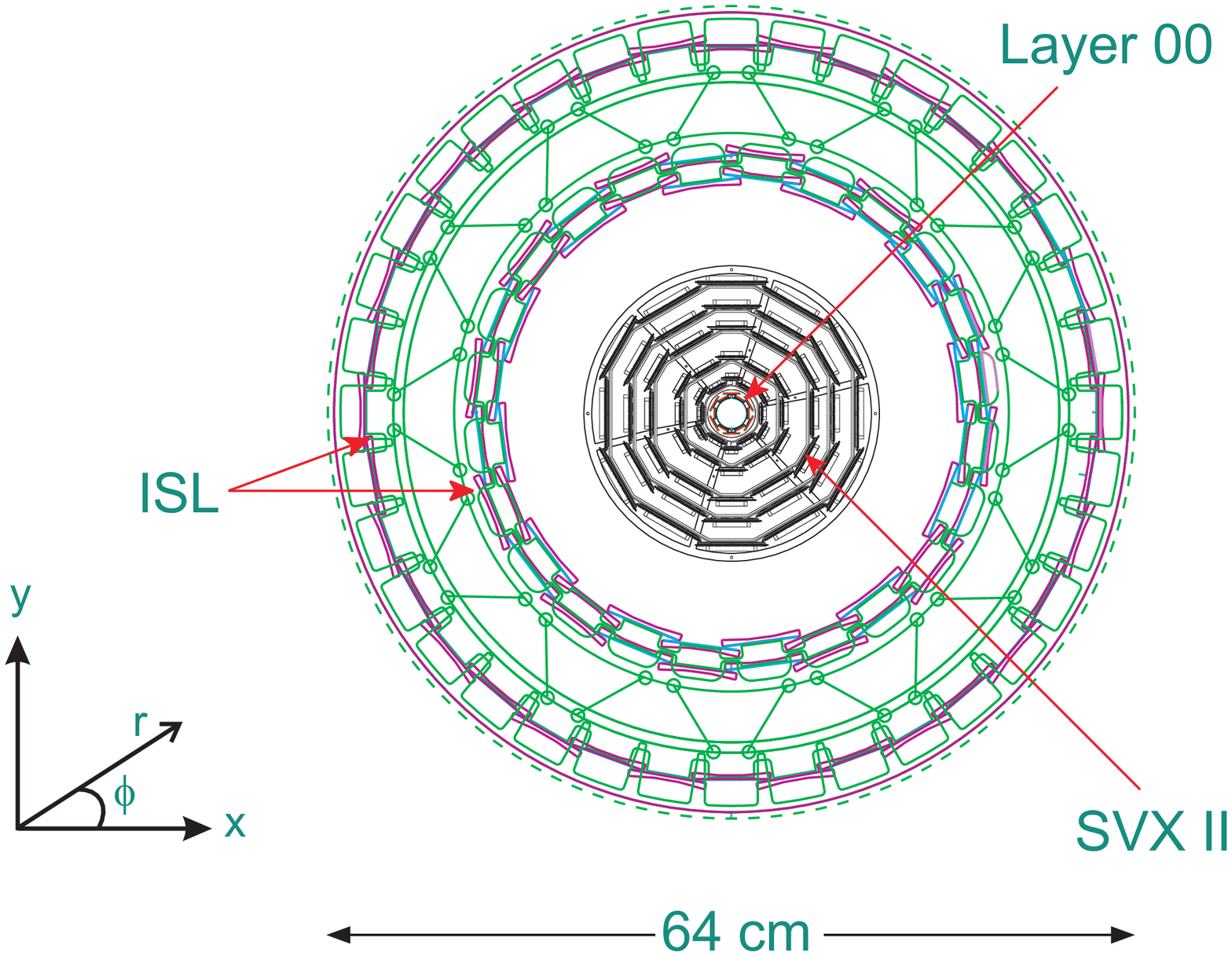}
\hspace*{0.04\textwidth}
\includegraphics[width=0.47\textwidth]{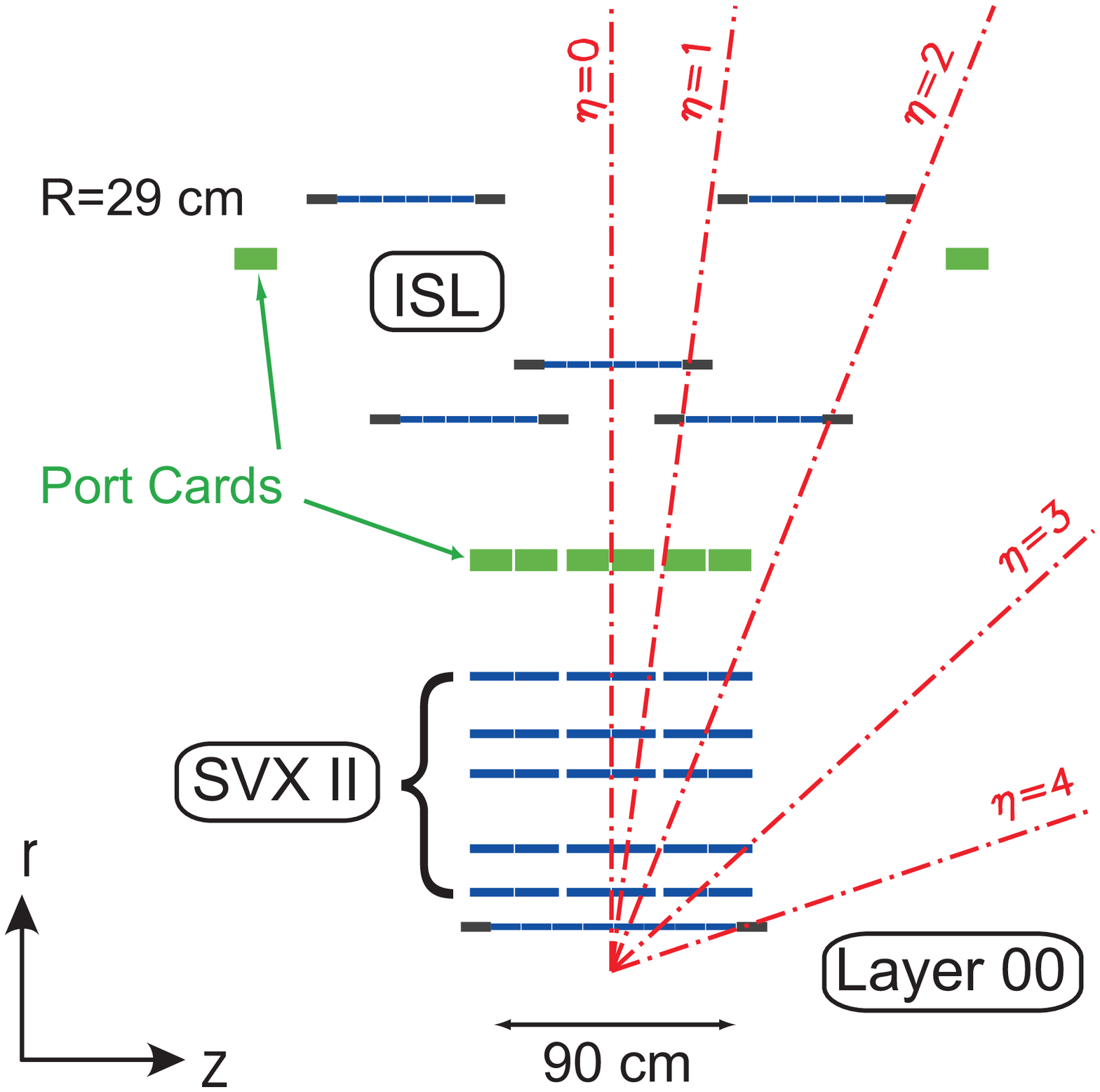}
\caption{
Schematic layout of the CDF silicon detectors showing $x$-$y$ (\rphi, left)
and $y$-$z$ (\rz, right) views. Note that the $z$ axis is compressed for 
illustration purposes.
}
\label{fig:Sili_layout}
\end{figure}

\begin{figure}[h]
\centering
\includegraphics[width=0.47\textwidth]{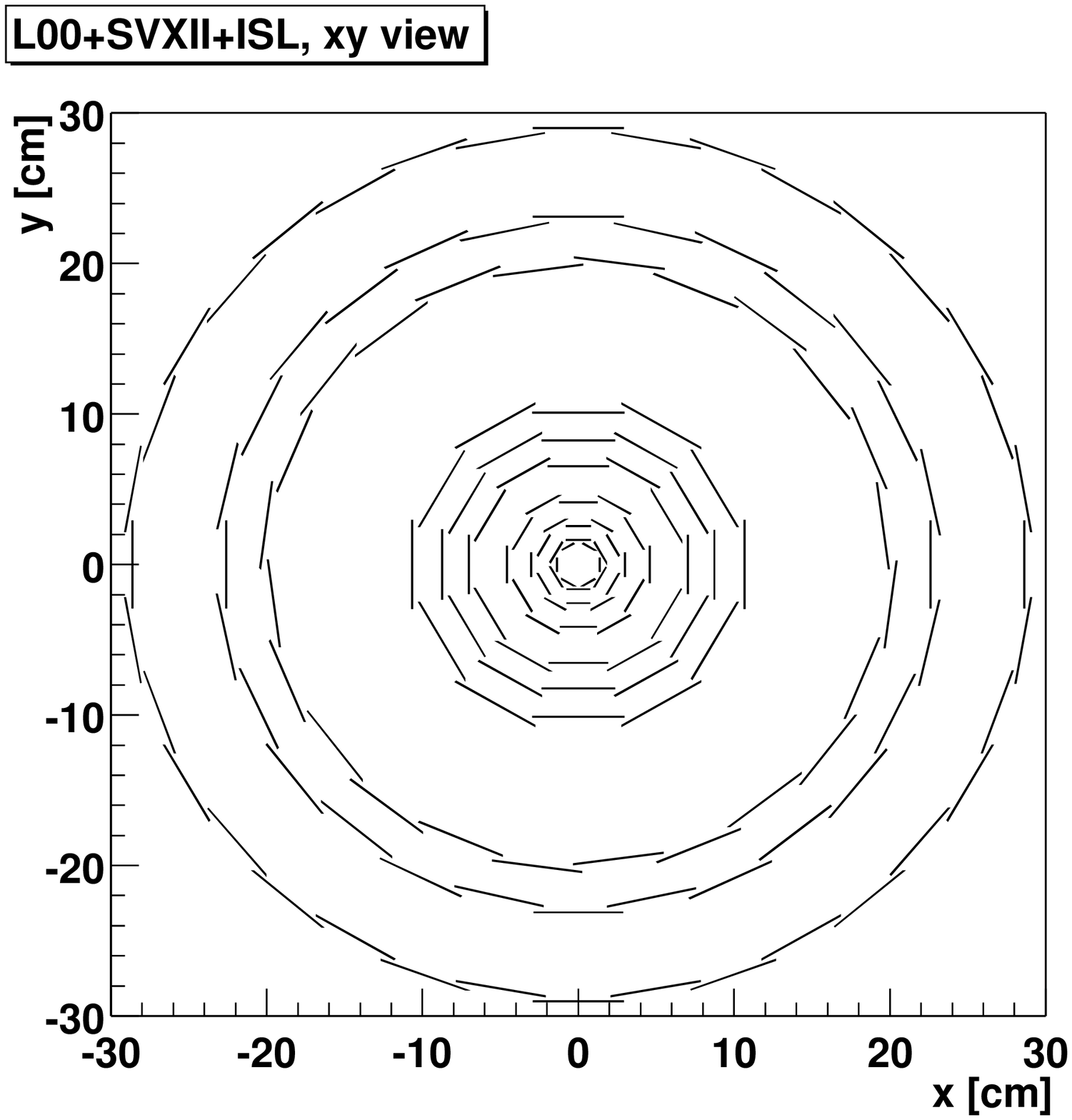}
\hspace*{0.04\textwidth}
\includegraphics[width=0.47\textwidth]{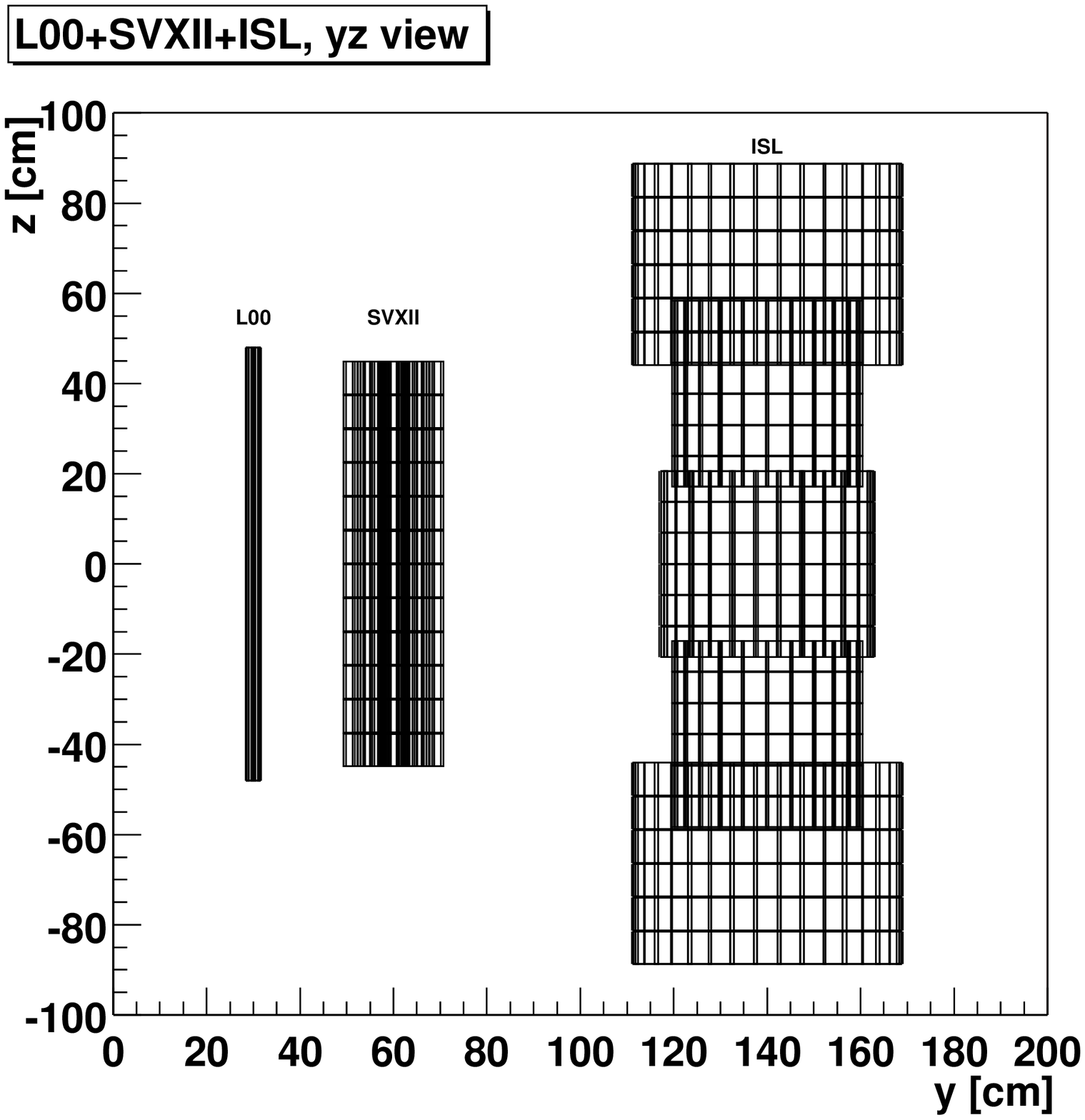}
\caption{
Dimensions, in cm, of the CDF silicon detector system. Shown are 
$x$-$y$ and $y$-$z$ views. In the $y$-$z$ view, each square corresponds to
one sensor and each subdetector has been displaced along the $y$-axis
for illustration purposes.
}
\label{fig:Sili_coverage}
\end{figure}

The CDF silicon detector system consisted of three sub-detectors, all
with barrel geometry: Layer 00 (L00)~\cite{l00_tdr1,l00_tdr2}, the
Silicon Vertex detector (SVX-II)~\cite{svxii_tdr1,svxii_tdr2} and the
Intermediate Silicon Layers (ISL)~\cite{isl_tdr}. Unless otherwise
stated, detector refers to the CDF silicon detector.  The design of
the system was driven by the goal of providing excellent
spatial resolution in the measurement of charged-particle tracks.
These measurements were crucial for the reconstruction of the displaced
secondary vertices and therefore, identification of events with
bottom-quarks. Figs.~\ref{fig:Sili_layout} and \ref{fig:Sili_coverage}
present the schematic layout of the CDF silicon detector, and
Table~\ref{SpecsTable} summarizes some of the basic parameters.  The
design had eight silicon layers to provide tracking which is robust
against failure or degradation of individual components.


\begin{table}[ht]
\caption{Summary of L00, SVX-II and ISL basic parameters.}
\label{SpecsTable}
\center
\begin{tabular}{l|D{.}{.}{-1}|l|l}
\hline\hline
Name & \textrm{Radius (cm)} & Readout & Manufacturer \\ \hline
L00 (narrow)   &  1.35 & \rphi         & SGS Thomson, Micron \\
L00 (wide)     &  1.62 & \rphi         & Hamamatsu    \\ \hline
SVX L0         &  2.54 & \rphi,\rz     & Hamamatsu \\
SVX L1         &  4.12 & \rphi,\rz     & Hamamatsu \\
SVX L2         &  6.52 & \rphi,+1.2$^\circ$ & Micron \\
SVX L3         &  8.22 & \rphi,\rz     & Hamamatsu \\
SVX L4         & 10.10 & \rphi,-1.2$^\circ$ & Micron \\ \hline
ISL L6 Central & 22.00 & \rphi,1.2$^\circ$ & Hamamatsu \\
ISL L6 Fwd/Bwd & 20.00 & \rphi,1.2$^\circ$ & Hamamatsu \\
ISL L7 Fwd/Bwd & 28.00 & \rphi,1.2$^\circ$ & Micron \\ \hline
\hline
\end{tabular}
\end{table}

The basic structural unit of a sub-detector was a \emph{ladder}, which
consisted of several silicon microstrip sensors bonded in series
(3 sensors for L00 ladders, four in SVX-II ladders and six in ISL ladders).
Strip width and multiplicity depended  on the \emph{layer}, or distance 
from the beam pipe. The sensors were made from high-resistivity n-type silicon
with a nominal thickness of 300 $\mu \mathrm{m}$. Sensors in L00 were 
single-sided, providing \rphi\ information, while sensors in the other layers 
were double-sided, providing both \rphi\ and \rz\ information. The sensors in 
SVX-II layers 0, 1, and 3 used double-metal readout for a 90$^\circ$ strips on 
the \rz\ side. The other double-sided layers used small-angle stereo strips.

The readout was carried out through aluminum strips AC coupled to the 
implant strips, which are of p-type for the \rphi\ and n-type for the \rz\ 
or small-angle stereo side. A full ladder was read out from both ends 
through SVX3D readout chips (described in Section~\ref{sec:svx3d}) mounted 
on electrical hybrids. These hybrids were located outside (for L00) or 
inside (for SVX-II and ISL ladders) of the tracking volume.  A circuit 
board called the \emph{portcard} was located at the periphery of each 
support structure or bulkhead and formed an interface with the hybrids 
and readout chips with the rest of the data acquisition system 
(Section~\ref{sec:daqtrig}).

Layer 00 was a single-sided silicon microstrip 
detector whose sensors could be biased to higher voltages than the
double-sided sensors. It was mounted on a carbon fiber support
structure which was in turn mounted directly on the beam pipe, and had
an inner radius of 1.15~cm and outer radius of 2.1~cm.  Its main
purpose was to improve the track impact parameter resolution which was
otherwise limited by multiple scattering in the additional material of
the SVX-II readout and cooling infrastructure;
a secondary purpose was to prolong CDF silicon detector lifetime by
providing a backup to SVX-II layer-0. 
Layer 00 consisted of one layer and had 72 ladders with 13,000 readout
channels in total. 

The SVX-II detector was built in three cylindrical barrels each 29~cm
long. Each barrel contained five layers of double-sided silicon
microstrips placed along the beam axis, with radial coverage from 2.5
to 10.7~cm.  Carbon fiber reinforced Rohacell foam~\cite{Rohacell}
provided support to the ladders, and
beryllium bulkheads provided additional support and alignment on each
end. Therefore the detector consisted of six bulkheads
($z$-segmentation), each with 12 wedges ($\phi$-segmentation)
consisting of 5 layers ($r$-segmentation). In total, it had 360
ladders with 405,504 channels in the system.  One side of each
microstrip sensor provided tracking information in the \rphi\ plane,
with strips oriented parallel to the beam direction, while the other
side had strips oriented either perpendicular to the beam axis,
providing 90$^\circ$ information, or at an angle of $\pm 1.2^{\circ}$
with respect to the beam axis, providing small-angle stereo
information.  Three of the five SVX-II layers had 90$^\circ$ sensors,
while the remaining two layers had small-angle stereo strips, as
detailed in Table~\ref{SpecsTable}.  The readout chips and electric
hybrids were mounted on the surface of the SVX-II silicon
sensors. SVX-II was read out in a strict $\phi$-wedge geometry in order
to feed the secondary vertex trigger, described in Section~\ref{sec:svt}.

The ISL was located between SVX-II and the COT drift chamber. It
consisted of one central ($\left|\eta\right|<1$) layer of silicon at a
radial position of 22~cm and two forward ($1<\left|\eta\right|<2$)
layers at 20~cm and 28~cm. Mechanical support for the ladders was provided by
carbon fiber rings. ISL had 148 double-sided ladders of 55 cm length
each with a total of 303,104 channels.
An ISL ladder was composed of three microstrip sensors bonded 
together. Like SVX-II, one side of each sensor provided tracking 
information in the \rphi\ plane while the other side provided tracking 
information in the \rz\ plane with $\pm 1.2^{\circ}$ stereo
angle. Also like SVX-II, the readout chip hybrids were  
mounted on the sensors.

\begin{figure}[hbt]
\centering
\includegraphics[width=0.47\textwidth]{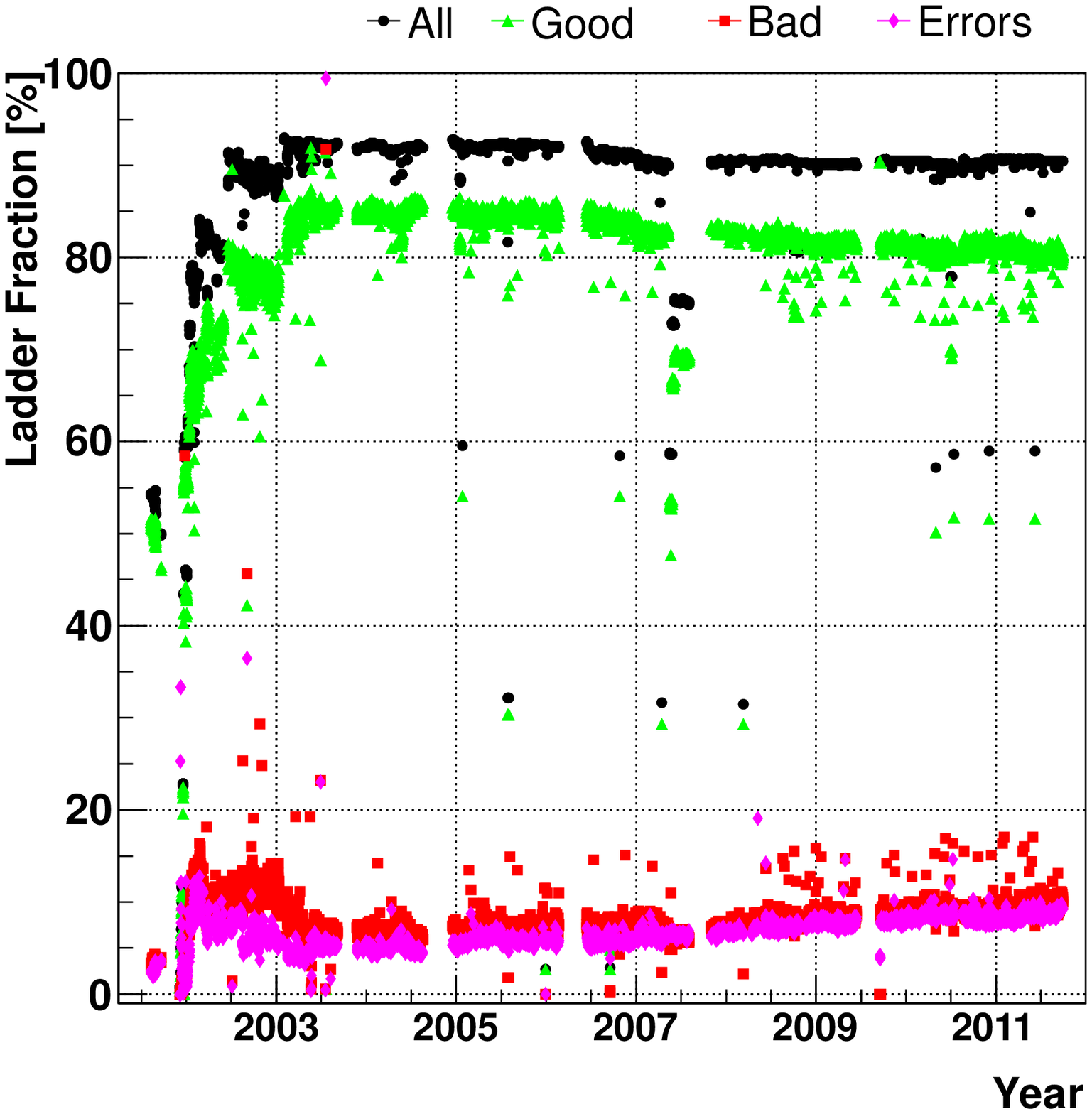}
\includegraphics[width=0.47\textwidth]{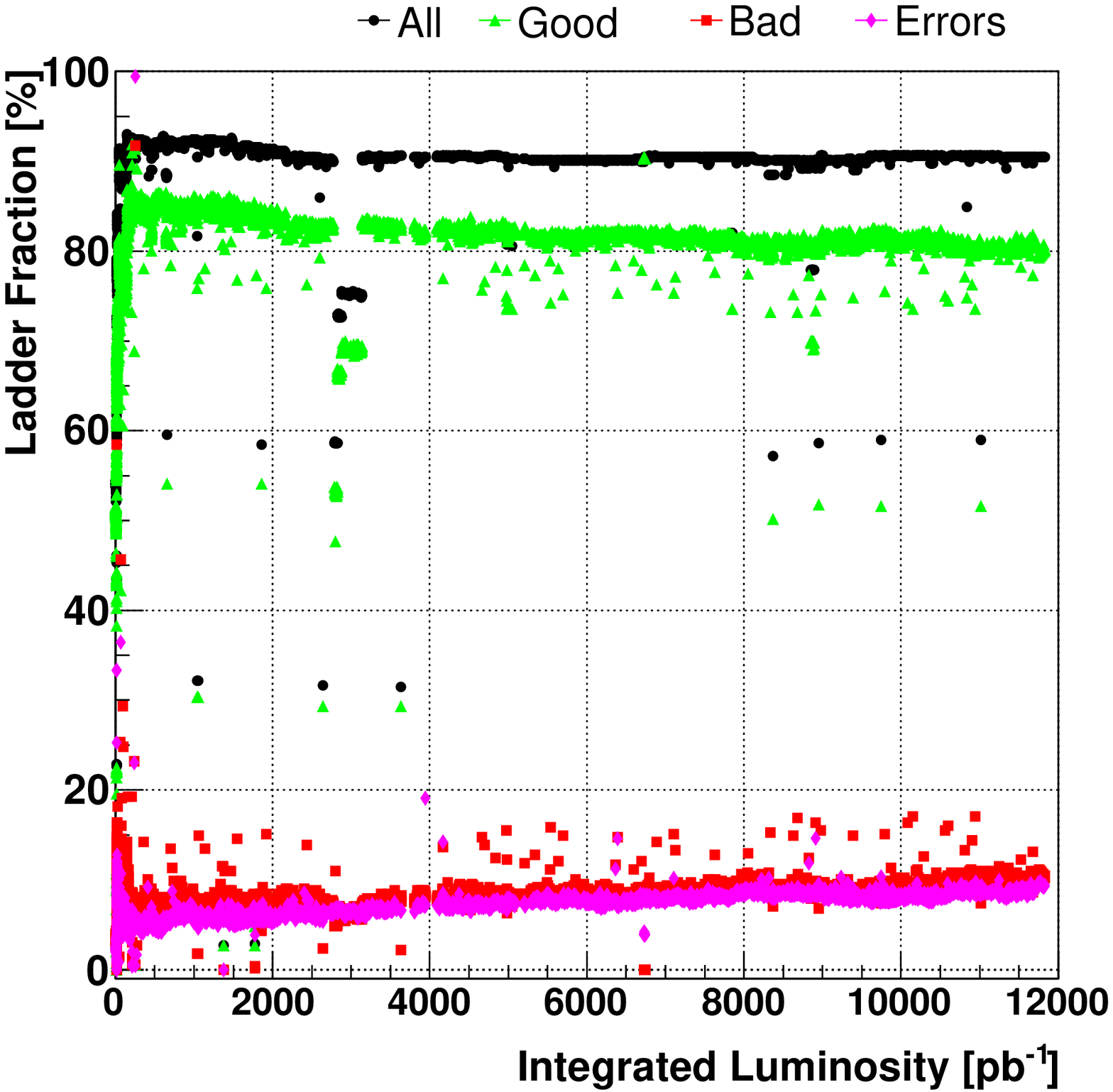}
\caption{
Fraction of ladders which were powered (black circles), 
considered good (green triangles) and bad (red squares) versus time (left) and versus integrated
luminosity (right). A ladder is considered good if it has less than
1\% digital errors. 
Also shown is the average digital error rate (pink diamonds).  The actual fraction of
good ladders is larger as it does not included ladders whose digital errors
are corrected by the offline reconstruction software.
}
\label{fig:det_digiErr}
\end{figure}

Fig.~\ref{fig:det_digiErr} gives a historical account versus time (left)
and integrated luminosity (right) of the fraction of detector ladders
included in data taking since start of commissioning in 2001.
Aside from variation during the start-up period, stable detector operation is observed 
over the long term data taking period. The rate of corruption in data transmission out 
of the detector (referred to as "digital errors") rises with time. Some of this corruption 
was recovered with offline processing by using knowledge of the data structure to identify 
and correct erroneous bits.


%
%
\section{The silicon detector data acquisition system}
\label{sec:daqtrig}
The data acquisition (DAQ) system of the silicon detector was
responsible for reading out and digitizing the charge collected by the
722,432 silicon strips. The DAQ worked in coordination with the CDF
trigger system that selected events (proton--antiproton collisions) of
interest~\cite{cdftdr}.  The DAQ system was comprised of
radiation-hard readout ASICs mounted on the detector, feeding optical 
data links and a
chain of VME boards that coordinated the DAQ process and that collated and
processed the data. The unique feature of the DAQ system was the
integration of the silicon detector with the secondary vertex trigger
(SVT) which had never been attempted at a hadron collider.

The first half of this section
(Sections~\ref{sec:daq:CDFDAQ}-~\ref{DAQ:offdetector}) describes the
CDF DAQ and then details the components of the silicon detector DAQ.
The second half
(Sections~\ref{DAQ:SVX3DExperience}-~\ref{DAQ:DAQPerform}) describes
the commissioning and operations experience, which includes the
unexpected behaviors of the SVX3D chip, noise from L00, and effect of
radiation on the DAQ.

\subsection{CDF timing and trigger}~\label{sec:daq:CDFDAQ}
The CDF trigger and DAQ systems were synchronized to the Tevatron
beams. The Tevatron divided the proton and antiproton beams into 3
trains, separated by 1.4\,$\mu$s, and each train was composed of 12
bunches separated by 396\,ns. In total there were 36 bunches of
protons and antiprotons. It took 21\,$\mu$s to complete one revolution of
the Tevatron. The orbits of the proton and antiproton beams were set
to collide every 396\,ns~\footnote{The Tevatron proton--antiproton
  collision rate was intended to be upgraded to 132\,ns, for which
  the CDF DAQ system was designed, but this change was not implemented.} 
at the two points where the CDF and D0 detectors were located.

The CDF clock signals were derived from the Tevatron clock system.
The most fundamental clock was derived from the 53\,MHz Tevatron radio
frequency (RF) system\rlap.~\footnote{The central Tevatron RF system fed
  the Tevatron's accelerating RF cavities.} The Tevatron also sent a
signal corresponding to the first proton bunch, with a period of
21\,$\mu$s (1113 Tevatron RF clock periods)
in phase with the RF clock.  Further
details on the Tevatron beam structure and clocks are available in
\cite{Holmes:2011ey} and references therein.  The fundamental CDF
clock was derived by dividing the Tevatron RF clock by 7, which gave a
period of 132\,ns and phased with the first proton bunch clock.
Additional clock signals were derived for valid bunch crossings and the 
gaps between trains.

At a hadron collider, only a small fraction of events are from
interesting physics processes. The CDF trigger was responsible for
identifying these interesting events in real time. 
CDF employed a three-level trigger system, where each level used more
refined information than the previous level to select events.  The
first level (L1) ran synchronously with the CDF clock and had a fixed
latency of 5.5~$\mu$s.  It reduced the event rate from 1.7\,MHz to
less than 35\,kHz, and was implemented with custom
hardware~\cite{cdftdr}. 
When events were selected by L1, the data for non-silicon detectors 
were stored in one of four buffers, pending processing by the second 
level (L2). 
L2 was an asynchronous trigger, comprised of
dedicated hardware and software, that selected a subset of the L1
triggered events.
It reduced the peak rate of accepted events to $\sim$800\,Hz. 
The third level (L3) was a software trigger that ran a fast version of
the offline event reconstruction on a computer farm using all data
from the CDF detector.  
It selected a subset of L2 triggered events for permanent storage at a
rate of $\sim$150\,Hz.  Overall, the CDF trigger selected $\sim$1 in
11,000 collisions for permanent storage.

\subsubsection{SVT}
\label{sec:svt}
A unique feature of the CDF's L2 trigger was the ability to select
events with a displaced vertex which were characteristic of bottom
quark hadron ($b$-hadron) decays. This method of selecting hadronic $b$-decay
events was more efficient than previous leptonic triggers that relied on
the rarer semi-leptonic $b$-quark decay. This displaced vertex trigger,
known as the Silicon Vertex Trigger (SVT), significantly increased
CDF's yield of $b$-hadrons for analysis. 

The SVT used data from the CDF silicon detector and COT to
perform precision tracking quickly.  Tracks were found by combining
information from the COT-based \emph{extremely fast tracker}
(XFT)~\cite{XFT} and SVX-II axial layers to patterns stored in look-up tables.  
The resulting tracks in the \rphi\ plane were used to calculate the 
2D distance ($L_{xy}$) of a track pair intersection from the primary vertex.
A key development of the SVT hardware was the custom chip-based
pattern recognition (associative memory). CDF was the first detector
at a hadron collider to implement a displaced vertex trigger.  Further
information on the SVT can be found in \cite{svt,SVT2,svtupg} and
references therein.
The demands of SVT to reconstruct tracks and identify
tracks displaced from the interaction point
drove the SVX-II design and led to the wedge and barrel layout, the
tight construction alignment tolerances, and the SVX-II DAQ design that is
discussed in this section.  

\subsection{CDF silicon DAQ architecture}
SVX-II was designed for SVT, which required a specialized DAQ system
to provide silicon strip data in 20-40~$\mu$s. The ISL and L00 that came as
extensions to the CDF silicon detector project inherited the SVX-II
DAQ. Thus the SVX-II DAQ defined the entire CDF silicon detector DAQ.
SVT demanded SVX-II provide data after every L1 accept decision (L1A) by
the trigger system.  This required deadtimeless readout to guarantee 
silicon data were always available. Also, to reduce the time to deliver 
and process the data, only information from silicon strips which collected 
a significant amount of charge relative to the nominal noise signal was 
propagated to SVT. Therefore the silicon readout volume, and thus readout 
time, was driven by the underlying physics processes that drove the 
occupancy of the detector. It was only by meeting these design challenges 
that allowed the combination of SVX-II and SVT to be integrated into the 
CDF trigger.

\begin{figure}[htb]
\centering
\includegraphics[width=0.85\textwidth]{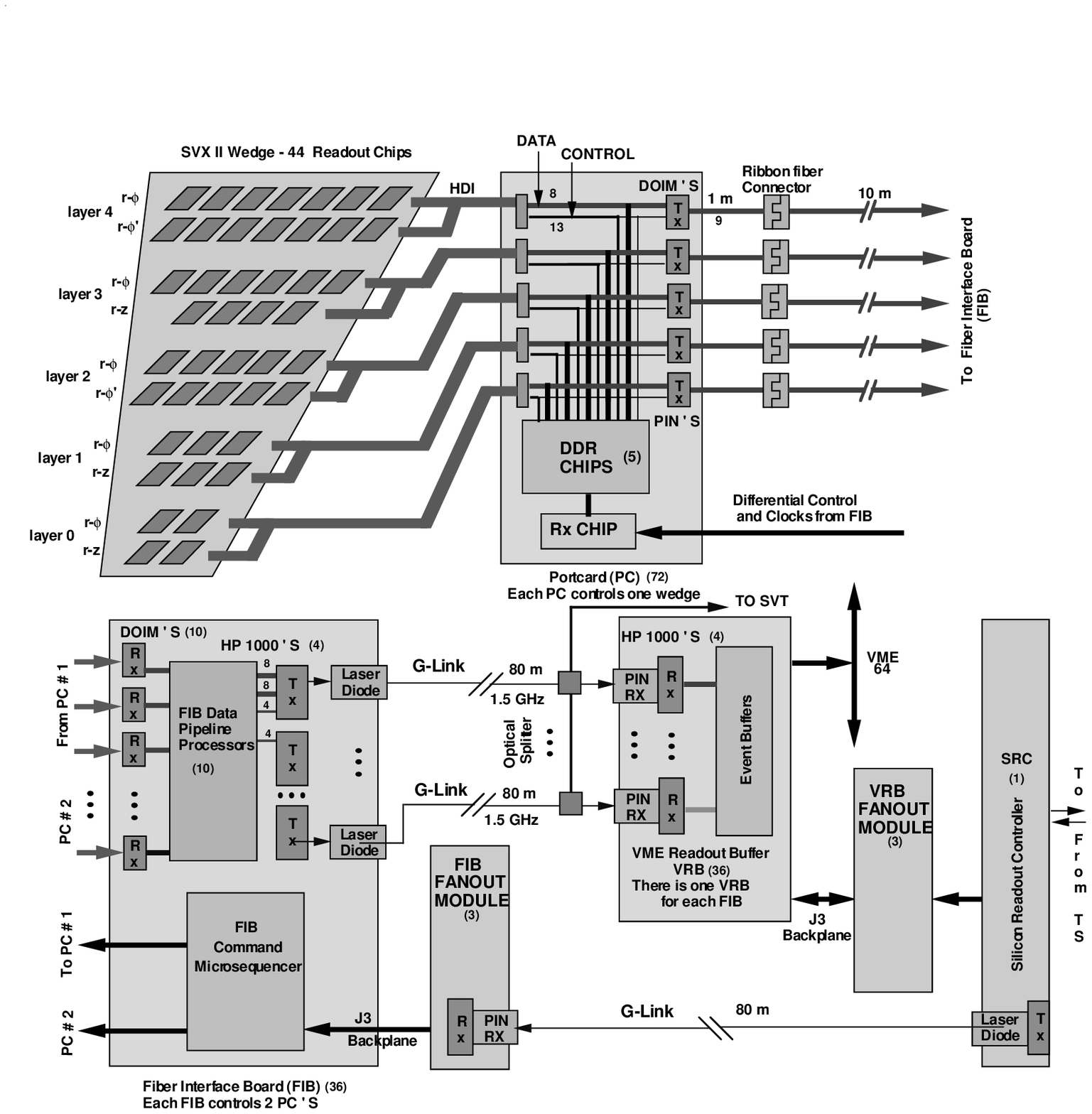}
\caption{ A schematic diagram of the silicon detector DAQ system. The
  SVX-II wedge and portcard, at the top half of the figure, were known
  together as the on-detector electronics as they were located
  directly on the silicon detector.  The components, in the lower half
  of the figure, were known as the
  off-detector electronics. The FIB and FIB fanout were located in the
  CDF collision hall. The VRB, VRB fanout and SRC were in the CDF
  counting room which was located above the CDF collision hall.  ISL
  and L00 DAQ were identical except the SVX-II wedge was substituted with a ISL
  and L00 wedge, respectively.}
\label{fig:daqschema}
\end{figure}

Fig.~\ref{fig:daqschema} shows a schematic diagram of the
silicon DAQ system. 
%
The process was coordinated by a central controller that interfaced with 
the trigger system, broadcasted commands to the individual ladders,
and controlled data flow through feedback signals from the VME readout 
buffers. The digitized data
from each ladder was transferred in parallel to the allocated readout
buffer.  The data were buffered until a L2 decision arrived and was
either transmitted to L3 or discarded. The Silicon DAQ was synchronized
to the CDF clock.

\clearpage
\noindent The electronics were divided into an on-detector part, 
mounted directly on the silicon detector, and an off-detector part, 
located in VME racks in the CDF
collision hall and in the CDF counting room (Fig.~\ref{fig:daqschema}).


\subsection{On-detector electronics}
The on-detector electronics were responsible for acquiring and
digitizing the charge from the silicon strips and then transmitting the
data to the off-detector electronics. The on-detector electronics
consisted of the SVX3D readout chip, the portcard, and DOIM optical
data links that are described below.

\subsubsection{The SVX3D readout chip}
\label{sec:svx3d}
The SVX3D ASIC was responsible for acquiring charge from the silicon
strips and digitizing them.  It was a custom, radiation-hard,
deadtimeless, 128-channel device~\cite{svx_3D_chip_1,svx_3D_chip_2},
capable of recording charge simultaneously from all 128 channels every
132\,ns and had an 8-bit wide output running at 53\,MHz.  All 128
channels were digitized simultaneously using an 8-bit grey-coded
modified Wilkinson type analog-to-digital converter (ADC) that 
included event-by-event dynamic pedestal subtraction (DPS). With DPS, 
which is described later in this section, it was no
longer necessary to read out every strip for an offline pedestal
subtraction.  Therefore the SVX3D could implement data reduction logic
(sparsification) to remove channels that were below a set threshold 
to reduce further the readout time. In this way, the
readout time was dictated by the occupancy of the silicon detector. The
SVX3D's deadtimeless operation, DPS, and sparsification were essential
for SVT.

The SVX3D was manufactured using the Honeywell CMOS 0.8\,$\mu$m
radiation-hard process. From irradiations up to 4\,MRad with $^{60}$Co
sources and 15\,MRad with a 55\,MeV proton source, the chip noise in the
innermost layer of SVX was expected to increase by 17\% after 8\,fb$^{-1}$
(3.1\,MRad)~\cite{svx3d_rad}.

The operations of the chip were divided into an analog front end (FE)
and a digital back end (BE).  The FE was responsible for acquiring
charge from the silicon strips and buffering them into a circular
analog pipeline.  The BE digitized the charge from the pipeline and
any channels above a programmed threshold were sent to a readout FIFO. 
The FE and BE were driven independently by the CDF clock and
Tevatron RF clock, respectively, which allowed the deadtimeless operation.

The chip operation began with an initialization phase which set
various operational parameters such as the signal polarity, chip
identification number (chip ID), and readout mode. After the
initialization phase, the FE changed to the acquisition mode, and the
BE alternated between digitization and readout, until a new
initialization was performed.

Each FE channel consisted of a charge integrator coupled to a
47 stage circular analog capacitor pipeline; 42 pipeline cells were allocated
for the L1 latency, four for L2 buffers, and one reserved to measure the
pipeline pedestal. Every CDF clock cycle, the FE charge integrator
acquired charge from the silicon strips and transferred it to an empty
cell in the analog pipeline. The FE was reset at every bunch crossing. 
Whenever a L1 decision to keep an event arrived, the appropriate pipeline 
cell was marked and skipped over until it was digitized. Under normal 
operations, up to four pipeline cells could be marked.

A marked analog pipeline cell (capacitor) was digitized by the BE.
The voltage across the marked pipeline capacitor, subsequently referred
to as the {\em strip voltage},  went to the input of
a comparator. The other input of the comparator was a voltage ramp
shared by each channel's comparator.  At the start of digitization,
the common voltage ramp started and an 8-bit grey-code counter started 
to increment. A channel's comparator changed state as soon as the
voltage ramp was larger than its strip voltage, which triggered a
latch to store the current value of the 8-bit counter. Therefore
channels with smaller strip voltages would trigger their latches earlier.
The common pedestal subtraction was implemented by delaying the
counter until the first 33 comparators had changed state. The choice
of 33 channels came from a study to optimize pedestal removal with
signal efficiency. DPS implicitly assumes a constant
chip-wide pedestal\footnote{This assumption was not true for L00 and
  DPS was turned off for L00 readout (Section~\ref{DAQ:L00Noise}).} 
and insulates the strip charge measurement from
environmental noise. 

The SVX3D implemented three modes of data reduction: read
out of strips above a set threshold (\emph{sparse}), sparse strips and
adjacent below-threshold strips (\emph{nearest neighbor, NN}), and no
data reduction (\emph{read-all}). Sparse mode had the smallest data
volume and therefore smallest readout time but NN was chosen as it 
allowed addition of neighbor strips below threshold. Additional 
information provided in the NN mode also proved useful for correcting 
single bit errors. Except in the read-all mode, the data volume was 
driven by the underlying physics that drove the occupancy of the 
silicon strips. 

Multiple chips were chained together to read out a single
silicon sensor; two for the innermost (narrowest) layer and up to 16
for the outermost (widest) layer.  Chip initialization and commands
were transferred serially from the first chip in the chain to the last
chip in the chain. The first strip of the first chip and last strip of
the last chip in the chip chain were always read.  Data from each chip
in the chip chain were transmitted one-by-one on a common data bus. The
data from each chip included the chip ID and the SVX3D channel number
of each read strip. There were 5644 SVX3D readout chips
in the CDF silicon detector which dissipated approximately 3\,kW of
power, thus active cooling was essential for stable operations
(Section~\ref{sec:cool}).

\subsubsection{Portcard}
The portcard was the interface between the on-detector and
off-detector electronics. It relayed SVX3D commands and trigger
signals from the DAQ boards and passed on power to bias the silicon
sensors and chip chains from the power supplies
(Section~\ref{sec:ps}). Data from the chips chains passed through the
portcard onto the optical data links (DOIMs)
(Section~\ref{SECTION:DAQ:DOIM}).
As the portcards were located within the silicon detector, they were
designed to have low mass to minimize the radiation length, 
to withstand radiation doses
of up to $\sim$200\,kRad ($\sim$10\,fb$^{-1}$), and with high heat
transfer capability~\cite{portcard}.

\subsubsection{Optical data link: DOIM}~\label{SECTION:DAQ:DOIM}
The Dense Optical Interface Module (DOIM) was the optical data link
used to transmit data from a chip chain to the off-detector
electronics~\cite{Hou2003166}. Each DOIM had a transmitter unit (TX)
located on the Portcard and a receiver unit (RX) in an off-detector
VME-transition module (Fig.~\ref{fig:daqschema}). The TX and RX were
connected by optical fibers. Each DOIM was capable of transmitting
8-bit wide data at 53\,MHz with an error rate of less than 1 in
$10^{12}$ words.
 
The DOIM TX housed twelve 1550\,nm InGaAsP edge emitting
lasers in a single package. Only 9 of the 12 were used: 8 to transmit
data and one as a data-strobe. The DOIM RX was a
InGaAsP/InP PIN diode array. It received the optical signal from the
TX and converted it back to an electrical signal. The DOIM TX was
tested for radiation hardness with 30\,MeV, 63\,MeV, and 200\,MeV protons 
with radiation doses up to 2\,MRad. The light degradation was measured to
be 10\% after 200\,kRads~\cite{chu}.

\subsection{Off-detector electronics}~\label{DAQ:offdetector}
The off-detector electronics were responsible for coordinating the
silicon DAQ process as well as processing and packaging the digitized
silicon strip data for the SVT and the CDF DAQ. 

The off-detector electronics were housed in eighteen 9U VME crates
using VME64~\cite{VMEStandard}. Of these eighteen crates, eight were
located in the CDF collision hall close to the CDF detector, while the
other ten were located in a counting room in the CDF assembly building
(Fig.~\ref{fig:daqschema}). The main difference between the two sets
crates was the use of two different custom J3 backplanes to 
accommodate different types of boards.  All together, there were 164 
VME boards. 

Because the data from the SVX3D chip chains were sparsified, the 
first strip of the first chip and last strip of the last chip of the
chip chain were always reported in order
to identify the start and end, respectively, of the chip
chain data stream.  The off-detector electronics appended its own
unique header to these data.  The combination of the header, chip ID,
and SVX3D channel encoded the unique location within the silicon
detector of each digitized charge.

\subsubsection{Silicon readout controller (SRC)}
The SRC was the master controller of the silicon detector DAQ and also
acted as the interface to the CDF DAQ and trigger systems. The SRC was 
housed in a
rack in the CDF counting room, which was also shared with the VME
readout buffers (VRB), and received the clock and the beam structure
from the central CDF clock fanout. It communicated with the Trigger
Supervisor (TS), which was the central CDF trigger processor. The SRC
also provided the central clock to the entire CDF silicon detector,
which was kept in sync with the CDF clock using a phase locked loop
(PLL).  The SRC commands, clock, and trigger signals were transmitted
by the SRC via a Transition Module (SRCTM) to the Fiber Interface
Board (FIB) crates in the CDF collision hall using a
GLINK~\cite{glink} optical link running at 53\,MHz.

The Silicon DAQ was originally designed to be driven by a single
SRC. However, the need to read out all channels of L00 every event
(Section~\ref{DAQ:L00Noise}) 
 required two
SRCs, one to drive SVX-II and another to drive ISL and L00
(Section~\ref{DAQ:TwoSRC}).
Implementation of the second SRC also helped mitigate the \emph{wirebond
  resonance} problem (Section~\ref{DAQ:Wirebond}).

\subsubsection{Fiber interface board (FIB)}
The Fiber Interface boards (FIB) were housed in eight crates located
in the four corners of the CDF collision hall. SVX-II and ISL/L00 had
four FIB crates each. The signals from the SRC were received by a FIB
Fanout (FFO) board in each FIB crate and distributed to the FIBs in
its crate via a custom J3 backplane. Each FIB communicated with two
portcards via a FIB Transition Module (FTM) on the backside of the FIB
crate.  It converted the high-level SRC commands into a sequence of
instructions suitable for the SVX3D chip chains, which were sent with
clock and trigger signals to the two portcards. The FTMs also housed
the DOIM RX that received the digitized SVX3D data, which were passed
to the FIB.  The FIB formatted the data stream, appended its own
unique header, and sent the data on four GLINKs to the VME Readout
Buffers (VRB) with a copy sent to the SVT through optical
splitters.

\subsubsection{VME readout buffer (VRB)}
The VME Readout Buffers (VRB) were located in the VRB crates in the CDF
counting room. Two VRB crates also housed the two SRCs. The VRB
buffered the data from the FIBs until a L2 decision was made by the
CDF trigger system, upon which the event was moved to the output
buffer and was collected by the Event Building system using the VME
Bus. The communication between the VRBs and the SRC was handled by
the VRB fanout system, which enabled the SRC to manage the buffer
provided by the VRBs.
 
The data from each VRB crate were transferred in parallel to the
event builder, which combined segments from the crates 
into an event record which was then passed to L3.
The SVX-II had 6 VRB crates that corresponded to the 6
SVX-II bulkheads.  The ISL and L00 originally had two and one VRB
crates, respectively. 
To cope with high instantaneous
luminosity (above $10^{32}~cm^{-2}s^{-1}$), 
it was necessary to reduce the size of the data segments
arriving from the VRB crates. In 2006,
the ISL and L00 VRB crates were mixed
and an additional VRB crate was added (Section~\ref{DAQ:LoadBalance}).

\subsection{DAQ commissioning}
Prior to installation, the VME based hardware and onboard detector 
electronics were thoroughly tested through the use of test stands and data 
emulation at various levels to verify the functionality and robustness of 
these systems.  However, due to time constraints, there was limited ability 
to test the two systems together after installation. Together with the 
unforeseen consequences of the environment in the collision hall, this led 
to several problems emerging in the course of the first few years of 
operation that required immediate attention to alleviate data corruption 
and potential damage to the detector. The wirebond resonance and L00 noise 
problems were severe and are described separately in
Sections~\ref{DAQ:Wirebond} and \ref{DAQ:L00Noise} respectively. 
It took from 2001 to 2003 to fully commission the silicon detector. 

\subsubsection{SVX3D commissioning}~\label{DAQ:SVX3DExperience} 
The SVX3D chip was thoroughly tested during its development. But a
number of unexpected behaviors, listed below, were encountered during
commissioning. The chip would latch to a state where the chip current
increased until it exceeded the power supply safety limit and forced a
power supply shut down (trip).  These behaviors were circumvented by
modifying the SRC, VRB and FIB firmware.

\paragraph{Abort digitize} 
The SVX3D chip had a feature to abort digitization before completion
if L2 had already rejected the event. However this feature made the
chip enter the high current state and trip. The SRC and FIB firmware were
modified to allow the SVX3D chip to always complete digitization which
stopped these failures.

\paragraph{Fifth L1 accept}
The SVX3D chip could accommodate up to four L1A signals without
releasing a cell in the analog pipeline (Section~\ref{sec:svx3d}). If
a fifth L1A arrived before a pipeline cell was released, the chip
would transition to either read-all mode or suppress all readout. 
The SRC firmware kept track of the number of unreleased pipeline cells, and 
was modified not to send the fifth L1A to the chips, and instead
to send an error signal back to the CDF DAQ.
This error signal forced a silicon CDF DAQ reset and re-synchronization.

\paragraph{Keep-alive}~\label{DAQ:KeepAlive} 
At least every 270\,$\mu$s, a command had to be sent to the chip
chains to prevent chips from entering the high current state and tripping.
The SRC firmware was updated to send these \emph{keep alive} signals
every 270\,$\mu$s in the absence of any commands. But the SRC state
machine was driven by the CDF clock and any glitches or interruptions
of this clock would also delay or interrupt the delivery of these
keep-alive signals, which would result in large portions of the
chips tripping off.  Given the sensitivity of the silicon detector to
any clock glitches, administrative procedures were implemented
requiring permission from either the CDF silicon detector project leader 
or the head of CDF detector operations before work was done on 
on the CDF clock or Tevatron clock.

\paragraph{AVDD2 errors}~\label{DAQ:AVDD2} 
There was a class of unrecoverable failures that affected 6\% of 
the SVX3D chips and could be reproduced only by
disabling one of SVX3D's analog voltage lines (AVDD2). The observed
symptoms were loss of communication with the FE, an increase in the
SVX3D's BE current, and loss of communication to chips beyond the
affected chip. This class of failure typically occurred after a
beam incident (Section~\ref{sec:beam}) or a large temperature change,
such as a cooling system failure. This type of failure became infrequent 
after 2003 (Figs.~\ref{fig:daq_svxChipAcct} and \ref{fig:daq_islChipAcct})
when operation procedures during shutdowns were changed to minimize
thermal cycles, coincident with a sharp decline in the frequency and 
severity of beam incidents.

\subsubsection{DAQ board enhancements}
Commissioning of the silicon detector, the SRC, VRB and FIB firmware 
were extended to circumvent the unexpected behaviors of the SVX3D readout 
chip, which are documented in Section~\ref{DAQ:SVX3DExperience}. In 
addition, minor problems appeared when the off-detector and on-detector 
components of the system were integrated. A few of these issues are 
described in detail to illustrate the type of problems encountered and 
solutions implemented.

Data were lost due to failures in the transmission of the clock signal 
from the FIB/FTM to the portcard.
Electronic components on the fiber interface input of the FIB were
replaced to increase the tolerance of varying duty cycles on a signal
that carried the clock information. The clock rate was known,
and failures in the transmission of its signal were overcome
by providing an identical backup clock signal. 
The firmware was also updated to increase
the allowed width of the front-end clock from about 28\,ns to about
34\,ns in order to avoid the loss of charge collection due to
inadequate integration time.  

As the readout was data driven with no fixed length, the FIB 
used the last channel of the last chip in the chain to identify the end
of a chip chain's data stream. Failure to detect this, which
could be caused by the chip, DOIM, or FIB error, could potentially
make the FIB wait for an indefinite amount of time. A timeout was added
to the FIB to terminate readout and append an error code to the data stream.

Data were also initially lost to a race condition in the data concatenation 
algorithm at the VRB level.  This condition shifted every other 4 bits 
in the data stream by 8 bits, leading to data corruption and the loss of 
events at the 1\% level.
Once the systematic shift was distinguished from random corruption, 
the VRB firmware was modified to eliminate this source of  data loss.

\subsection{Wirebond resonance: spontaneous loss of \rz\ sides in the double-sided ladders}~\label{DAQ:Wirebond}
Shortly after the beginning of data taking operations in 2002, 4\% of
the \rz\ side of SVX-II ladders were spontaneously lost during
operations.  In the SVX-II, the \rphi\ and \rz\ side hybrids of the
ladders were connected with a set of wire bonds, known as the
{\em jumper}. The jumper was perpendicular to the 1.4 T magnetic field
produced by the CDF solenoid (Section~\ref{sec:intro}). On every
readout sequence of the chips, a varying current flowed through the jumpers, 
which resulted in a Lorentz force that induced a kick on the
jumpers. This process usually did not lead to a resonant
condition, as the readout commands were typically randomly spaced.
However if the readout commands came at a fixed frequency, a resonant
Lorentz force could cause the wire to break from mechanical
fatigue. It had been shown that some resonant frequencies of the jumpers
were in the 10\,kHz range (which exactly matched the CDF L1A trigger
rate under certain conditions) and only a few kicks were necessary to excite the
jumpers~\cite{Bolla2004277}.  These resonant readout conditions
would arise when there were synchronous L1As from calibration
triggers, faults in the trigger hardware, and ladders with large and fixed
length readout. The silicon detector was removed from all calibration
triggers and faulty trigger hardware was replaced. L00 had large fixed
readout, discussed in section~\ref{DAQ:L00Noise}, and its 
separation from the SVX-II readout (Section~\ref{DAQ:TwoSRC}) was 
necessary to mitigate the resonances.

\subsubsection{Operational mitigation}
During the initial investigation, it was understood that the damage was 
correlated with the L1 trigger rate.  The maximum L1 trigger rate was set 
to 20\,kHz, where the typical peak rate was about 12\,kHz. A limit was 
implemented in the DAQ software, called the \emph{trigger handbrake},
that would halt data taking if the four-second-average of the L1 trigger 
rate exceeded the maximum rate.  After the wirebond resonance was 
discovered and the Ghostbuster protection system was commissioned, the 
maximum rate was raised to 35\,kHz. In addition, an administrative 
procedure required that every change to the trigger system was tested 
without the silicon detector and signed-off by the silicon operation group.

\subsubsection{Ghostbuster protection system}
Given that it only took a few kicks at a $\sim$10\,kHz resonance
frequency to excite a resonance, the \emph{Ghostbuster}~\cite{daq:gb}, 
already developed for SVT, was reprogrammed to detect the
onset of a resonance condition within $\sim$1~ms. The Ghostbuster paused
data taking as soon as a series of synchronous readout commands had
been detected. 
The FFO was modified to send readout commands to the Ghostbuster. 
The addition of the Ghostbuster was essential to allow CDF
and the silicon detector to acquire data at the highest L1
rates. Without the development of this board, the CDF physics program
would have been severely limited. After the introduction of the
hardware protection system, losses of the \rz\ side in ladders due to
resonant conditions were eliminated, except for two cases in 2005 and
2007 (Fig.~\ref{fig:daq_svxChipAcct}). After the commissioning of the
Ghostbuster, the trigger handbrake remained as a redundant 
limit on the L1A rate.

The Ghostbuster algorithm paused data taking when it appeared that
the timing of readout commands was within a narrow frequency band.
The Ghostbuster recorded the time interval between successive
readout commands.  A difference in successive intervals of less
than 1~$\mu$s was counted as a {\em tick}.
If the difference in successive interval lengths was greater than
1~$\mu$s, the tick counter was reset.
A resonance error was declared when the tick counter reached
a preset threshold, typically set to 11.
The threshold value was initially determined from a Monte Carlo simulation 
of the DAQ, and tuned to running conditions when necessary.
There was always a non-negligible chance that a set of consecutive 
random L1 triggers would look
like a resonance, and the threshold set point was a compromise
 between detector safety and limiting false resonance alarms.

\begin{figure}[htb]
\centering
\includegraphics[width=0.65\textwidth]{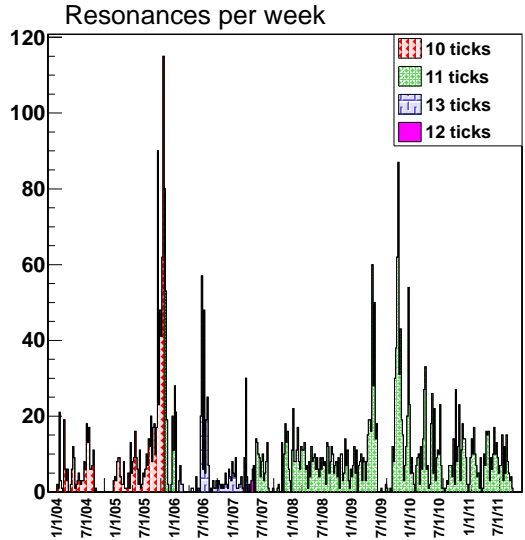}
\caption{
A historical account of the number of resonance per week detected by
the Ghostbuster. The maximum number of ticks, which triggered the
Ghostbuster resonance detection, varied from 10 to 13 ticks during Run
II, and eventually settled at 11. The spikes in the resonance rate are
mostly due to faulty trigger hardware. The typical resonance rate was
10 per week (1.4 per day). 
}
\label{fig:daq_svxReso}
\end{figure}

 Fig.~\ref{fig:daq_svxReso} shows the the number of
resonances per week detected by the Ghostbuster during all of Tevatron
Run II. The spikes in the
number of resonances per week were mostly caused by faulty trigger
hardware or long readout times for ladders.  The typical rate, neglecting 
those originating from faulty
trigger hardware, was about 10 resonance errors per week (1.4 per day)
consistent with stochastic operation.

As mentioned previously, any ladders with long fixed readout would
exacerbate the likelihood of resonances at some L1 rates. The readout
time could increase from corruption of the pedestal subtraction
algorithm or if the noise level had grown. In both cases, the
ladder readout time is no longer dictated by detector occupancy but rather
ladder noise. This could lead to approximately fixed long readout
times. Also some chips in a chip chain were switched to read-all mode
to fix some errors, which also increased the readout time. 
During 2009-2010, the noise level in a handful of ladders had grown large enough, and
consistent enough in length, that the number of resonances increased and gradually
forced the peak L1 trigger rate to be limited to 25\,kHz.  
The problematic noise growth was due to malfunctions in the chip or the sensor itself,
and was several times larger than the noise growth due to radiation damage observed
in most ladders.
The noise was often isolated to a few chips of the chain, and by
increasing the NN sparsification threshold of the affected chips, the noise was
suppressed without compromising the data from other chips in the ladder.
With this noise suppressed, the peak L1A rate was
increased to 32\,kHz counts during normal data-taking, without creating
resonances. 

\subsection{L00 noise}~\label{DAQ:L00Noise}
L00 was included to improve the precision of measuring displaced
vertices that was essential for the discovery for $B_{s}$
oscillations~\cite{cdfbsmixing}. Unlike SVX-II and ISL, the L00 readout chips
were not mounted directly on the L00 sensors to minimize the amount of
material and so reduce the effects of multiple scattering. Instead, a
fine-pitched cable connected the sensors to the readout chips.

After L00 installation, significant noise was observed on the L00
readout that manifested as large pedestals that varied across a chip
and with each event, with the largest variation at the edges of the
readout cables. These pedestals could not be
removed by DPS. An investigation concluded the
noise was picked up by the fine-pitched readout
cables~\cite{l00_tdr2}.  

L00 was forced to operate in read-all mode and the pedestals were
removed by an offline event-by-event correction.  
During the offline data processing, the recorded charge across a chip
was fit to Chebyshev polynomials to extract the pedestal. Tests using
simulation were performed to check for biases from fitting and none
were found~\cite{l00_tdr2}.        

\subsubsection{Two-SRC mode}~\label{DAQ:TwoSRC} 
The original silicon DAQ read out SVX-II, ISL and L00 together.  
A consequence of operating L00 in read-all mode was that it had the
largest data volume and was fixed length, which exacerbated the wirebond
resonances (Section~\ref{DAQ:Wirebond}). As L00 and ISL were not used by
SVT, they were separated from SVX-II DAQ and read out by a separate
SRC after a L2 accept. This improved the readout time and 
also mitigated the wirebond resonances.

\subsubsection{Load balancing}~\label{DAQ:LoadBalance}
During Tevatron Run II, the peak instantaneous luminosity
increased from $50 \times 10^{30}$\,cm$^{-2}$s$^{-1}$ to
$400 \times 10^{30}$\,cm$^{-2}$s$^{-1}$ and this had two consequences: 
higher detector occupancy and an increased trigger rate. Without 
improvements to CDF's
trigger and DAQ, CDF would not have been able to operate at this high
instantaneously luminosity. In particular, some data segments
arriving at the event builder were significantly larger
than others, and the event builder performance was 
limited by the largest of these data segments.
To mitigate this effect, the number of VRB crates was expanded for both 
silicon and non-silicon systems, and data rates were equalized across the 
crates.

As described in Section~\ref{DAQ:offdetector}, data from the CDF
silicon detector was buffered on VRBs until a L2 decision arrived.
Each VRB crate was read out in parallel and the total readout time was
dictated by the L00 VRB crate as it had the largest data volume. The
L00 VRB crate had a fixed event size of 28\,kB per event as a
consequence of its read all mode (Section~\ref{DAQ:L00Noise}).

The readout time was reduced by mixing ISL and L00 VRBs to balance the
data volume across VRB crates, thus reducing the peak data volume
per VRB crate. A configuration was found and implemented during the
2006 Tevatron shutdown. The maximum event size of a single crate
reduced from 28\,kB to 20\,kB. 

Despite the success of the load balancing, the increased instantaneous
luminosity of the Tevatron forced another re-optimization of the CDF
DAQ. For the silicon DAQ, balancing the load across VRB crates was no
longer sufficient. An additional VRB crate was added to the L00/ISL
readout and remained in operation until the end of Run II.

\subsection{Operational experiences and improvements}
The commissioning of the CDF silicon detector was completed
at the start of 2003 and the detector was included  safely in
normal data taking. This section describes the routine day-to-day
problems that persisted to the end of Run II. The issues that affected
the silicon DAQ on a daily basis were broadly categorized as effects
from single-event-upset and bit errors.
Procedures were developed to resolve common
problems but required constant vigilance by the detector operations
shift crew. Another issue, which persisted, was the full-detector
trips, where either subdetectors, or the complete silicon detector
would switch off. Although far rarer, typically four times per year, it
took 45-60 min to recover and resume data taking. Another large
component of detector operation was the daily maintenance of the 580
ladders and 5644 readout chips. With such a large number of
components, at least one readout chip and/or ladder required some daily
adjustment.

\subsubsection{Impact of ionizing radiation}~\label{DAQ:SEU}
A sizable fraction of the DAQ system was installed in the CDF
collision hall, which was subjected to radiation from the Tevatron's
colliding beams. As a result, the majority of electronics failure were
due to radiation induced single event upsets (SEU). A SEU is a change
of state caused by radiation striking a sensitive component in an
electronic device. The change of state is a result of the free charge
created by ionization in or close to an important logic element or
memory bit.

During Run II, the Tevatron substantially reduced radiation
rates, thereby reducing the radiation induced failures. At the end
of Run II, the rates were so low that they fell below the detectable
threshold of the radiation monitoring counters during the course of a
Tevatron store. Also, the introduction of the
Silicon-Autorecovery (SAR) in 2008 automated the detection and
recovery of these SEU failures of the DAQ and the power supplies. It
reduced interruptions to data taking from 10-20 min to less than
5 min.

\paragraph{Reinitialization of chip chains} 
During data taking, the current consumed by the analog FE of a SVX3D
chip chain (Section~\ref{sec:svx3d}) could spontaneously drop by
80-100\,mA, implying one chip in the chain was not
recording any data. This typically occurred at rate of once or twice a
day.  A program dedicated to monitoring the power supplies sent an alarm
to the DAQ if such a drop in chip current was detected, and the 
data taking was paused for less than a minute to
reinitialize the chip chain. Overall, this only had a minor impact on
data taking.

\paragraph{FIB bit errors and FIB FPGA burnout}
Each FIB had 16 FPGAs, and thus a higher rate per board of SEU damage
than the other VME modules.
Data corruption errors were resolved by reloading
the FPGA programs of the affected FIB, which typically occurred 1-2
times a day. On rare occasions (3-4 per year), one of the FPGAs would
enter a high current state and blow a fuse on the board.
In a majority of cases the affected
FPGA had to be replaced, and in the remaining cases, a reprogramming
of its firmware was needed.

\subsubsection{Bit errors in the data stream}~\label{DAQ:bit_err}
As the silicon DAQ is data driven, the data format had to be
self-describing to identify not only the amount of charge but also the
location where it was recorded. Therefore any
corruption of the data implied more than just errors in digitization.
During operations, several sources of bit errors in the data stream
were detected and immediately addressed. While some errors,
especially those from on-detector components, could not be repaired,
many could be corrected in the offline reconstruction. Operating
the SVX3D readout in nearest-neighbor mode guaranteed at least three
consecutive strips would be read, and the error-correction algorithm 
could exploit this feature to identify and correct single-bit errors.

\paragraph{Bit errors in the optical links}
One common instance of bit errors in the data stream was in
the DOIM system. At the start of Tevatron Run II, most DOIM bit errors
were traced to bad electrical contacts in the sockets that held
the RX in the FTM. Gold plating the pins
of those devices to establish a better connection with the sockets
eliminated this source of bit errors.

During Tevatron Run II, the typical DOIM bit errors manifested as bits
that were stuck low or high in the data stream, which corresponded to
either a malfunctioning RX unit or TX unit. Faulty RX units were accessible
and replaced from a pool of spares when necessary. Faulty TX units
were inaccessible and thus irreplaceable. Some TX failures could be
recovered by adjusting power supply settings to tune the TX unit light
output.  Another class of TX-related errors were linked to bad
connections in a circuit board which served as the interface between
the sensors and the power supplies.  These boards were located just
outside the tracking volume and were accessible only when the Tevatron
was shut down for at least a week.  In these instances a borescope and
a custom tool were used to push the circuit board back into place to
re-establish the electrical connection.

\paragraph{Bit errors in the readout boards}
The FIB occasionally caused bit errors in the data stream, most
commonly due to SEUs and component failures.
SEU related bit errors were resolved by reloading the programs to the
FPGAs on the boards. FIBs boards with failed components were replaced.
On rare occasions (less than once per year), VRB boards gave bit
errors that were traced to component or printed circuit boards
failures and were replaced.

\subsubsection{Full detector trips}
There were several incidents in which most or all SVX3D chips in
the silicon detector would go into a high current state leading to
power supply trips. These trips occurred in some or all of
the SVX-II, ISL, and L00 sub-detectors and occurred more frequently in the
winter season. On average, it took about 45-60 min to recover from
these incidents.

Only some of the sources of these trips were reproducible and the
remainder were hard to diagnose and resolve due to their rarity. While
the origin of all these trips was unknown, many potential causes
were identified and eliminated as sources of the problem. It was
observed several times that personnel working near the electronics
area could induce this problem, suggesting that loose or corroded
contacts may have been a source; the re-seating and replacing of many key
components proved inconclusive. It was also suspected that differences
in grounding levels between the racks could generate this problem, but
no evidence of a bad ground was found.  Another of these sources
was the corruption of the clock signal and consequent lack of
keep-alive commands sent to the chip, as detailed in
Section~\ref{DAQ:KeepAlive}. The underlying reasons for corrupted
clock inputs were not clear, and many full detector trips did not show
any indication of clock corruption.  

It was found that some of these trips did occur in coincidence with a
high-voltage power supply trip of one particular muon detector
chamber. The muon detector chamber was powered with 3500 V and had a
current draw of about 1 mA. The trip of its power supply during these
incidents was attributed to arcing between the high-voltage lead and
the chamber ground.  The hypothesis that an arc in a completely
unrelated subsystem would give rise to a massive power trip in the
silicon detector was tested by inducing an arc in the muon detector
chamber and observing the behavior of the silicon detector. The arc
was forced by closing in the high-voltage lead to the ground until a
spark was generated, and full subdetector trips in the silicon
detector were reliably reproduced each time the spark was induced. The
mechanism was believed to be electromagnetic pickup between the muon
chamber high voltage distribution cables and the clock crate during
the occurrence of the spark.  The electromagnetic field bursts were
observed by carefully-placed coils in the surroundings of the clock
crate. The pickup induced a change in the ground level of the clock
signal, which exceeded the specification for the silicon electronics,
resulting in an effective lack of clock signal during a period of time
of about a few microseconds; thus no keep-alives were sent during this
clock interruption. This problem was solved by reducing the voltage
applied to the defective muon detector chamber to 3200~\,V.

\subsection{Summary of DAQ performance}~\label{DAQ:DAQPerform}
The CDF silicon DAQ was very complex in order to meet the
challenges of providing data to SVT. Its performance during Tevatron
Run II was defined by three major phases: commissioning (2001-2003),
steady operations (2004-2008), end of Run II (2009-2011). The long
commissioning period was directly due to the large number of
unexpected problems that were apparent only after installation of the
silicon detector. The challenge of tackling these problems
simultaneously (in addition to a number of non-DAQ issues), while
attempting to take data simultaneously was considerable. However after
this commissioning period, the silicon DAQ entered a stable period and
efforts to optimize the performance and operations were carried out,
in addition to daily maintenance of the DAQ components. From 2009, the
dwindling pool of working spare components and the effects of
radiation were taking their toll. 

\begin{figure}[!htb]
\centering
\includegraphics[width=0.75\linewidth]{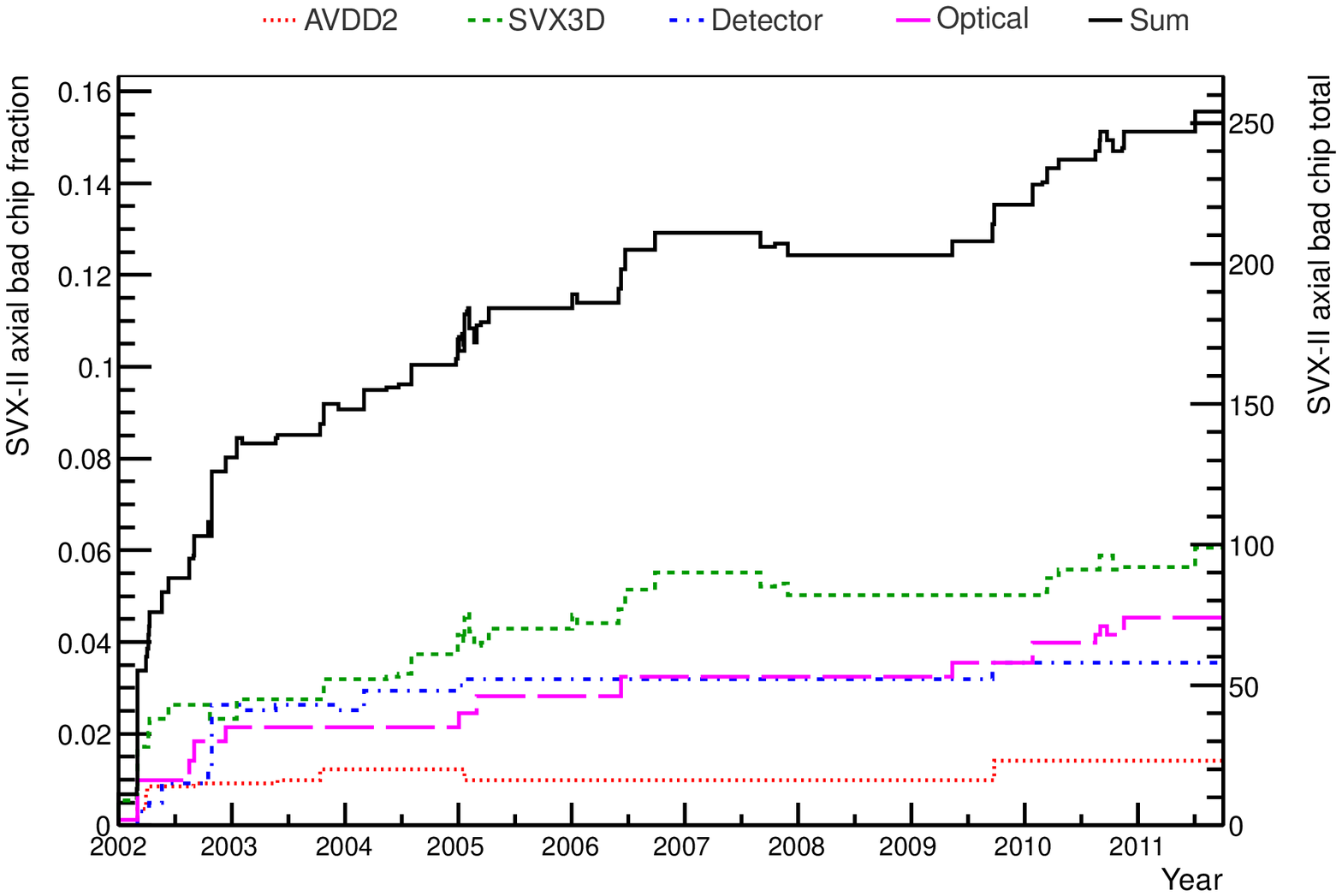}
\includegraphics[width=0.75\linewidth]{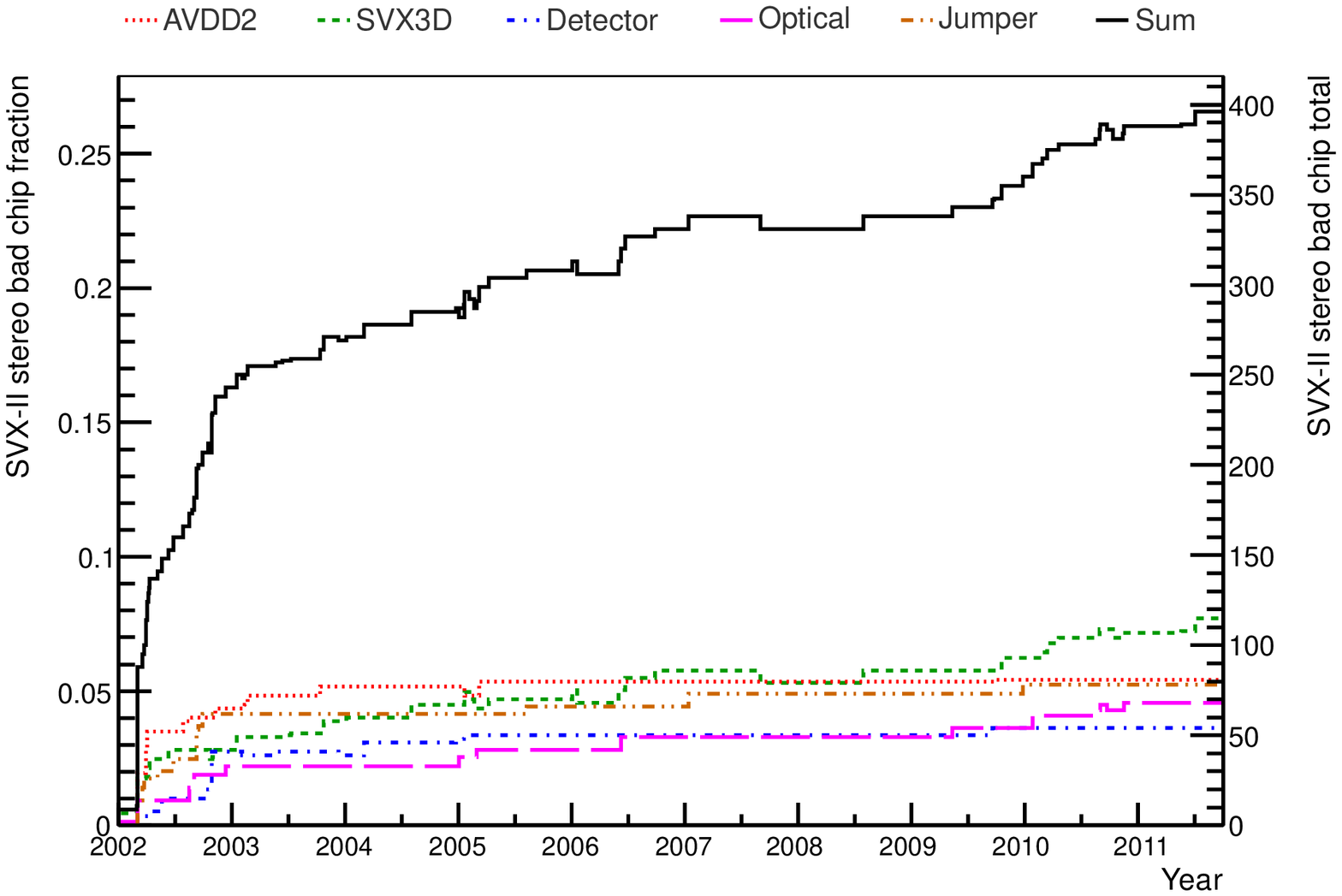}
\caption{ A historical account of the different errors and faults
  accrued by the SVX-II \rphi\ (top) and
  \rz\ (bottom) SVX3D chips during Tevatron Run II.
   Each error category is
  defined in Section~\ref{DAQ:DAQPerform}. At the end of Tevatron Run
  II, 84\% \rphi\ and 73\% \rz\ SVX chips were still operating
  without error.  }
\label{fig:daq_svxChipAcct}
\end{figure}

\begin{figure}[htb]
\centering
\includegraphics[width=0.75\textwidth]{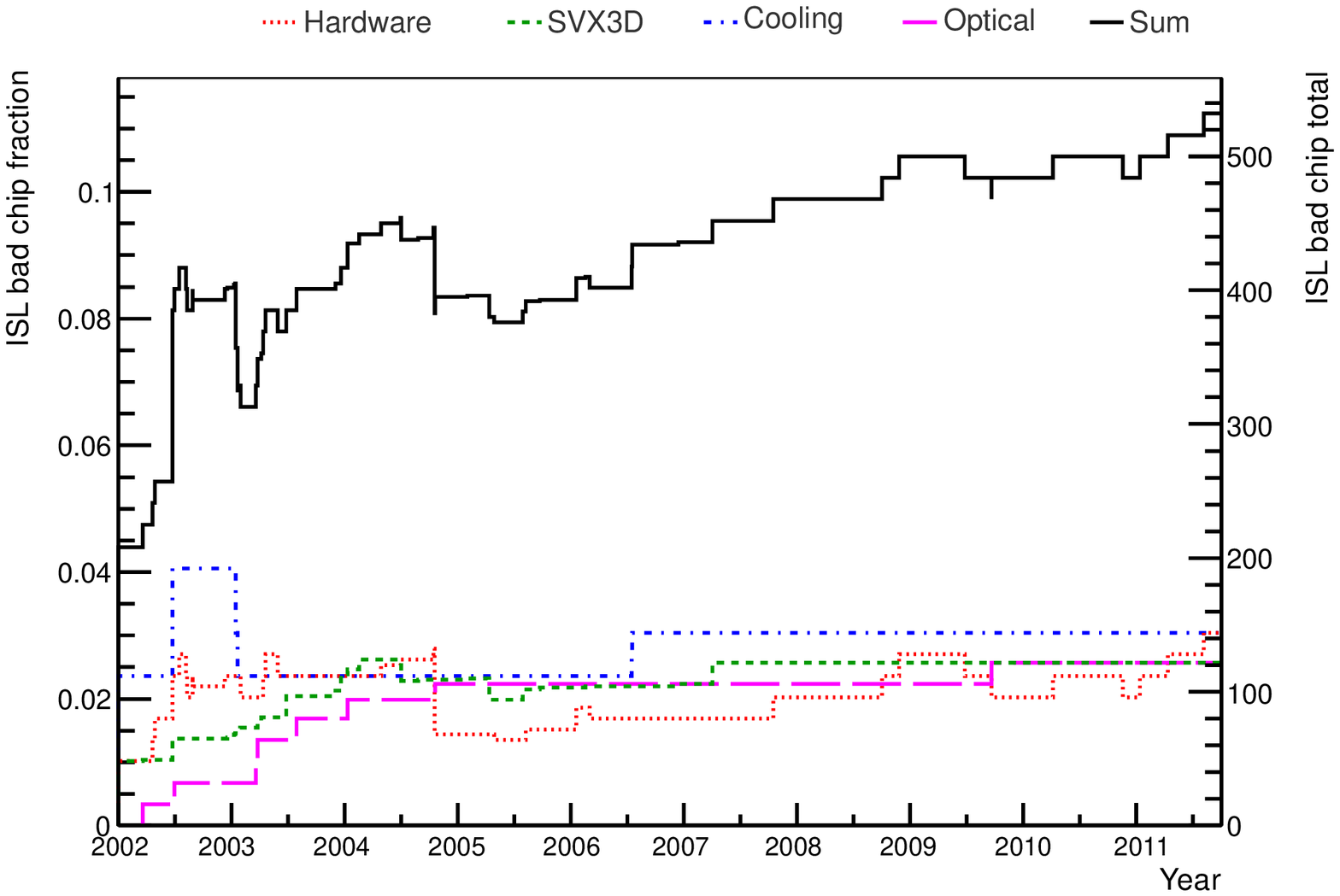}
\caption{ A historical account of the different errors and faults
  accrued by the ISL SVX3D chips during Tevatron Run II.  
  Each error category is defined in
  Section~\ref{DAQ:DAQPerform}. At the end of Tevatron Run II, 89\% of ISL
  SVX3D chips were still operating without error.}
\label{fig:daq_islChipAcct}
\end{figure}

Figs.~\ref{fig:daq_svxChipAcct} and \ref{fig:daq_islChipAcct} are
historical records of the different types of errors accrued by the
DAQ electronics during Tevatron Run II, expressed as the fraction of
bad readout chips. Fig.~\ref{fig:daq_svxChipAcct} shows the SVX-II
\rphi\ and \rz\ sides separately.  The definition of the different
errors are:
\begin{itemize}
\item \emph{AVDD2}: SVX3D errors diagnosed as AVDD2 type errors
  (Section~\ref{DAQ:AVDD2}) 
\item \emph{SVX3D}: SVX3D errors that are not AVDD2 type error
  (Section~\ref{DAQ:SVX3DExperience}) 
\item \emph{Detector}: Faults which originate in the silicon sensor 
\item \emph{Optical}: Errors which are from DOIM TX or RX
\item \emph{Jumper}: Ladders whose \rz\ side was lost from wirebond resonances
  (Section~\ref{DAQ:Wirebond}) 
\item \emph{Cooling}: Ladders turned off due to lack of ISL cooling
  (Section~\ref{sec:blockageISL})
\item \emph{Hardware}: Error and faults which do not match any of the
  categories defined above 
\end{itemize}

During the commissioning phase, the number of bad chips (ladder) grew as the 
different problem manifested, then stabilized in 2003 (steady state running).
The \rz\ plots in Fig.~\ref{fig:daq_svxChipAcct} shows that there were only 4 
additional jumper failures after the inclusion of the Ghostbuster and none of 
these was an immediate consequence of beam incidents or resonant conditions.
After the inclusion of the Ghostbuster, data were taken safely and reliably 
with peak L1 trigger rates in excess of 25\,kHz - essential for CDF physics.
The failure rate of chips after 2003 is far lower compared to the commissioning 
period. Most of the different failure categories stabilized. From 2009,
the chronic effects of radiation damage and aging were starting to take their 
toll on the silicon detector. The failures from SVX3D and optical were 
steadily increasing and half the total radiation dose was delivered between 
2009 and 2011. With the increased radiation, components started failing more often, 
shrinking the pool of spare components. These plots also highlight other 
problems of the silicon detector that affected the DAQ. 
Fig.~\ref{fig:daq_islChipAcct} shows a large rise and fall in the fraction 
of bad ISL chips during 2003 due to blocked cooling line and its eventual 
clearance; this is discussed in further detail in Section~\ref{sec:cool}.

At the end of Run II, 84\% (73\%) of all SVX-II \rphi\ (\rz) and 89\% of all 
ISL SVX3D chips continued to function without error and did not compromise the 
silicon detector tracking performance. This is an impressive feat as the CDF 
silicon was designed to be replaced after the first 2-3 fb$^{-1}$, about three 
years. It survived four times the radiation dose and lasted three times longer 
than the original design.


%
%
\section{Power supplies}
\label{sec:ps}
The CDF silicon detector used power supply modules manufactured by
CAEN.  A total of 114 custom modules were housed in 16 SY527
mainframe crates located in the corners of the CDF collision hall.
The crates were elevated 2-7 m off the floor due to space
constraints in the collision hall.  This location had a distinct
disadvantage; the supplies and crates were continuously exposed to
radiation, which not only shortened the life of many internal
electronic components, but also resulted in single event upsets that
required a crate reset and in single event burnouts that necessitated
additional hardware protection for the detector.  Investigation or
replacement of a problematic power supply required access to a
radiation controlled area for 1-2 h.

\subsection{System overview}
One power supply module provided low voltages (2~V and 5~V) to the
portcard, low voltages (5--8~V) to the SVX3D chip chains, and high
voltage (up to 500~V) to bias the sensors of one wedge of the silicon
detector.  The low voltages were set via potentiometers on the side of
the supply, while the high voltage was set via software on the SY527
crate.  All channels had a maximum voltage setting. If the channel
voltage exceeded its maximum, an ``overvoltage'' error was met and the
power supply cut power to the affected channel. These limits were also
set via potentiometers on the modules. Any adjustments to the low
voltage or the maximum voltage settings had to be done with the supply
inserted into the crate and the crate powered. Because of this need
and the location of the potentiometers, changing these settings were
not possible in the collision hall, rather it had to be done on a test
stand, using a specially modified crate.

The SY527 crate had an RS-232 interface for connection to a computer,
as well as a Lemo input to reset the crate remotely.  In addition, the
crate had a proprietary serial communications port, which connected to
a V288 high speed CAENet VME Controller. The V288 connected to another
controller card in the same crate which had an ethernet port for
external communications.  A VME crate was used 
to communicate with all the crates in the collision hall this way, and
a Java program provided a graphical interface of the power supply
controls. A PC monitored and logged parameters, including power supply
voltages, currents, and the time of last communication for each power
supply crate. This PC logged parameters for the cooling system as well
(see Section~\ref{sec:cool}).

\subsection{Decreasing low voltages}
\label{subsec:droopylv}
In 2005, after over 4 years of data taking, low voltage channels of 
several power supplies were found to
drift erratically from their nominal values. Specifically, filter 
capacitors in a particular regulator circuit of each low voltage channel 
gradually lost their capacitance, causing the low voltage supplied by the 
circuit to drop over time, typically a tenth of a volt over the 
course of three months. Left unchecked, this could cause the readout 
electronics to stop working.

The solution was to replace all of the filter capacitors (32 in total)
on the supply. This repair was done on-site at FNAL, and typically
only took a few hours. The swaps were usually not urgent (the readout 
electronics would still work at $0.5$~V below nominal), and were done
when another problem in the detector required an access to the collision hall.
\begin{figure}[htb]
\centering
\includegraphics[width=0.65\textwidth]{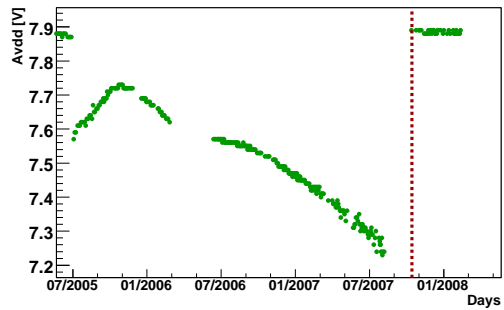}
\caption{
An analog low voltage line powering a series of SVX3D chips of an SVX-II 
ladder is seen dropping gradually over time. The vertical 
dashed line marks the replacement of the faulty power supply.
}
\label{fig:ps_LvDrop}
\end{figure}
Fig.~\ref{fig:ps_LvDrop} shows the analog low voltage channel of an SVX-II 
ladder drifting over a period of two years. The vertical dashed line marks the 
time when the faulty power supply was replaced, restoring the low voltage to 
its nominal value. The gaps in the plot correspond to Tevatron maintenance 
shutdowns. During the 2007 (2009) shutdown 47 (21) supplies out of a total of
112 were repaired and replaced. Most of the remainder were replaced 
gradually a few at a time utilizing the spare pool in hand.

\subsection{Single event upsets}
\label{subsec:psradvulnerable}
A consequence of the power supplies being located in the
radiation environment of the collision hall was that both the crates
and supply boards were subject to single event upsets. This typically
only required the crate to be reset.  Even so, every reset cost a
few minutes of data acquisition time.

The SY527 crates had a Lemo input for remotely resetting the crate;
the reset cable ran from the crate to the counting room outside the
collision hall.  Thus, the shift crew could reset the crate when
necessary, although the procedure, from identification of a problem to
manually turning the supply on again, took about 10 min.  In order
to eliminate any delay due to human intervention, 
an automatic system was developed to detect when a particular crate
has stopped communicating its voltage and current readings.  Once this
state was detected, the automatic reset system sent the reset signal
to the crate. The supplies had to be turned on again once the crate
had rebooted.  Recovery from an automatic crate reset was automated
with the development of Silicon Auto-Recovery, described in
Section~\ref{subsec:SAR}, which reduced the downtime from such
incidents from 10 min to less than 5 min.

The supplies themselves were also subject to radiation induced
effects.  Specifically, certain power metal oxide semiconductor field
effect transistor (MOSFET) components of the L00 supplies underwent
single event burnouts (SEBs)~\cite{ps:seb}, causing the supply to
output its maximum bias voltage.  If this were to happen with the
detector connected, it could result in permanent damage to the silicon
sensors. Fortunately these SEBs were first observed during
commissioning, before the detector was connected.

In order to prevent potential damage, compact voltage fuses, called 
{\em crowbars}, were developed and installed.  The crowbars interrupted 
the current if the bias voltage exceeded a specified voltage. They were 
placed between the high voltage detector cable and the supply itself.
The initial crowbars protected the sensors from bias voltages above 150~V.  
In 2008, when radiation damage necessitated operating at larger bias voltages, 
new crowbars with a voltage limit of 450~V were installed.

\subsection{Operational experience}
Ten years of experience in operating the power supplies  helped the 
detector operation crew identify potential problems and react to them before 
they caused significant downtime. Many improvements were made,
mostly to the procedures used to test supplies after repair or work 
in the collision hall.

\subsubsection{Testing procedures}
All the repaired supplies were run through a series of tests, designed to mimic 
the operating conditions in the collision hall, before their installation in 
the detector. These included long periods of being powered on (burn-in), as 
well as rapid power cycling. The former were crucial to detecting intermittent 
problems.

The testing procedure consisted of connecting the supply to a set of
static impedance loads (a loadbox), and turning the supply on for
approximately 24 h. It was followed by a test that turned the
module on and off every 2 min for approximately 24 h. The currents 
and voltages read out in each cycle were analyzed to ensure stability.

If a module was forcibly switched off for its protection (``tripped'') during 
the first 24 h or any currents or voltages were
unstable during the second, the module was sent for additional
inspection or repair. The safety features of the module were also tested
to verify it tripped properly under limiting conditions of voltage and
current and when the supply enable signal was absent.

In addition, the power supplies were tested in the collision hall just before
their deployment in the detector. The procedure to check out a supply in the 
collision hall consisted of connecting it to a loadbox and turning it on. 
Currents and voltages as read back from the module were recorded, and voltages 
were measured and compared at the loadbox. The safety features were tested 
again to make sure that the supply tripped off at intended voltage and current 
conditions. Then, the supply was connected to the sensors, and turned on. Given 
the location of the crates, the RS-232 interface provided the optimal method 
of testing, as it allowed easy control of the module using a laptop on the 
collision hall floor.

The loadbox was made by the Computing Division ESE at FNAL and
incorporated different load resistances for L00 and SVX-II/ISL supplies
with capability to switch between them. It utilized an ADC to
automatically measure the voltages on each channel. With the loadbox,
the collision hall check-out procedure was quick, reliable and
reproducible.

\subsubsection{Damage to power supplies during transportation}
When power supply modules could not be repaired on site, they were
sent back to the manufacturer in Italy. Shipping the modules back to
Fermilab damaged more than half of the shipped power supplies.
In 2008, a short study was done to examine the shipping
method adopted by the manufacturer and implement 
improvements to prevent future breakage.

In order to streamline the transit and prevent breakage, a set of
procedures was devised for the manufacturer to follow when shipping
supplies back. First, instead of routing via CAEN's business office in
New York, the modules were required to be shipped on a direct flight
from Italy to Chicago.  Second, the supplies were to be shipped on a
standard shipping pallet, requiring the use of a forklift to prevent
the box from being mishandled.  Finally, an accelerometer inside the
box and shock/tilt sensors affixed to the outside of the box were used
to monitor the detailed motion the supplies were subjected to while
in transit. Using these procedures, every supply that was shipped back
to Fermilab arrived in working order.


%
%
\section{Cooling system}
\label{sec:cool}

\subsection{System overview}
The sensors had to be kept chilled at all times in order to prevent
migration of radiation-induced defects, which shorten the working
lifetime of the detector, and to reduce the sensor leakage current,
which increases with radiation damage.

\begin{figure}
\centering
\includegraphics[width=0.9\linewidth]{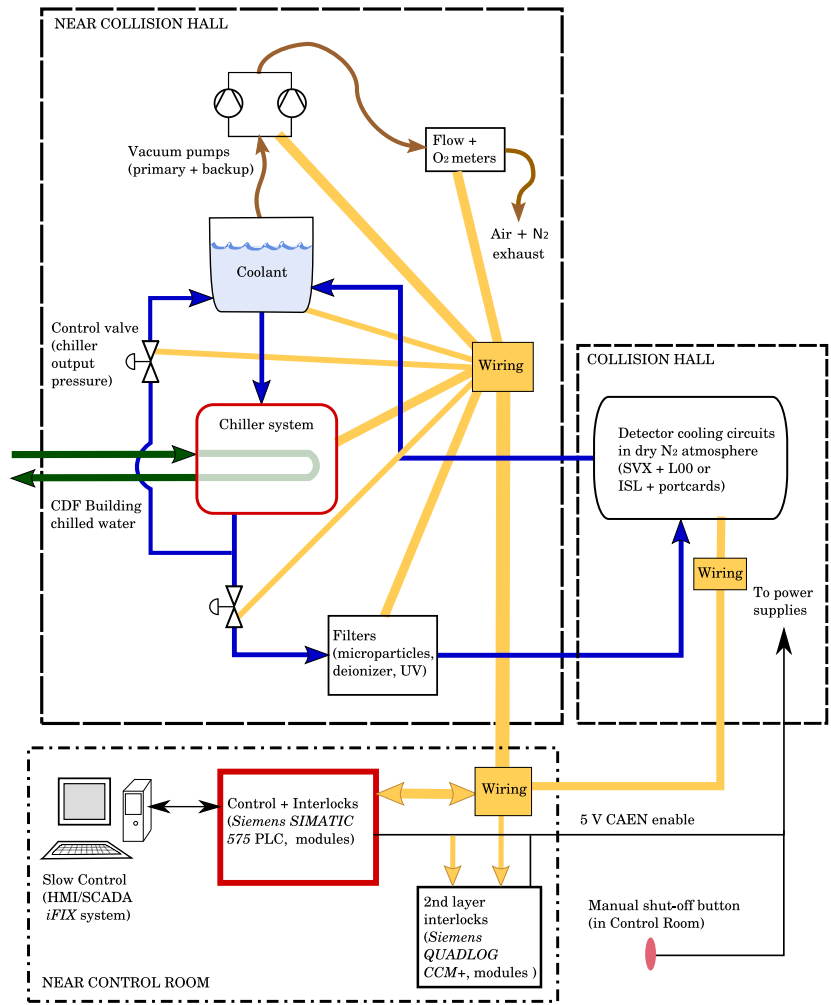}
\caption{Schematic of the SVX-II/L00 cooling subsystem. The subsystem 
for ISL can be described in the same manner, though both subsystems share the 
same PC and control/interlocks crates. 
\label{cooling_diagram}}
\end{figure}

The SVX-II, ISL, and L00 detectors were cooled by two closed-circuit
liquid cooling systems.  One system, shown in
Fig.~\ref{cooling_diagram}, was used to cool SVX-II and L00. A second
separate system was used for ISL, although some of the cooling control
electronics and interlocks were shared.

Due to the higher radiation levels close to the beam line, the operating
temperature of L00 and SVX-II was chosen to be lower than ISL.
The temperature of the coolant out of the chiller was $-10$\degreesC~
for SVX-II/L00 and $+6$\degreesC~ for ISL. The coolant for SVX-II/L00
was a mixture of 30\% ethylene glycol and 70\% water by weight, and
the coolant for ISL was distilled water. Both coolants warmed up by
a degree or two in the piping between the chillers and the detector.
The total cooling load for SVX-II/L00 during operation was
approximately 5~kW, and that of ISL is approximately 4~kW.  About half
of this heat load was produced by the silicon detector and the
remainder was heat transferred from the warmer surroundings.  Most of
the heat generated by the silicon detector was produced by the SVX3D
readout chips.  All of the portcards were cooled by the ISL system 
because the temperature of the ISL system was within the range of the
optimal performance of the light transmitters.  In order to prevent
water from condensing on the sensors and electronics, dry nitrogen 
flowed through the silicon detector volume.

In the SVX-II ladders, the electrical hybrids were cooled through thermal
contact to beryllium support bulkheads, with integrated cooling channels.
For L00 and ISL ladders, and for the portcards, cooling was achieved
through thermal contact to aluminum tubes glued to the mechanical supports.
The SVX-II and
ISL sensors were not in close thermal contact with the coolant,
however, nor were their temperatures directly monitored.  Based on the
measurements of the ambient temperatures and the bulkhead temperatures
and on thermal models, we estimated the temperatures of the sensors while the 
detector was powered to
be between 0 and 10\degreesC~for SVX-II, and between 15 and
25\degreesC~for ISL. The L00 sensors,
however, were in close thermal contact with the coolant tubes, and
their temperature was about -5\degreesC~when powered.

In order to prevent damage to the electronics if a coolant pipe
leaked, all cooling pipes were operated below atmospheric pressure so
that a leak in a pipe or a fitting would draw nitrogen into the
cooling system rather than leak coolant into the detector volume.  The
cooling system had vacuum pumps and air separators in order to
maintain the sub-atmospheric pressure in the system at all times.  Two
vacuum pumps were available per system, with one running and one piped
in as an immediately available spare, switchable with electronically
controlled valves.

The coolant was circulated by pumps on the two chillers: one for
SVX-II/L00, and one for ISL.  A third chiller was available
to be used as a spare in case of failure of either chiller.  
This system was used successfully in November 2004, when the spare
chiller was used for two days while a leak in the ISL chiller
was identified and fixed.

Each of the SVX-II/L00 and ISL chiller circuits had a set of filters --- a
microparticle filter, a UV sterilizer to limit biological activity in
the coolant, and a resin-cartridge de-ionizer.  The conductivity of
the coolant was regulated to be approximately 0.6~$\mu$S/cm.  The pH of each
of the two coolants was monitored via weekly samples drawn from the air
separators and both stayed near a pH of 6.

To ensure the safety of the silicon detector, a series of interlocks
prevented the power supplies from being turned on when insufficient
cooling was available, and coolant flows were switched off if an
unsafe situation existed.  If the pressure in any of the cooling lines
rose to within 1~psi of atmospheric pressure, electrically
controlled solenoid valves on the coolant supply lines to the affected
detector subsystem were shut automatically.  If dewpoint sensors
detected that the dewpoint was within $3$\degreesC~ of the minimum
temperature in the detector volume, then flows were shut off.

The interlock system had the ability to disable power to the detector.
The CAEN power supply modules required a voltage of +5~V in a Lemo
connector in order to enable detector power. Dropping of this +5~V
had the effect of switching the power off.  If there was insufficient
flow of coolant to the detector or if temperature sensors indicated
that a coolant temperature was too high then the detector power
supplies are turned off via the +5~V control lines. The electronics
crates that housed the CAEN modules monitored the temperature of the
electronics and the status of the crate fan pack, and would shut off
if a failure was detected.

These interlocks were controlled by a Siemens SIMATIC 575 Programmable
Logic Controller (PLC)~\cite{Siemans575PLC}, attached to two crates
containing modules that read out temperature, pressure, and flow
transducers. A third crate provided control for the solenoid valves,
the vacuum pumps, the chillers, and the 5~V power supply enable lines.

A second layer of interlocks was provided by a Siemens QUADLOG CCM+
PLC~\cite{SiemansQuadlogPLC}.  This controller had an independent
readout of the coolant flows and pressures and also monitored the
state of the power supply interlocks.  If power supplies were
permitted to be turned on but coolant flows were too low ($<1$~LPM)
--- which could only happen if the first PLC's interlocks had failed
--- this interlock system would turn off the power to the silicon
power supply and FIB racks (section~\ref{sec:daqtrig}).  The QUADLOG
system also protected against over-pressure conditions --- again in the
event of a failure of the first PLC system --- by shutting down
coolant flows.

During a power outage, backup systems kept the cooling system
functioning at a reduced level.  Power was supplied to the PLCs and
control electronics from an uninterruptible power supply (UPS), which
was backed up with a diesel generator. The vacuum pumps were powered
by the diesel generator but not the UPS, and therefore they did not
pump during the time required for the diesel generator to start up at
the beginning of a power outage, but resumed pumping shortly into the
outage.  CDF's building water chillers did not operate during a power
outage, and so a dedicated air-cooled backup chiller, which was
powered by the diesel generator, was able to supply chilled water to
the SVX-II/L00 chillers.  The ISL chiller's compressor did not operate
during a power outage, but a backup coolant pump maintained coolant
circulation.

\subsection{Operational experience}
\label{sec:opexpcool}
\begin{figure}
\centering
\includegraphics[width=0.9\linewidth]{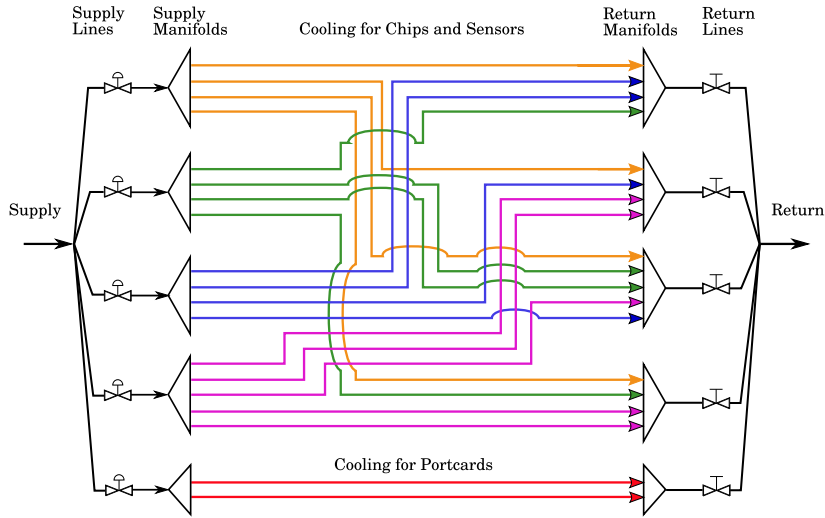}
\caption{
Diagram of the connections of the cooling lines of the east half of ISL showing
the supply and return manifolds located in the detector frame. Flow which
enters the detector via one of the four supply lines was be shared among
several return lines. Electronic supply valves and manual return valves are
shown.  
\label{ISL_East_circuits}}
\end{figure}

The cooling and interlock system for the CDF silicon detector had a
high reliability and failed infrequently. The main goals of protecting
the silicon detector and preventing any damage to other detectors were
well fulfilled. 
However, a few major incidents affecting the ISL cooling structure
inside the detector revealed the inadequacy of the interconnected
branching scheme of the cooling piping (see Fig.~\ref{ISL_East_circuits}). 
Isolation of leaky segments
(Section~\ref{Leaks}) was very difficult and required a multi-week
shutdown of the detector. During these down times a great deal of work
was invested to investigate and repair leaks. In contrast, the
SVX-II/L00 cooling system, which had a simpler geometry, was more
stable and performed better.

\subsubsection{Blockage of cooling flows in ISL}
\label{sec:blockageISL}
When the ISL detector was commissioned, it was discovered that 35\% of
the lines in ISL were not cooling.  Further investigation using long
borescopes showed that the coolant flow was blocked by epoxy found at
aluminum right-angle elbows in these cooling tubes.  In 2002 and 2003,
these blockages were opened by shining Nd:YAG laser light to vaporize
the epoxy. A pulsed laser operated at an average power of 10--40~W was
used, guided by an optical fiber with a 400~$\mu$m core and a
20~$\mu$m Al jacket. The fiber was attached at the end to a device
holding a prism at a right angle so that the laser light could be
aimed at the epoxy plugs just beyond the bends of the
elbows.

This operation was delicate and difficult, as the coolant piping had
an inside diameter of 4~mm and the elbows were approximately two
meters away from the accessible end of the tubing.  One cooling line
was found to be extremely difficult to open and during the attempts
the prism holder became detached from the fiber and remained lodged
inside the pipe.  The flow in the line was not restored, and a
concern of leaks developing due to stagnant coolant with corrosive
ions building up made it prudent to plug the narrow tube.  Aluminum
plugs were inserted at the manifolds where the cooling supply and
return lines divide into four or five narrower tubes.  A second prism
holder was stuck in another narrow tube and repeated attempts to
remove it failed. This reduced the flow, but did not block it.

The successful opening of the blocked ISL flows raised the
fraction of working ISL cooling lines from 65\% to 96\%.

\subsubsection{Degradation of coolant and corrosion}
In 2005, the SVX-II chiller setpoint was lowered from
$-6$\degreesC~ to $-10$\degreesC~ in order to extend the longevity of
the silicon detector.  After this change to the operating
temperatures, there was an incident during routine maintenance work on
the ISL vacuum pump that accidentally fired the safety interlocks of
the cooling system. This stopped the flow of ISL coolant for at least
30 minutes. The ISL cooling system also serves
as the cooling for the portcards for the entire silicon
detector and during this period, the coolant in the SVX-II portcard
cooling lines began to freeze due to their proximity to the SVX-II
cooling system which was at -10\degreesC.  This constricted the flow
of coolant in the SVX-II portcard lines to less than the accepted
minimum rate for the safety interlocks which in turn did not permit
the SVX-II detector to be powered. The frozen coolant in the SVX-II
portcard line was melted by raising the SVX-II coolant temperature
from -10\degreesC~ to +6\degreesC\ and turning the ISL on to raise the
ambient temperature.  After two hours, flow was reestablished to the
SVX-II portcard lines which allowed the SVX-II detector to be
powered. In order to prevent freezing incidents, which risk portcard line
ruptures, 10\% of ethylene glycol by weight was added
to the ISL coolant.

In 2007 the pressure in an ISL portcard supply line rose beyond the
operational tolerance due to leaks in the aluminum manifold which
distributes the coolant to the portcards.
Sufficient flow could not be maintained to cool the silicon sensor
readout electronics and the east half of the detector had to be
switched off.  The investigation found that the pH of the ISL coolant
had dropped to approximately 2.0. The conductivity had risen from
2~$\mu$S/cm to around 3000~$\mu$S/cm. Unfortunately, the conductivity
meters at that time saturated at values far below 3000~$\mu$S/cm. An
analysis based on ion chromatography revealed that the ethylene glycol
had degraded into light organic acids, mainly formic acid at a
concentration of 12.5~g/l (0.265~moles/l). Another analysis ruled out
the possibility of microbial-induced degradation.  There was evidence
that warming up the ISL detector to $13$\degreesC~ during a two
month shutdown in the summer of 2006 accelerated the acidification in
correlation with the rise in conductivity.  Several system components
corroded faster than others with this degraded coolant.  Outside the
detector, the brass valve stems of the solenoid flow control valves
had corroded, causing failures in the valves weeks before the
operational collapse.  The portcard manifolds were made with aluminum
5052 piping welded using aluminum 5356 filler~\cite{aluminumhandbook}.
The filler material had corroded more quickly than the piping
material.

The system was successfully repaired during a shutdown in the summer
of 2007.  The repair work involved inspection of the system with a
borescope. The affected manifolds were located at a distance of 1\,m
inside the cooling tubes. With the use of a custom-made tool, Scotch
DP190 epoxy~\cite{ScotchDP190Expoxy} was laid down on the welded area
of the leaking manifolds.  In order to prevent corrosion the coolant
was replaced by deionized water, and the pH and conductivity were
monitored frequently in order to identify and mitigate hazards
quickly.  Through the remainder of  Run II, the pH and
conductivity were stable, and the affected portions of ISL were
cooled.

The +6\degreesC\ cooling water supplied to the portcards and the heat
generated by the portcards when powered were sufficient to keep the water
from freezing during normal detector operation.  When the portcards
were not powered, however, a freezing hazard existed.  Protection
against freezing (the original impetus for adding glycol to the ISL
coolant) was implemented via an interlock that inhibited flow to the
SVX-II when the temperature measured in any portcard coolant circuit
fell below 1\degreesC.

\subsubsection{Leaks in ISL}\label{Leaks}

The overall leak-rate of the ISL coolant system increased
steadily after installation.  The leak rate, as determined by the amount 
of time it took to leak up to atmospheric pressure when the vacuum pumps 
were valved off, increased by a factor of five between 2007
and 2009, prompting a third intervention to extend the longevity of
the ISL cooling system.  The aluminum ladder-cooling tubing was found
to be leak-tight, except for a few smaller leaks in two tubes.
The epoxied joints between polyethylene tubing and the aluminum
manifolds where the small aluminum cooling pipes join were found to be
the leakiest in the system.  Additional epoxy was applied to three of these
joints, and the flow performance improved, but the overall nitrogen leak rate
remained at comparable levels as measured by the exhaust flow rate out of the
vacuum pump. The flow rates and cooling performance were monitored closely
for the remainder of CDF's data-taking run, and did not degrade to the
point of requiring a change in the operation of ISL. 


%
%
\section{Particle beam incidents and monitoring}
\label{sec:beam}
The particles from standard Tevatron running conditions (physics runs
with proton--antiproton collisions) were responsible for the vast
majority of the radiation dose to the CDF silicon detector
sensors and components.  However, beam instabilities and sudden beam
losses were an unavoidable part of running a large accelerator, and
posed a threat to particle detectors.  As described in
Section~\ref{sec:detdesc}, the silicon detector was the closest to the
beam and suffered larger consequences than the other CDF
sub-detectors.  Beam incidents early in Run II resulted in large and acute
radiation fields that permanently damaged about 4\% of the readout
chips in the detector.  A two-pronged approach was taken to reduce the
possibility of additional damage from beam incidents: a thorough
review of past incidents and a strict beam monitoring system.

\subsection{Particle beam incidents}
\label{sec:beaminc}

When a beam incident occurred, a thorough review of the problem, in
collaboration with the Fermilab Accelerator Division, often resulted
in stricter testing of any hardware involved in the incident, as well
as procedural changes in accelerator operation. The following list
briefly describes the main types of beam incidents and the measures
taken to lessen their impact.

\subsubsection{High beam losses} 

Particles leaving the outer halo of the beams at CDF (beam losses)
were measured by counting hits in scintillation counters located on
both sides of the detector. The counters were gated to exclude hits
coincident with proton--antiproton collisions at the center of the
detector.  High losses at any time were indicative of higher radiation
fields in the detector volume.  More importantly, sudden changes in
the losses indicated potential instabilities in the Tevatron beam.
Monitoring software, described in Section~\ref{sec:radprot}, would
automatically ramp down the silicon sensor bias voltage after
dangerous beam conditions were detected. Under exceptional
circumstances when the radiation increased dramatically, fast hardware
protection systems would issue an abort which immediately removed beam
from the Tevatron (Section~\ref{sec:radprot}).

\subsubsection{Kicker magnet pre-fires} 
\label{sec:beam:kickerPrefire}
A set of 10 \emph{abort kicker magnets} were used to remove the
circulating proton and antiproton beams from the Tevatron by steering
them into a beam dump.  The abort kickers have a finite rise time, so
any beam which passed through those magnets as their field ramped up
were not cleanly extracted into the beam dump.  The beam train
structure contained an unpopulated 1.4\,$\mu$s gap, known as the
\emph{abort gap}, to allow the abort kickers to ramp up without
affecting circulating beam.

Normally, beam aborts were synchronized with the abort gap so that the
kickers would reach nominal field before the first bunches arrived to
be sent to the dump.  Occasionally, one of the 10 thyratrons that
powered the individual abort kickers would trigger spontaneously.
When such a \emph{pre-fire} was detected, the other abort kickers were
fired intentionally, without synchronizing to the abort gap, in order
to abort the beam as quickly as possible.  Any beam that passed
through the pre-fired and other kicker magnets before they had reached
nominal field could continue traveling with a distorted orbit,
possibly hitting accelerator components and creating secondary and
tertiary showers of particles at the experiments.  The location of the
abort kickers relative to the detector made CDF susceptible to large,
acute doses from proton initiated showers.  One such incident in 2003
resulted in the loss of about 4\% of the silicon readout chips.  After
a thorough review of the incident, a new collimator was installed to
intercept protons that would strike the CDF detector due to an abort
kicker pre-fire.  Although pre-fires continued to occur several times
per year, after the installation of the new collimator, the silicon
detector did not sustain significant damage from such incidents.

\subsubsection{Quenches}
A quench is the sudden transition of a superconductor from a state
with zero electrical resistance to a normal state with small, but
finite, electrical resistance. For superconducting magnets like those
in the Tevatron, a quench could be caused by a temperature rise of the
current-carrying superconducting cable above its critical temperature.
This could be caused by localized beam losses in the magnet or a loss
of cryogenic cooling.  An automated quench protection system protected
the magnets from potential damage caused by the sudden ohmic heating,
generated by the large current powering the magnets when resistance
became normal.  The quench protection monitors (QPMs) monitored the
resistive voltage across a string of several magnets.  When a quench
was detected, the QPM simultaneously energized heaters within each
magnet to enlarge the quenched region and enabled switches to bypass
current out of the affected magnets. In addition, it initiated a beam
abort, to reduce the impact of the orbit distortion caused by the
decaying magnetic field of the quenched magnets.

A key component of protecting the silicon detector was the Tevatron
QPM being able to detect a quench and abort the beam as early as
possible.  The early Tevatron QPM operated at 60\,Hz, leading to
quenches possibly remaining undetected for up to 16\,ms ($>$760 beam
revolutions) between QPM measurement cycles.  Indeed, this shortcoming
was at the heart of an incident in 2003 that caused considerable
damage to the Tevatron and to the experimental detectors.  When a
movable experimental detector (Roman Pot) suddenly moved toward the
beam, high beam losses scattered from the device caused a very fast
quench of nearby superconducting magnets that likely went undetected
for the entire gap between QPM measurement samples.  The beams
circulated for most of that time with highly distorted orbits that
showered the experiments and accelerator components.  A stainless
steel collimator had a groove bored into its surface over half of its
1.5\,m length.  A review of the incident led to a higher bandwidth
upgrade of the QPM system (to 5\,kHz) completed in 2006 which allowed
quenches to be detected and beam aborted in $\sim 500$\,$\mu$s (25 beam
revolutions).

\subsubsection{Separator sparks}
In the Tevatron, the proton and antiproton beams circulated within a
single beam pipe. Electrostatic separators kicked the beams onto
distinct helical orbits so that head-on collisions occurred only at
the CDF and D0 interaction points. These 26 separators were stainless
steel, parallel-plate electrodes, 2.5\,m long, with a 5\,cm gap
operating with a gradient of up to $\sim$40\,kV/cm. Occasionally, a
high-voltage breakdown (spark) occurred between the plates, or between
a plate and the surrounding shell. The effects of a separator spark
depended on which separator broke down, and when the spark occurred
during a Tevatron cycle. Such breakdowns caused a momentary kick to
the beams resulting in orbit distortions that caused beam loss spikes,
emittance growth, and a small drop in instantaneous luminosity.  The
orbit distortion caused by a separator spark could be large enough to
drive beam into collimators and cause quenches of nearby
superconducting magnets. Improvements in the high voltage conditioning
of the separators reduced the overall spark rate compared to early
Run~II conditions. During the latter years of Tevatron operations,
only one or two stores per year terminated prematurely from a
separator spark.

\subsection{Beam monitoring and detector protection}
\label{sec:radprot}
CDF had a slow-reacting (of the order of seconds) software monitor and
a fast-reacting (of the order of 10~$\mu$s) hardware protection system
to help prevent damage to the silicon detector from the incidents
mentioned in Section~\ref{sec:beaminc}.

The software program, called {\sc TevMon}, collected several
measurements that described beam conditions, some of which were
provided to CDF by accelerator monitoring systems.  These included
beam losses, RF station voltages, instantaneous luminosity, abort-gap
beam current, and abort kicker magnet voltages.  A variable that
entered the warning range caused an audible alert for the shift crew,
which indicated degraded beam conditions that may warrant their
attention.  When {\sc TevMon} reached the alarm state, the bias voltages of
the silicon detector were turned off automatically.

The fast reacting hardware protection systems, the beam condition
monitoring (BCM), consisted of four Beam Loss Monitors (BLMs).  These
were two ionization chambers on the east and two on the west side of
the detector, about 4.3~m from the nominal interaction point (IP) and
at a radius of about 20~cm from the beam axis. The location of the
BLMs from the IP was necessary due to their size, which made it
impossible to put them inside the detector.  The BLMs were read out
every 210~$\mu$s (10 beam revolutions), and a circular buffer of 2048
measurements was kept. The BLMs monitored the radiation accumulated
over the last minute and the radiation rate. An accumulated dose
greater than 19 rads (0.19 Gy) over the previous minute would issue an
\emph{integrated dose alarm}, whereupon the shift crew would pass on
the alarm to the Tevatron operators and adjustments to the Tevatron
operation would be made.  If the radiation rate exceeded 12 rads/s
(0.12 Gy/sec), the \emph{radiation abort alarm} would fire, whereupon
the Tevatron would automatically issue an abort and the beams would be
dumped.

\begin{figure}
\centering
\includegraphics[width=0.9\textwidth]{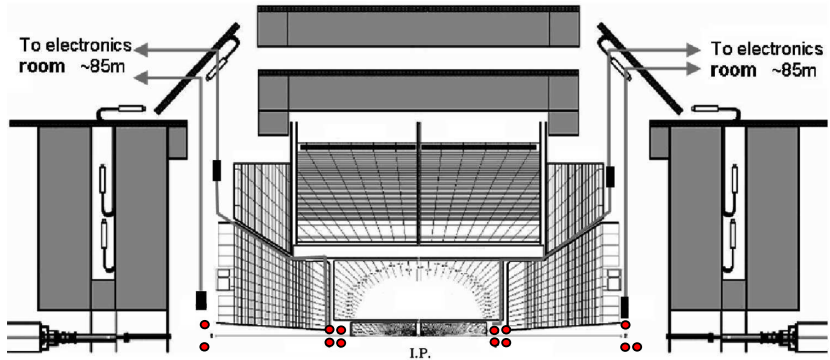}
\caption[Diamond locations within CDF]{
The locations of the diamond-based BCM system in the CDF detector. The
picture shows the upper hemisphere of the CDF detector. The proton
beam circulated from left-to-right along the horizontal axis,
colliding with the anti-proton beam that circulated in the opposite
direction. The collision occurred at the interaction point, indicated
by ``I.P.'' in the figure. A pair of four red dots, symmetrically placed
on either side of the IP, indicate the sensors inside the tracking
volume, and the red dots further away from IP along the beamline
indicate sensors outside the tracking volume. 
}
\label{fig:DiamondLoc}
\end{figure}

A closer examination of beam incidents showed the BLMs lacked the
timing resolution and dynamic range to foresee the conditions leading
to an \emph{radiation abort}. A system of smaller sensors, closer to
the beam and with a faster readout system, could abort the beam more
rapidly to improve the safety of the CDF silicon detector. This led to
the installation of a diamond-based BCM system~\cite{DiamondPaper}.  A
total of thirteen diamond sensors were installed in the CDF detector,
at the locations indicated in Fig.~\ref{fig:DiamondLoc}.  Two groups
of four diamond-based sensors were located inside the tracking volume,
with each group mounted in a support structures at a distance of
$1.7$~m from the nominal IP, arranged as the
sides of a 4\,cm by 4\,cm square.  Five more diamond-based sensors
were installed outside the tracking volume, on the previous BLM system
support structure; two diamonds on the west side and three on the east
side.

The BCM system was configured to abort the Tevatron beams when at
least four diamond sensors measured a current of at least 500\,nA and
the CDF solenoid was fully energized, as the diamond dark current was
affected by the external magnetic field.  These settings were
determined from a six-month study period to optimize the thresholds,
and allowed quick response to potential beam incidents, while
minimizing the number of false aborts. In general, the response of the 
diamonds was found to be significantly faster than that of the 
BLM-based devices.


%
%
\section{Sensor readout calibration}
\label{sec:calib}
The algorithm that forms clusters from individual hits in the silicon
detector uses the measured pedestal mean and RMS values for each individual
strip to separate actual signals from noise fluctuations. These values were
measured bi-weekly for each channel during Run II with special calibration
runs and recorded in the experiment database. A standard silicon calibration
required two runs taken in read-all mode: one with DPS turned off and one
with DPS turned on.

\begin{figure}
\centering
\includegraphics[scale=0.50]{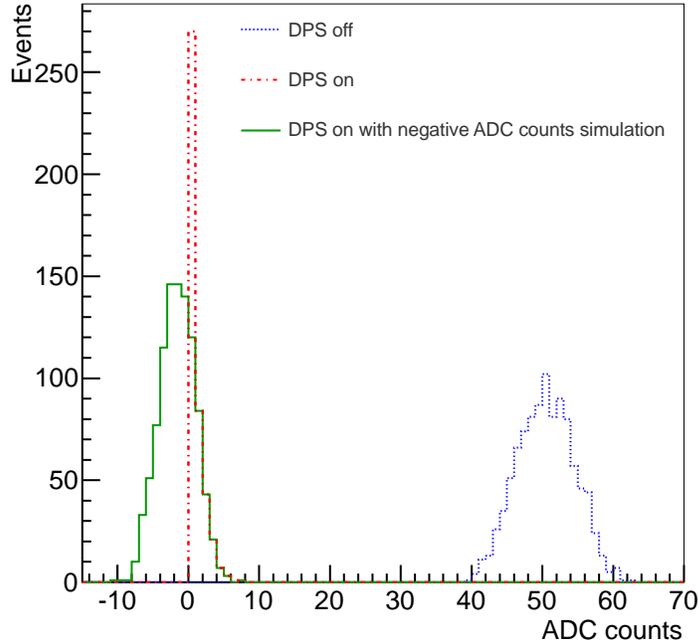}
\caption{Effect of Dynamic Pedestal Subtraction on pulse height
  distribution of a single channel in 1000 events.  
  The DPS-off distribution is shown as dotted blue, DPS-on as dot-dashed red,
  and the solid 
  green histogram is an extrapolation of what the DPS-on distribution would
  be if negative ADC counts could be recorded.  The peak at 0 ADC counts
  in the DPS-on distribution is suppressed by choice of scale.}
\label{dps}
\end{figure}

The analysis of data from these runs started with accumulation of the ADC
pulse height distributions for individual channels. Section~\ref{sec:svx3d}
describes the DPS circuitry, in particular that the start of the ADC counter 
is delayed until a fraction $f_d$ of the 128 channels in the chip have a
voltage below a common voltage ramp. This fraction $f_d$ can vary from chip
to chip but is close to 33/128 channels. A comparator switches state when
the fraction of channels exceeds $f_d$, and a finite propagation delay
$t_d$ exists between the time the comparator switches state and the start 
of the ADC counter. Because of the delay, the effective pedestal values in 
this scheme are negative for most of the channels and the distribution 
from DPS-on data is not sufficient to determine the pedestal mean and RMS. 
Instead, the expected DPS-on distribution is calculated from the measured 
DPS-off distribution with a simulation of the DPS algorithm that uses the 
values of $t_d$ determined for each chip during commissioning. The effect 
of the DPS circuit on the pulse height distribution of a single channel is 
illustrated in Fig.~\ref{dps}, in which the value of $t_d$ is exaggerated 
for illustration purposes.

The pedestal and noise values for each individual channel were stored
in the offline calibration database. Channels were flagged as {\em dead} or
{\em noisy} based on their expected occupancy. Noisy channels are not
included in the default offline clustering algorithm.  The calibration
constants stored in the offline database are used for data
reconstruction as well as detector simulation.


%
%
\section{Detector monitoring and operations support}
\label{sec:mon}
The operation and maintenance of the silicon detectors required dedicated
personnel and software for fast problem response. The CDF Silicon
Group was comprised of approximately ten on-call experts who provided
24-hour support to the CDF operations group and performed regular
maintenance to ensure optimal performance of the detector. The 
group was led by two sub-project leaders. Two experts were assigned to 
each of the following subsystems: DAQ, power supplies, and cooling. Another
expert periodically performed calibrations of the detectors and 
ensured their quality. In addition, three to four experts provided
online monitoring of the detector and gave prompt feedback to the group 
of any developing hardware problems. The group strived to optimize and 
automate most aspects of the silicon detector operations over the years.

Many software packages were developed for detector operations and 
monitoring. They were broadly divided into three categories: stand-alone 
java applications used by the shift crew, scheduled jobs to acquire data 
on power supplies and the cooling and DAQ systems, and Perl-based CGI 
scripts that provided real-time information to the Silicon Group.

\subsection{Silicon auto-recovery}
\label{subsec:SAR}
The SVX3D readout chips (see Section~\ref{sec:svx3d}) required proper 
initialization sequence as soon as they were powered. The process of 
turning on power to the silicon detectors required synchronized actions
in both the power supply control and the DAQ.
Over time, a set of well established procedures to recover from common
failure modes was developed. However these manual recovery procedures
reduced the data-taking efficiency, especially when a large fraction 
of the detector channels needed to be turned on after failures, such as a 
power supply crate reset (see Section~\ref{subsec:psradvulnerable}).
 
In order to automate and speed up the recovery of power to the detector 
channels, a \emph{Silicon Auto-Recovery} (SAR) software tool was developed.
SAR detected channels that lost power during data-taking and sent a 
request to the power supply control to turn these channels back on.
Once the power to the chips was restored, SAR took the corresponding
FIB through the initialization process which also initialized the chips. 
During this process, data taking was suspended.  After automating 
both the power supply crate reset and recovery, the average experimental
downtime due to lost communication (see Section~\ref{subsec:psradvulnerable}) 
with the power supply crates was reduced from 10 min down to less than 
5 min.

\subsection{SVXMon}
\emph{SVXMon} was the monitoring application used for both online
and offline diagnostics of the CDF silicon detector problems. It ran 
continuously during data taking as part of a set of CDF
monitoring applications. It accumulated various statistics and presented a 
coherent set of silicon performance plots to the shift crew. On special 
occasions, it sent automated requests to reinitialize 
DAQ components showing problems.

SVXMon was a highly configurable program capable of presenting both very 
general and very detailed views of the silicon data. For each silicon 
strip, SVXMon accumulated the number of hits and pulse height distribution. 
These were used to create plots of occupancies, average pulse heights,
distribution shapes, etc., with various degrees of detector granularity. 
The monitoring application had a large number of configuration parameters 
which evolved over time to optimize the information useful to detect and
understand error conditions in the silicon detectors.

\subsection{IMON}
\label{sec:imon}
\emph{IMON} was an application used to monitor currents in the silicon 
detectors. 
IMON displayed each ladder of the detector as a set of color-coded boxes 
(one for each channel of the ladder). If the measured current was within a 
pre-set range, then the box showed up green. If it was just outside the 
good range, it showed up yellow. If it was far outside that range, it turned 
pink\rlap.\footnote{Other colors were used to indicate ladders which had 
  tripped, turned off, had lost communication or were ignored by the DAQ.} 
When a ladder turned pink, it alerted the shift crew so they could take action.

As the sensors degraded due to exposure to radiation, they drew more
current.  Eventually, a sensor drew enough current to send it over the
pre-defined ``good" range; this was normal and simply required
adjustment of the good range for that particular sensor by a member of
the Silicon Group. A typical bias channel drew an additional $30~\mu$A
of current for every 500~pb$^{-1}$ of delivered integrated
luminosity. The currents drawn by the chip chains did not change. When
one exceeded the normal limits, it typically needed to be
reinitialized.

Although the concept for adjusting the bias current limits was simple,
the actual procedure was tedious. Limits for the bias currents were
adjusted and documented in a database. The rate of increase varied
widely across the channels due to the sensor type, radiation exposure,
and other causes.  With close to 500 bias channels, 25--50 adjustments
were made every week.  Changing the limits by hand could take several
minutes per channel.  For bias channels which had increases consistent
with normal aging, the limits were adjusted by a monitoring program.
After automatic adjustment software was made operational, it took care of
approximately 95\% of the limit adjustments, drastically cutting down
the workload of the Silicon Group.

\subsection{ADCMon}
The Silicon Group had the responsibility to ensure good quality of the
data collected by the silicon detectors. \emph{ADCMon} was an
application developed to ease that task. ADCMon read the raw
information recorded by the DAQ and provided the distributions of
charge in ADC counts for each silicon ladder. Two different versions
of ADCMon were implemented: online and offline. The online version
operated in the CDF control room during data taking and provided data
to SVXMon. It provided the charge distribution of the last 500 events,
as well as the statistics accumulated during the entire run of data
taking. The offline version was executed with a delay of less than one
day from the end of the run and was useful to understand long term
behavior of the silicon ladders. It generated a table of histograms
representing the charge distribution for events of a given run in
comparison with a reference run. It also provided information about
the percentage of bit errors in digital transmission of data.

Optical data transmission bits could permanently get stuck in a high or low
state due to radiation damage, cable or electronics 
malfunction, as described in Section~\ref{DAQ:bit_err}.
This could lead to a lower resolution of the charge distribution. 
Discrepancies in the shape of charge distributions or absence of data 
could be caused by a FIB or power supply failure. Severely 
underdepleted ladders could show an observable drop in the high end
of the charge distributions. A visual evaluation of the offline results 
required 5--10 min per day. 

\subsection{iFIX}
\label{subsec:iFIX}
Some components of the silicon detector needed to be monitored independent
of data taking. A system to do this was developed based on the commercial
automation software \emph{Proficy HMI/SCADA iFIX}, licensed by 
GE Fanuc~\cite{ge_fanuc}.
Data were stored at intervals on the order of a few seconds, with recent readings
ranging from a few hours to a few days being
displayed in the CDF control room.  Older data were migrated to permanent
storage. This information could be retrieved from iFIX computer nodes.

\subsubsection{Slow control of cooling system}
An iFIX node was connected to the PLC system through a VME module to a 
Siemens SIMATIC 505 Crate, where the I/O modules resided (see 
Fig.~\ref{ISL_East_circuits} in Section~\ref{sec:opexpcool}). 
Information from the cooling related devices was available in the iFIX software
which provided displays of real-time readings of the devices (temperature 
sensors, pressure gauges, flowmeters, valves, etc). Changes to parameters of 
selected devices could be made through this system. The system-wide and 
sub-component interlock status were also displayed. In particular, the most 
valuable quantities monitored in this system were the temperatures
of the cooling lines inside the detector, as well as the flows and pressures.

\subsubsection{Audible alarm system}
Alarm conditions were defined for selected variables, such as the cooling 
variables which could potentially trigger the interlocks. In the case of 
temperatures, pressures and flows, low and high warning and alarm limits were
defined. Other variables in the alarm list included the high voltage and trip
status of power supplies. Any alarm going off was followed by an audible voice alarm.

\subsubsection{Monitoring of rack modules}
The status of the AC power to the racks hosting the FIB and CAEN crates in the 
collision hall was monitored, and remote power cycling of racks was possible. 
In addition, monitoring and alarming on low-voltage status was available for 
small devices in the racks, such as fans inside the crates. This was also 
true for the racks hosting the VRB modules outside the collision hall.
We also monitored the power supply output voltages for the VRB and FIB crates.

\subsubsection{Reset of silicon power supplies}
The automatic reset of CAEN power supply crates (explained in 
Section~\ref{subsec:psradvulnerable}) was monitored in the iFIX system. 
Additionally, the silicon operations team 
could manually reset any power supply crate from this system. 


%
%
\section{Monitoring radiation damage effects}
\label{sec:age}
\subsection{Depletion voltages}
Periodically, the depletion voltage of each sensor was measured to
monitor the radiation effects on the sensor, and the operating voltage
increased to ensure that the sensors remained fully depleted.  The
operational definition of depletion voltage for the CDF silicon
detectors was the bias voltage at which the charge collection
saturates.  Specifically, the voltage at which the charge collected
was 95\% of the maximum value.  Operating at bias voltages larger than
a depletion voltage defined this way ensured the best detector
performance.

Two different methods to measure the depletion voltage were used: the
\emph{noise scan} was used for double-sided sensors that had not
undergone type inversion, and the \emph{signal scan} was used for all
sensors.  In each case, the bias voltage was changed and 
either the signal from charged tracks or the noise in the
sensor was measured.  The results of both methods are presented here in 
the context of operating the detector.
Further discussion of the observed effects
of radiation damage in the sensors is reserved for a future article 
dedicated to this topic.

\subsubsection{Noise scans}
The noise scan measured the average noise for each ladder as a
function of bias voltage.
For each bias voltage setting, data were taken in read-all mode and
the ADC count distribution for each strip was recorded.  For each
ladder, the noise was determined by taking the RMS of the ADC
distribution for each strip and then averaged over the strips on the
p-side (\rphi) and the n-side (\rz) separately.

\begin{figure}[ht]
\centering
\begin{tabular}{cc}
\includegraphics[width=0.47\textwidth]{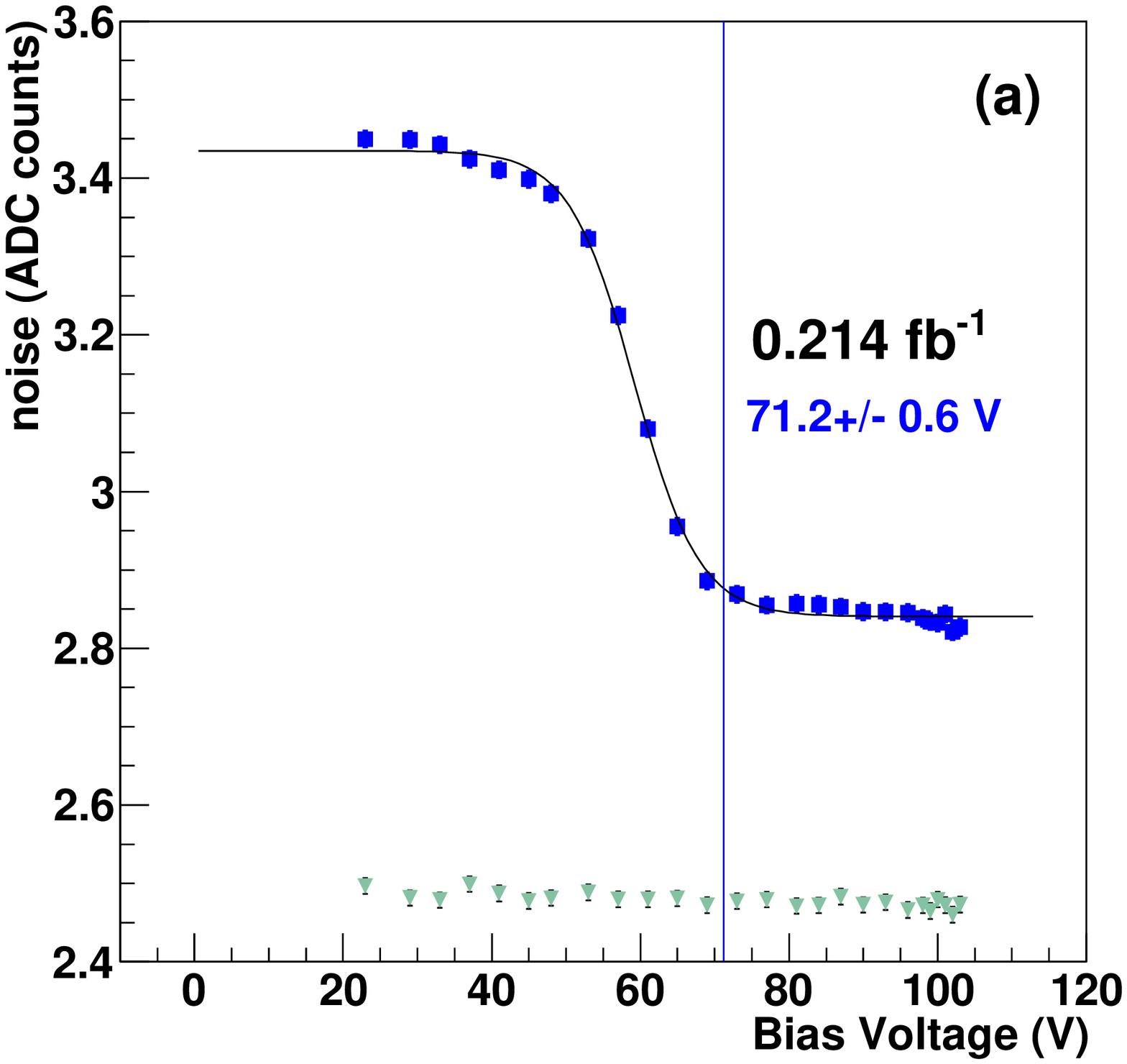}
& \includegraphics[width=0.47\textwidth]{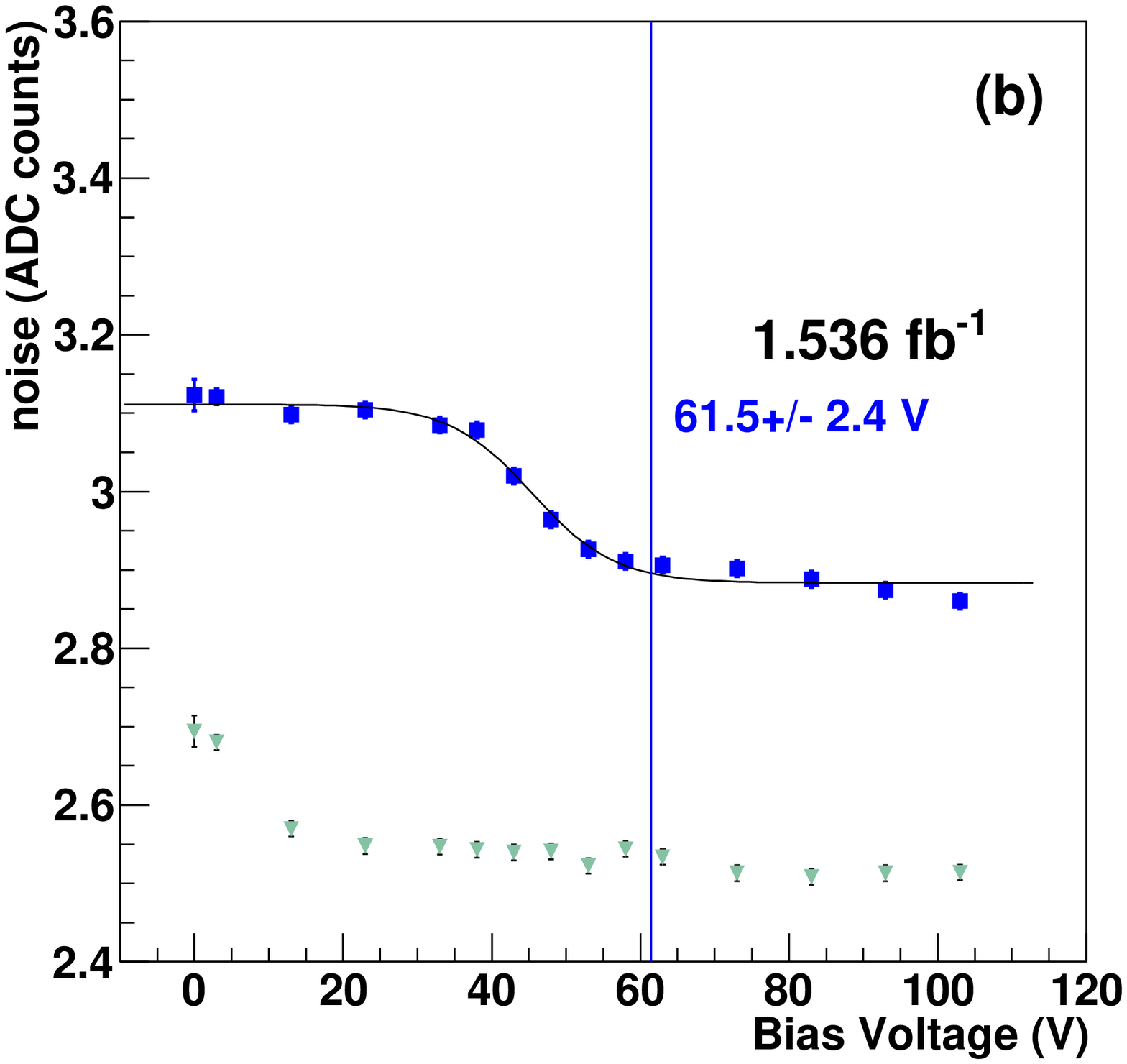} \\
\includegraphics[width=0.47\textwidth]{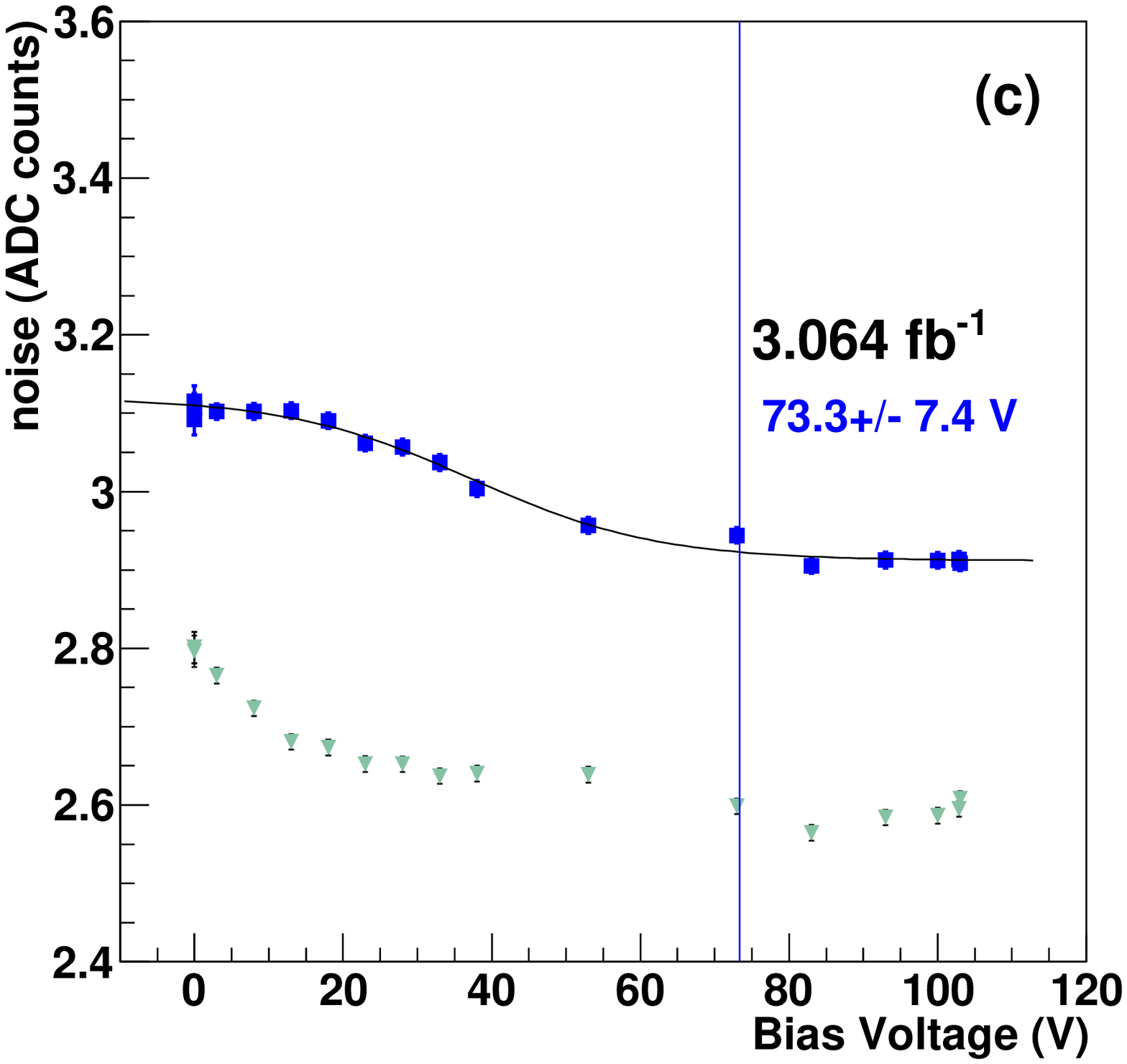}
& \includegraphics[width=0.47\textwidth]{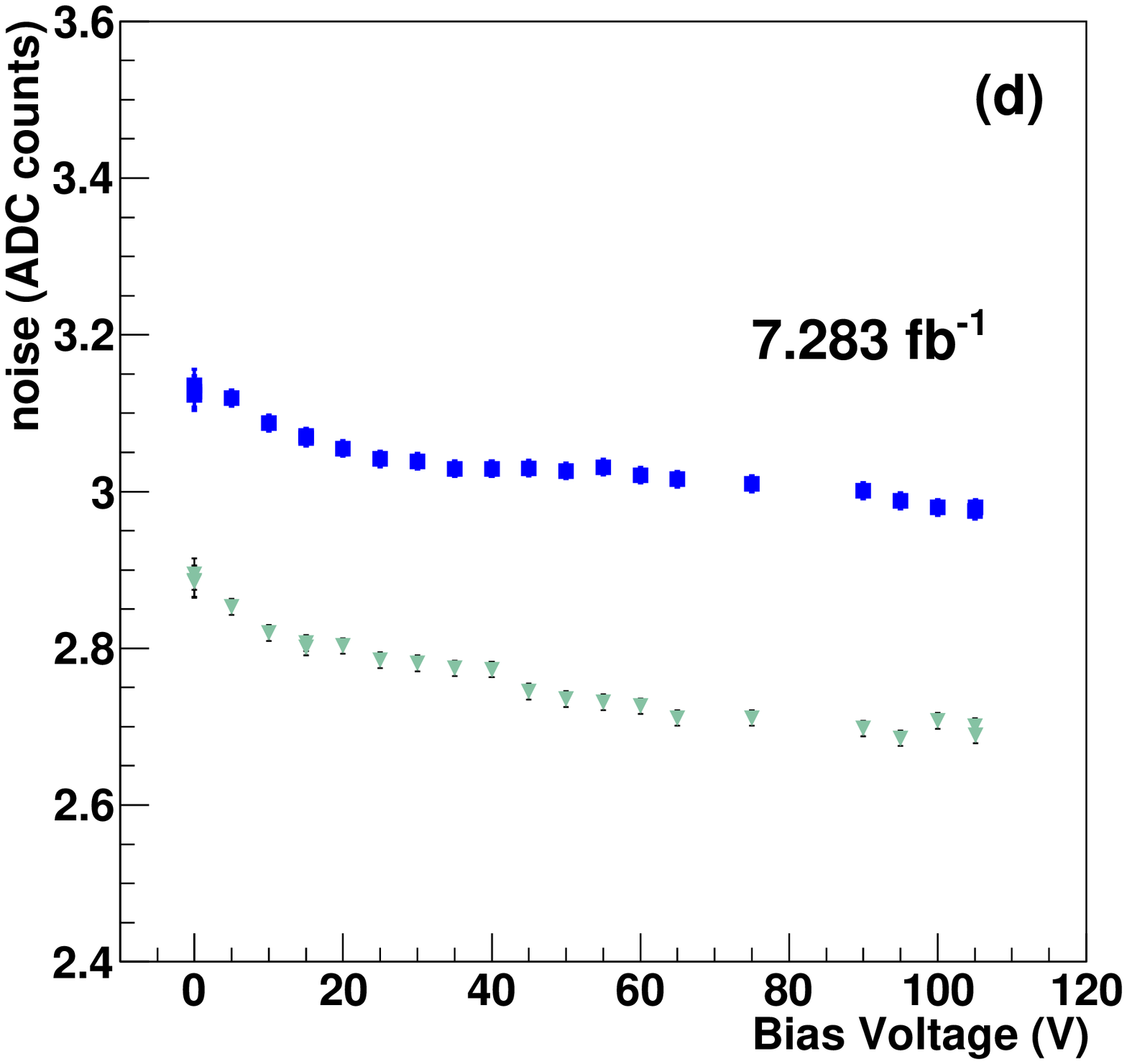} \\
\end{tabular}
\caption{ Noise scan measurements for a single ladder from layer 1 of
  SVX-II at four different integrated luminosities (given in
  fb$^{-1}$).  The blue squares are the measured noise on the n-side.
  The pale green triangles are the measured noise on the p-side of the
  sensor.  The extracted depletion voltage is indicated with a vertical line.}
\label{noisescan1}
\end{figure}

\begin{figure}[h]
\centering
\includegraphics[width=0.47 \textwidth]{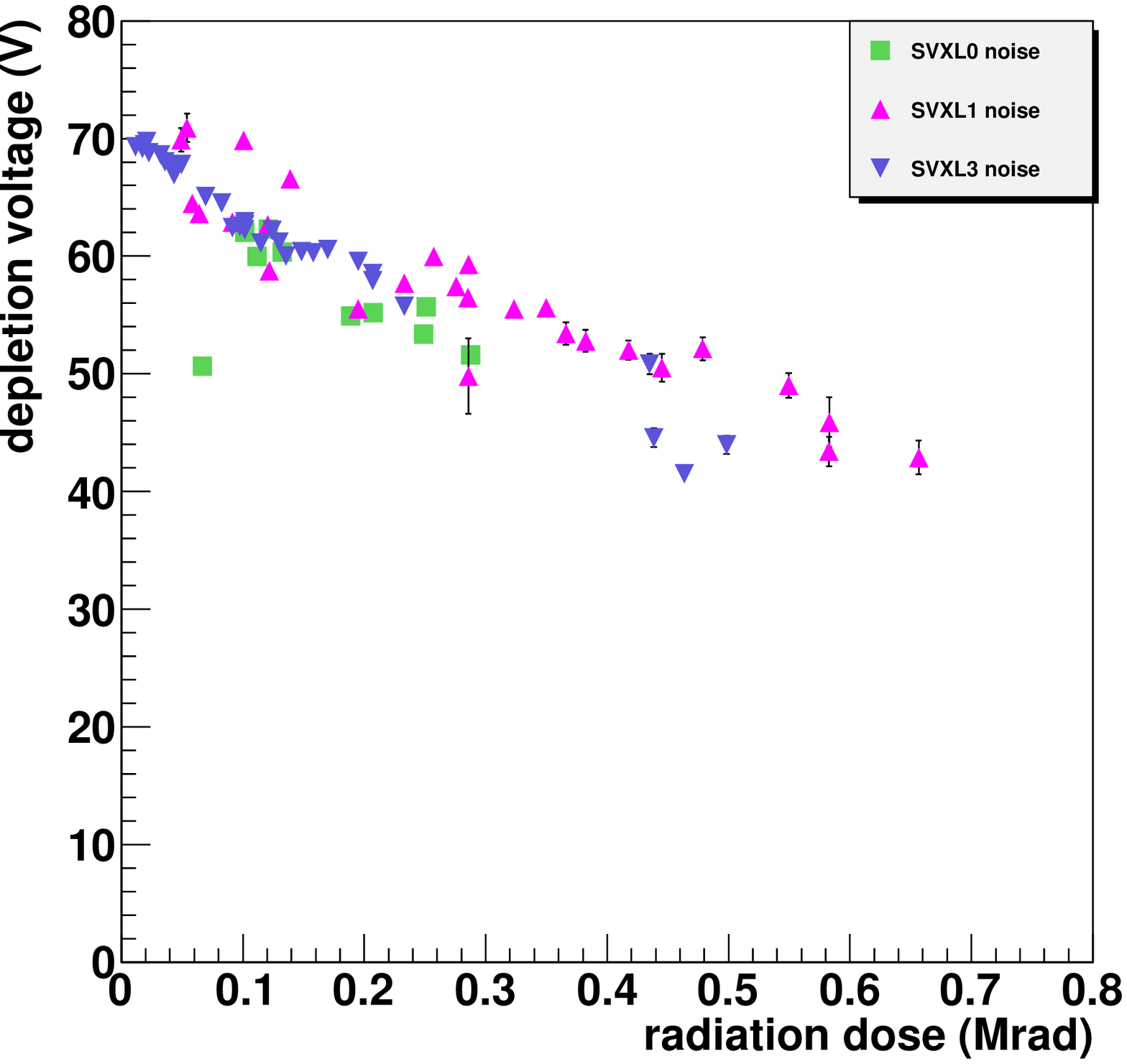}
\caption{The average depletion voltage determined from noise scans is plotted
for the Hamamatsu sensors in SVX-II (see Table~\ref{SpecsTable}) as a function 
of radiation dose.}
\label{AllNoise}
\end{figure}

The depletion voltage for each ladder was determined by fitting the
n-side noise as a function of the bias voltage to a sigmoid function
\begin{equation}
\textrm{noise}=A+\frac{B}{1+\exp\left[-C(V-D)\right]}
\label{eq:sigmoid}
\end{equation}
where $A,B,C$~and~$D$ are fit parameters and the variable $V$ is the
bias voltage.  The depletion voltage was identified as the voltage at
which the function is equal to the sum of lower plateau of the sigmoid
function and 5\% of the height of the fitted sigmoid function, or
V$_{dep}=D(ln 19)/C$.

Noise scans for a typical ladder taken at different integrated luminosities 
are shown in
Fig.~\ref{noisescan1}.  In the early scans, the separation between the
two noise levels for the n-side was large, and the depletion voltage
was easily determined, as seen from Fig.~\ref{noisescan1}(a).  As the
sensor became irradiated, the underdepleted noise level decreased
while the depleted noise level increased, and
it became increasingly difficult to determine the depletion voltage
using this method. Specifically, noise scans where the two noise
levels of the n-side were separated by less than 0.2~ADC counts did
not give a reliable determination of the depletion voltage.
Fig.~\ref{noisescan1}(d) shows the measured noise after inversion of
the sensor, for which the p-side noise and n-side noise have similar
behavior. After the inversion, the overall noise level increased with
radiation dose as expected, but the shape of the curves remained the
same.

The noise scan method was used to monitor the depletion voltage before
the inversion of the sensors.
Fig.~\ref{AllNoise} shows the average depletion voltage for ladders in 
layers 0,1, and 3. Noise scans where the two noise levels of the
n-side were separated by less than 0.2~ADC counts are not included in
the plot.  In order to compare the different layers, integrated
luminosity was converted to the equivalent dose of the radiation field
measured inside the CDF detector with over 1000 thermal luminescent
dosimeters in 2001~\cite{TLD}, summarized in Table~\ref{tab:raddose}.  
The behavior of the three different
layers was remarkably consistent considering that the integrated dose
received by the layer depends on the distance from the interaction
region.  The sensors from layers 2 and 4, which are from a different
manufacturer than the other layers, were not included in this analysis
because they developed complicated noise profiles and the simple data
analysis described above did not give reasonable quantitative results
and signal scans were used instead to monitor the depletion voltage.

\begin{table}
\begin{center}
\begin{tabular}{ccc}
\hline
Layer & r (cm) & dose/L (kRad/fb$^{-1}$) \\
\hline
L00 narrow & 1.35 & $994\pm199$ \\
L00 wide & 1.62 & $756\pm151$\\
SVX-L0 & 2.54 &$385\pm77$\\
SVX-L1 & 4.12 &$186\pm37$\\
SVX-L2 & 6.52 &$94\pm19$\\
SVX-L3 & 8.22 &$66\pm13$\\
SVX-L4 & 10.10 &$49\pm10$ \\
\hline
\end{tabular}
\end{center}
\caption{The radiation dose per unit luminosity measured by the TLDs in the 
CDF tracking volume~\cite{TLD}, extrapolated to the location of the individual 
silicon layers .}
\label{tab:raddose}
\end{table}

\subsection{Signal scans}
The signal scan provided the best evaluation of the depletion voltage,
and in many cases the only one. The charge collected by the sensor
increased with increasing bias voltage as the depleted region in the
sensor grew, until the sensor was fully depleted and the charge
saturated. The scan had to be performed with colliding beams and
required approximately two hours per layer. To minimize the amount of
lost physics data, the scans were done when the instantaneous
luminosity was low.

Data were acquired with a specific trigger selecting collision events
containing at least two tracks.  From these events, tracks traversing
the silicon layer under study were identified using the COT and
remaining silicon layers.  If one and only one cluster existed within
150~$\mu$m of the location where the extrapolated track crossed a
sensor, the total charge of that cluster was recorded in a histogram.
A reasonable fit result required at least 1000 tracks per ladder per
bias voltage setting, with additional tracks per point needed below
20~V.

\begin{figure}
\begin{tabular}{cc}
\includegraphics[scale=0.35]{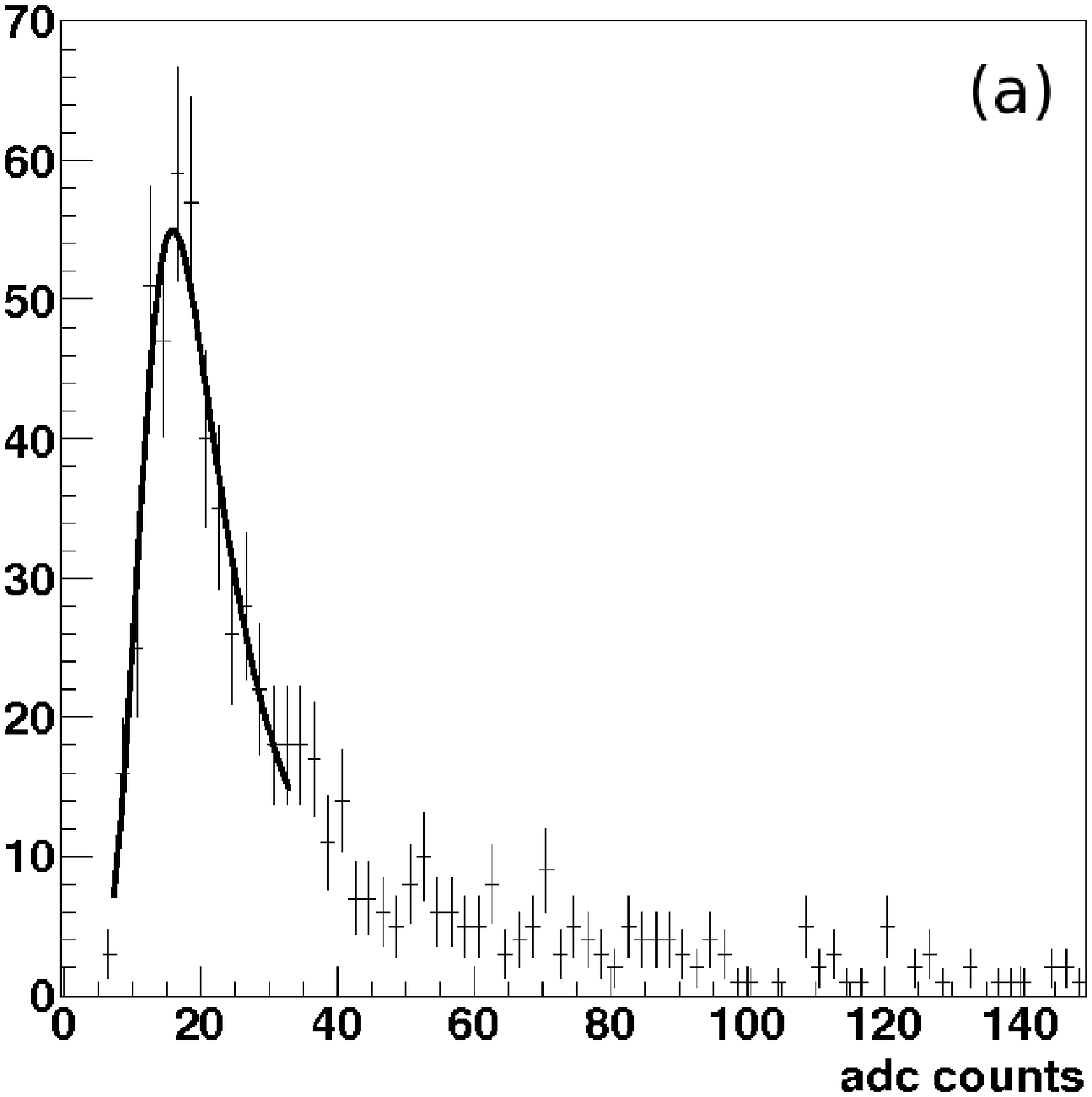} &
\includegraphics[scale=0.35]{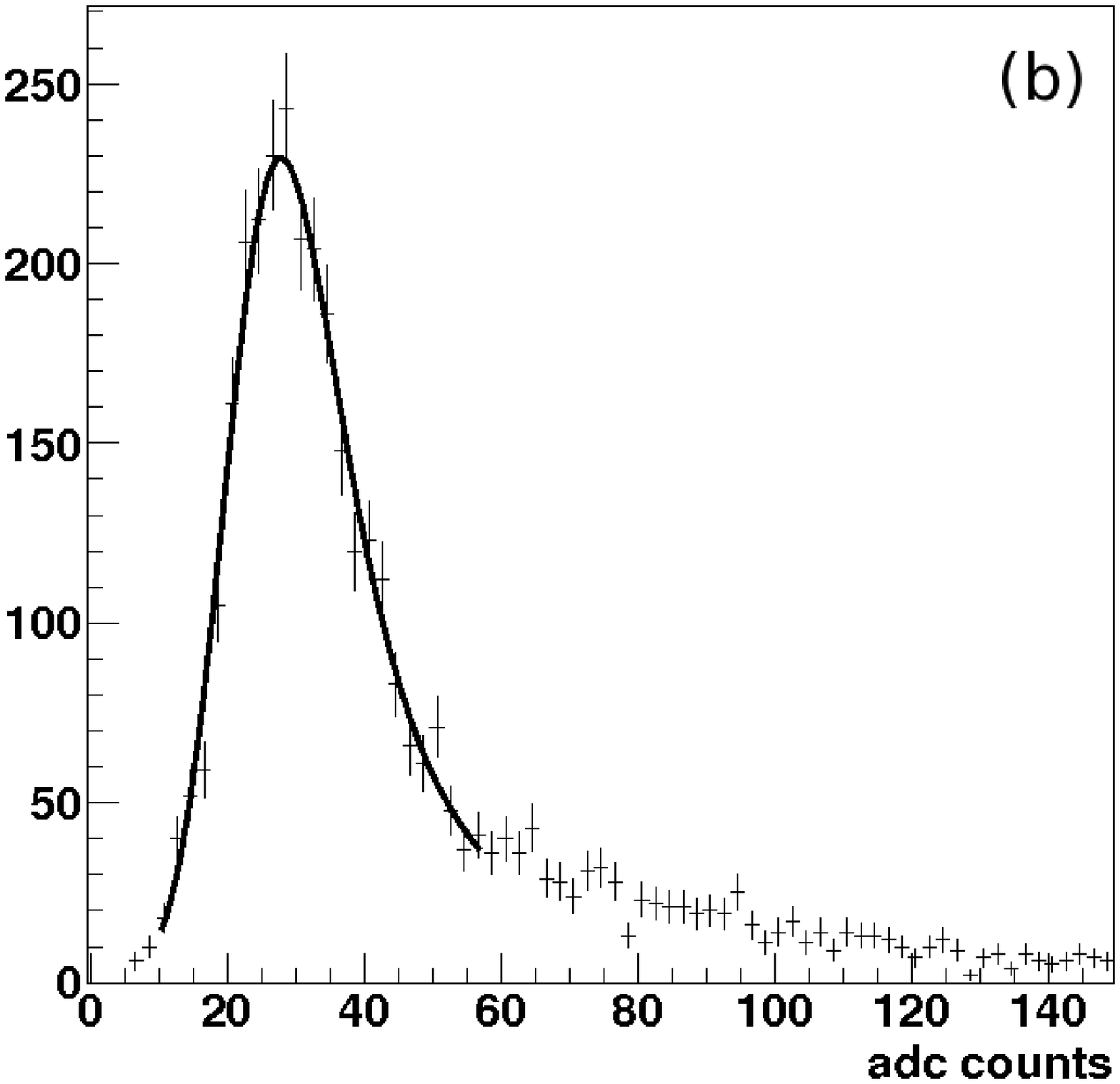} \\
\includegraphics[scale=0.35]{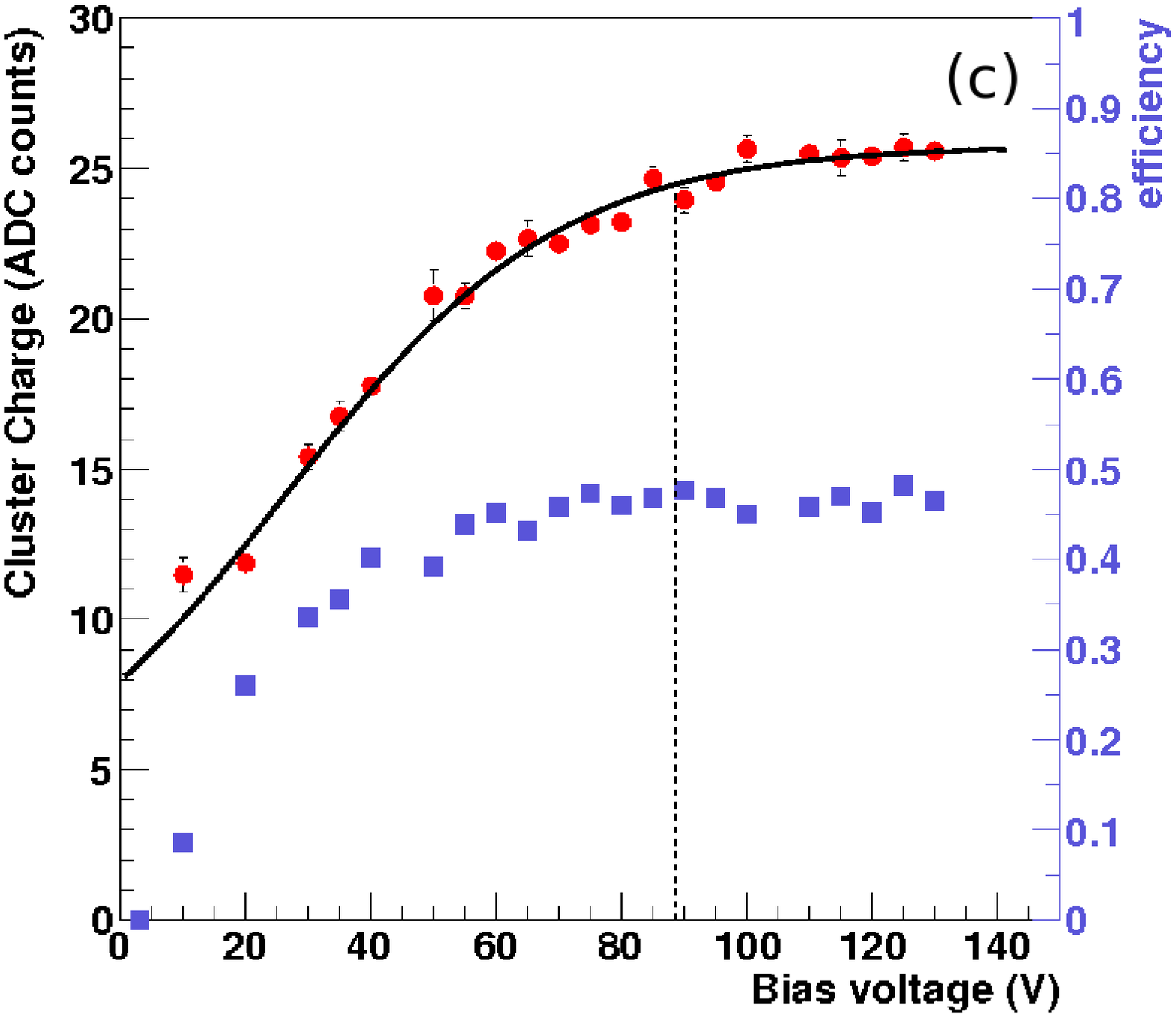} &
\includegraphics[scale=0.35]{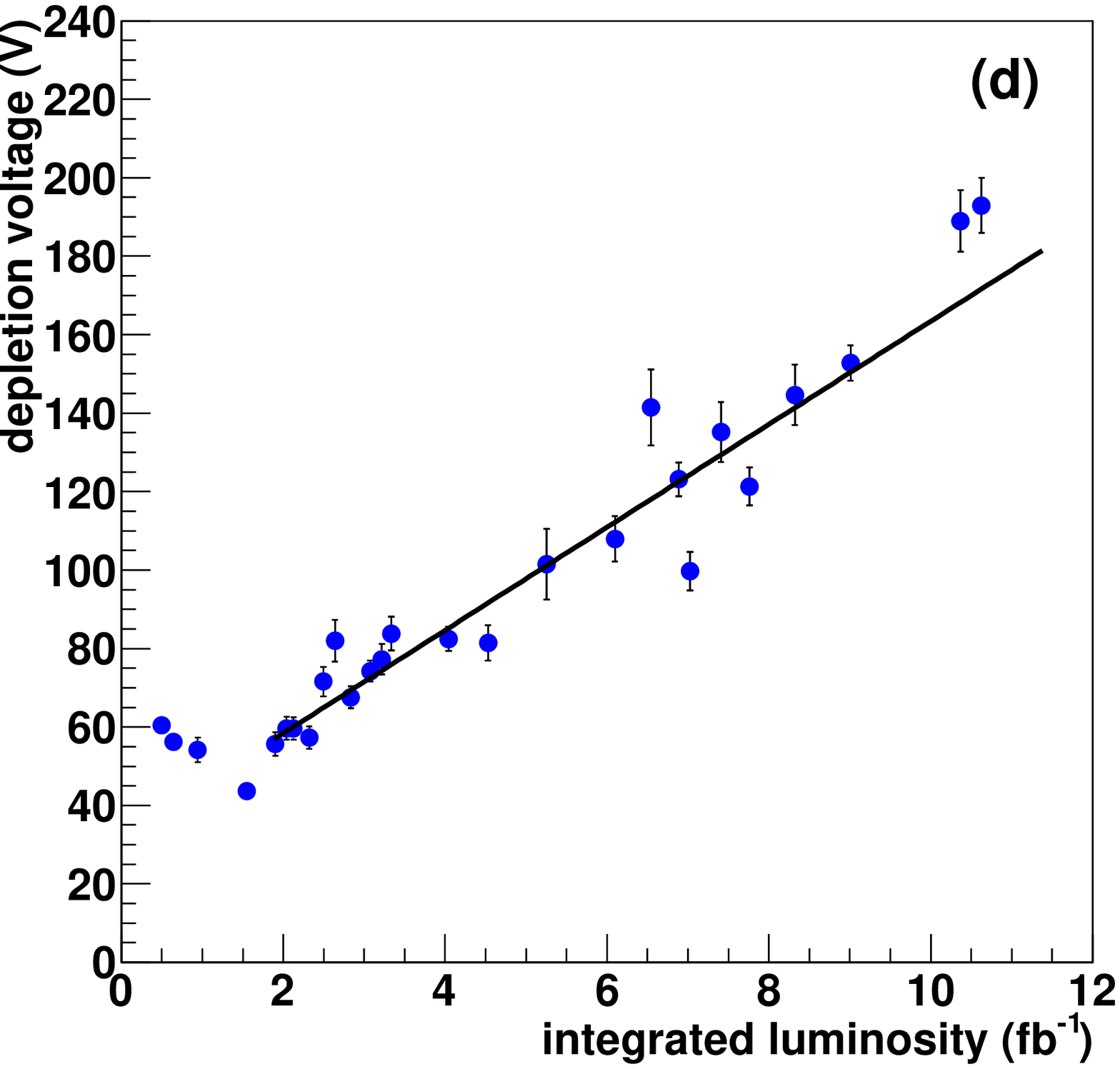} \\
\end{tabular}
\caption{The upper plots show the measured cluster charge distribution
for a single L00 ladder at a bias voltage of 30~V
(a) and 130~V (b) after 4 fb$^{-1}$ of integrated luminosity.  
Plot (c) shows the peak of the cluster charge distribution (red
circles) and the efficiency (blue squares) as a function of bias voltage.
The dashed line indicates the depletion voltage extracted from the sigmoid fit.
Plot (d) shows the measured depletion voltage for this ladder as a function
of integrated luminosity, and the linear fit used to extrapolate to
higher luminosity values.}
\label{DVmeas}
\end{figure}

The distribution of cluster charges was fit to the convolution of a
Landau function and a Gaussian function in the region around the peak.  
The upper plots
of Fig.~\ref{DVmeas} are examples of this distribution at two
different bias voltages.  The most probable value of the fitted
function was plotted as a function of bias voltage, and these points
were fitted to a sigmoid function of the same form as Eq.~\ref{eq:sigmoid}.
The measured depletion voltage was the bias voltage
at which the function value is 95\% of the total charge is collected,
or $V_{dep}=0.95*(A+B)$.  An example is shown in
Fig.~\ref{DVmeas}(c).  Also shown on this plot is the efficiency,
defined as the fraction of tracks for which a cluster is found.  Due
to the limited data samples of these special runs, the track
selection was quite loose, and the absolute value of this
efficiency does not reflect normal sensor performance during data taking.

The non-zero value of the cluster charge for bias voltages below 10~V
is a measure of the effective clustering threshold for that ladder.
All L00 strips were read out every event, while for SVX II and ISL,
only strips above 9~ADC counts plus the neighboring strip on either
side were read out. In the offline reconstruction, clusters were
formed from a set of neighboring strips that satisfied one of
the following criteria:
\begin{itemize}
\item A single strip has pedestal-corrected ADC count larger than
      4 times the measured RMS of its pedestal distribution.
\item Two neighboring strips each have pedestal-corrected ADC
      counts larger than 3 times the measured RMS of their respective
      pedestal distributions.
\item Three neighboring strips each have pedestal-corrected ADC counts
      larger than 2 times the measured RMS of their respective
      pedestal distributions.
\end{itemize}

At low bias voltages, very few clusters are above threshold and the 
cluster threshold algorithm determines the shape of the cluster charge 
distribution for these few clusters above threshold.  The mean of this 
distribution is thus a monitor of the effective threshold of the 
offline clustering algorithm. Although the initial rise of the cluster 
charge is hidden by the clustering threshold, the increasing efficiency 
indicates increasing mean cluster charge as more and more clusters are 
found above the threshold.  Because the strip noise increased with 
radiation dose, these clustering thresholds crept upward with integrated 
luminosity, and the signal scans allowed them to be monitored.

Close monitoring of L00 and the inner layers of SVX was
essential after inversion to keep the operating bias voltages above the
depletion voltage.  Operating voltages were increased on a sensor by sensor
basis after extrapolating the linear trend in the measured depletion
voltages several months into the future.

\begin{figure}[h]
\centering
\includegraphics[width=0.47\textwidth]{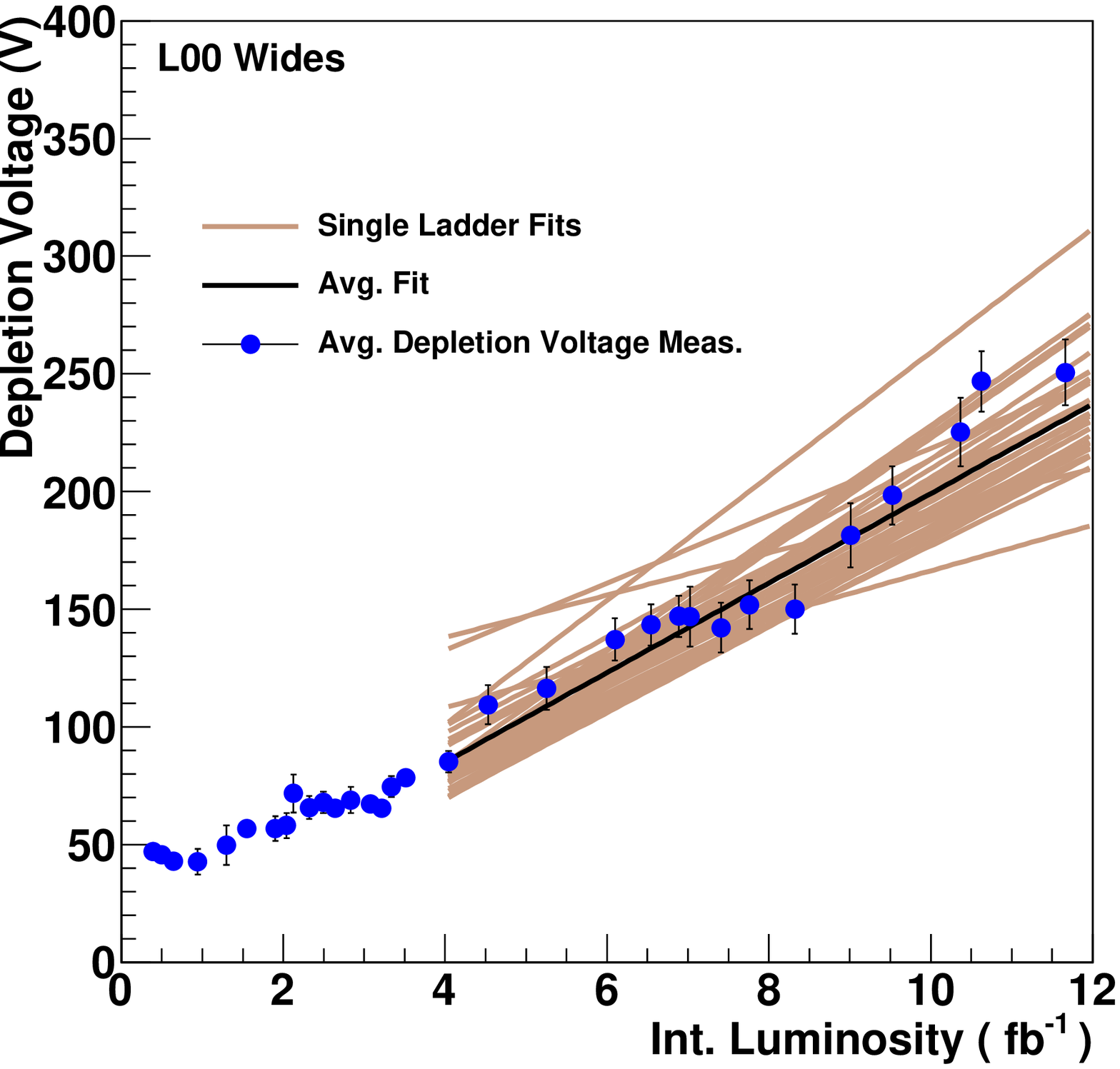}
\includegraphics[width=0.47\textwidth]{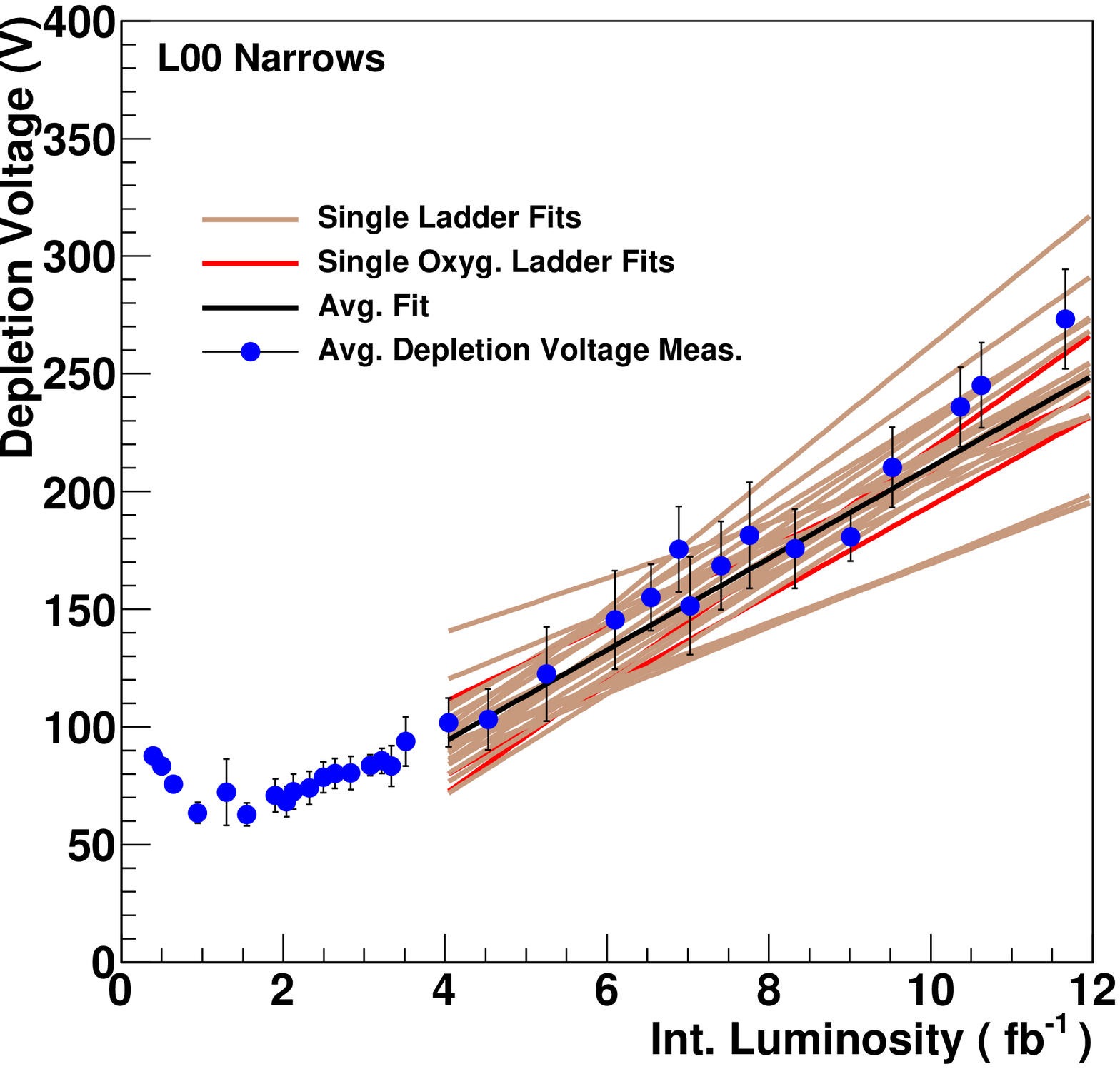}
\caption{Summary of depletion voltage measurements and fits
for L00 wide ladders (left) and narrow ladders (right).}
\label{inversionextrapolationL00}
\end{figure}

\begin{figure}[h]
\centering
\begin{tabular}{cc}
\includegraphics[width=0.47\textwidth]{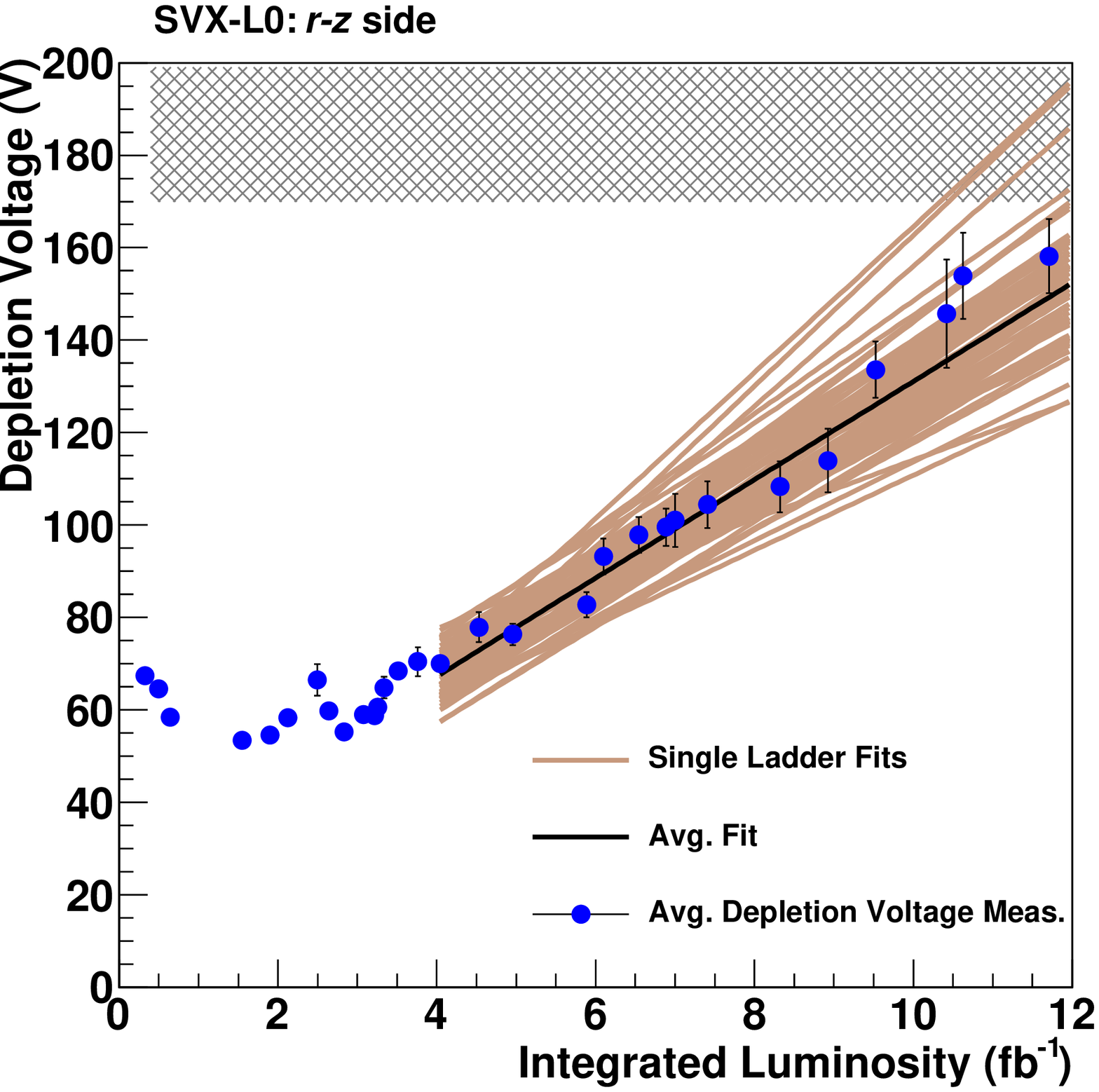} &
\includegraphics[width=0.47\textwidth]{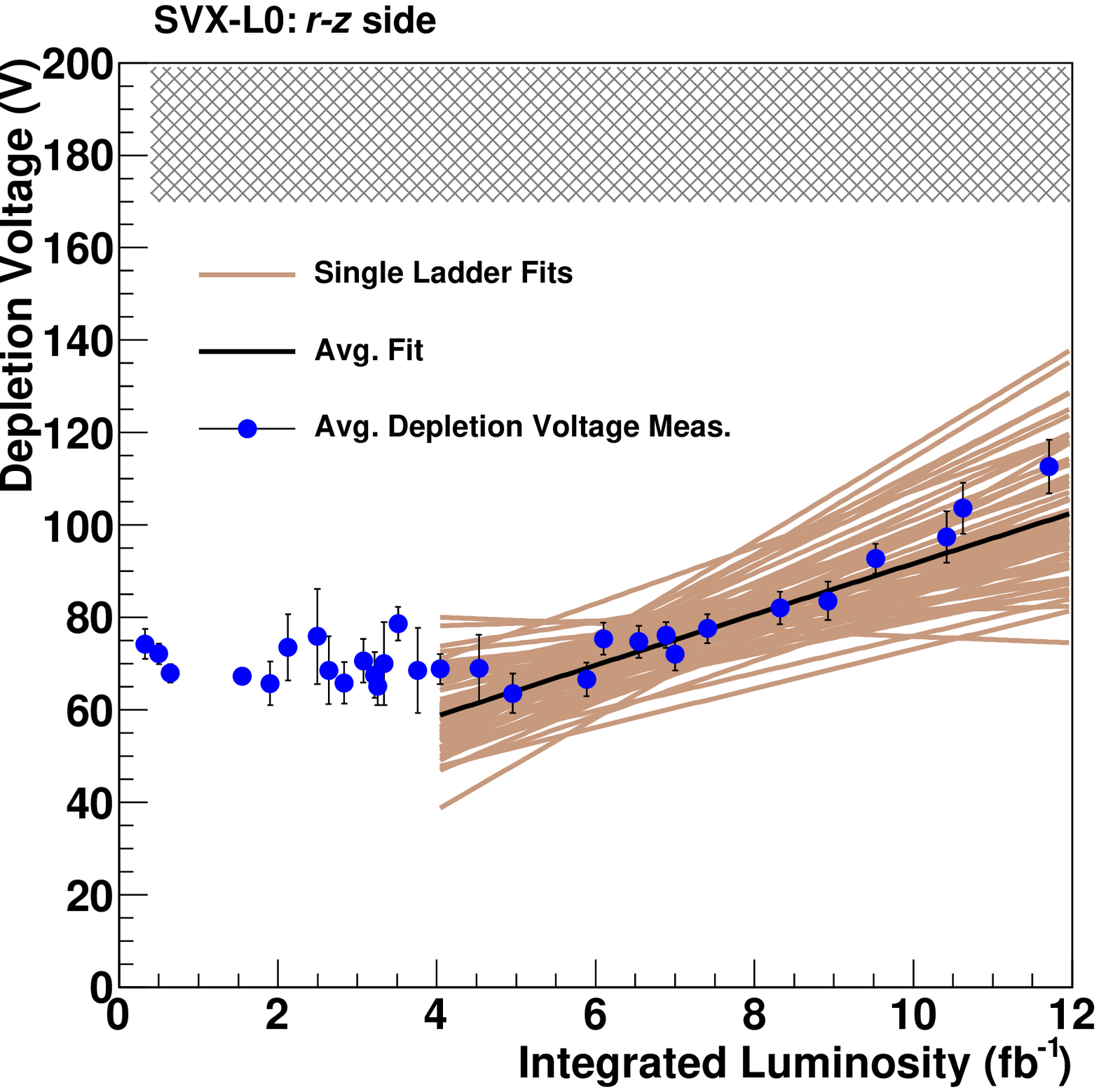}
\end{tabular}
\caption{Summary of depletion voltage measurements and fits
for  the \rphi\ side or p-side (left) and the 
\rz\ side or n-side (right) of SVX-L0 sensors.}
  \label{inversionextrapolationL0}
\end{figure}

Fig.~\ref{inversionextrapolationL00} shows the linear
fits for the individual L00 ladders as gold or red lines and the average 
over all the ladders as a black line and blue points.  The predicted depletion
voltages for all L00 ladders lie well below the power supply limit of
500~V and the sensor breakdown region that starts at 650~V, and they
were fully depleted through the end of Run~II.
Fig.~\ref{inversionextrapolationL0} shows the linear fits
for the \rphi\ side (p-side) of the first layer of SVX-II
(SVX-L0).  The fits for individual ladders are shown as red lines
and their average as a black line.  The blue points are the average measured
depletion voltage for all ladders.  The power supply limit for these sensors 
is 250~V and sensor breakdown was expected 
in the range 170-270~V, indicated with a shaded region.

In agreement with the projections, roughly one third of the SVX-L0
ladders were not fully depleted for an operating voltage of 165~V at
the end of Run~II.  The
performance of these underdepleted ladders was only slightly compromised:
 the charge collected on the
p-side was reduced, while the charge collected on the n-side was unaffected.
Because of the risk of damage to the sensor, we decided not to
operate any ladders above 165~V until the hit efficiency of the ladder
began to decrease.  Only one ladder reached
this condition by the end of the run.  That is, the hit efficiency of all 
but one of the underdepleted ladders was still maximal at the end of the 
run despite the reduced charge collection on the \rphi\ side (p-side).

\subsection{Surprises in the behavior of irradiated sensors}
In the traditional model for the electric field in a reverse-biased silicon 
sensor, the field increases linearly through the bulk when the bias voltage is 
applied. There is however evidence that trapped charge in heavily irradiated 
sensors dramatically affects this simple picture of the field inside the 
sensor.  In CDF, we have observed evidence that, after irradiation, the field 
was non-uniform and extended from both sides of the sensor. This resulted in 
a much longer lifetime for SVX-L0 operating with {\em safe} bias voltages than 
had been anticipated.

In undamaged CDF sensors, the electric field was highest at the p-side of a 
reverse biased pn diode junction, and decreased linearly through the bulk 
material. At bias voltages less than the depletion voltage, the electric field 
at the n-side was zero and essentially no signal was induced in the n strips 
as particles passed through the sensor.  The general understanding of radiation 
damage to silicon sensors when the CDF detector was built (summarized in the 
{\em Hamburg model}~\cite{Moll:1999kv}) was that radiation induced crystal 
damage made the bulk material increasingly more p-type.  At sufficiently high 
dose, the n-type bulk material was expected to become effectively p-type (this 
is referred to as {\em type inversion}) and the junction side of the detector 
was expected to move from the p-side to the n-side. After type inversion, 
the electric field in the sensor was expected to be highest at the n-side 
diode junction and to decrease linearly through the bulk material. It was 
expected that essentially no signal would be recorded on the p-side at bias 
voltages less than the depletion voltage.  

This behavior is now understood to be a consequence of the properties of 
damaged silicon with an applied bias voltage.  In heavily irradiated silicon, 
the trapping of leakage current charge carriers dramatically affects the 
electric field inside the sensor.  Leakage current is generated approximately 
uniformly throughout the thickness of the sensor.  Electrons carry charge 
toward the n-side and holes carry charge toward the p-side.  This means that 
the density of moving electrons is highest near the n-side and the density of 
moving holes is highest near the p-side. Because the equilibrium number of 
trapped charges depends on the density of moving charges as well as trapping 
probabilities and trap lifetimes, the density of trapped electrons is highest 
near the n-side of the sensor and the density of trapped holes is highest near 
the p-side of the sensor.  These trapped charges create an electric field with 
maxima at both sides of the sensor.  The importance of trapped charges to the 
static field in heavily irradiated sensors was pointed out by Eremin, 
Verbitskaya, and Li in 2002~\cite{Eremin:2002wq}. Swartz~{\em et al.} have 
tuned a two trap model to fit CMS pixel beam test results, including 
temperature dependence~\cite{swartz}.

\begin{figure}
\begin{center}
\begin{tabular}{cc}
\includegraphics[scale=0.3]{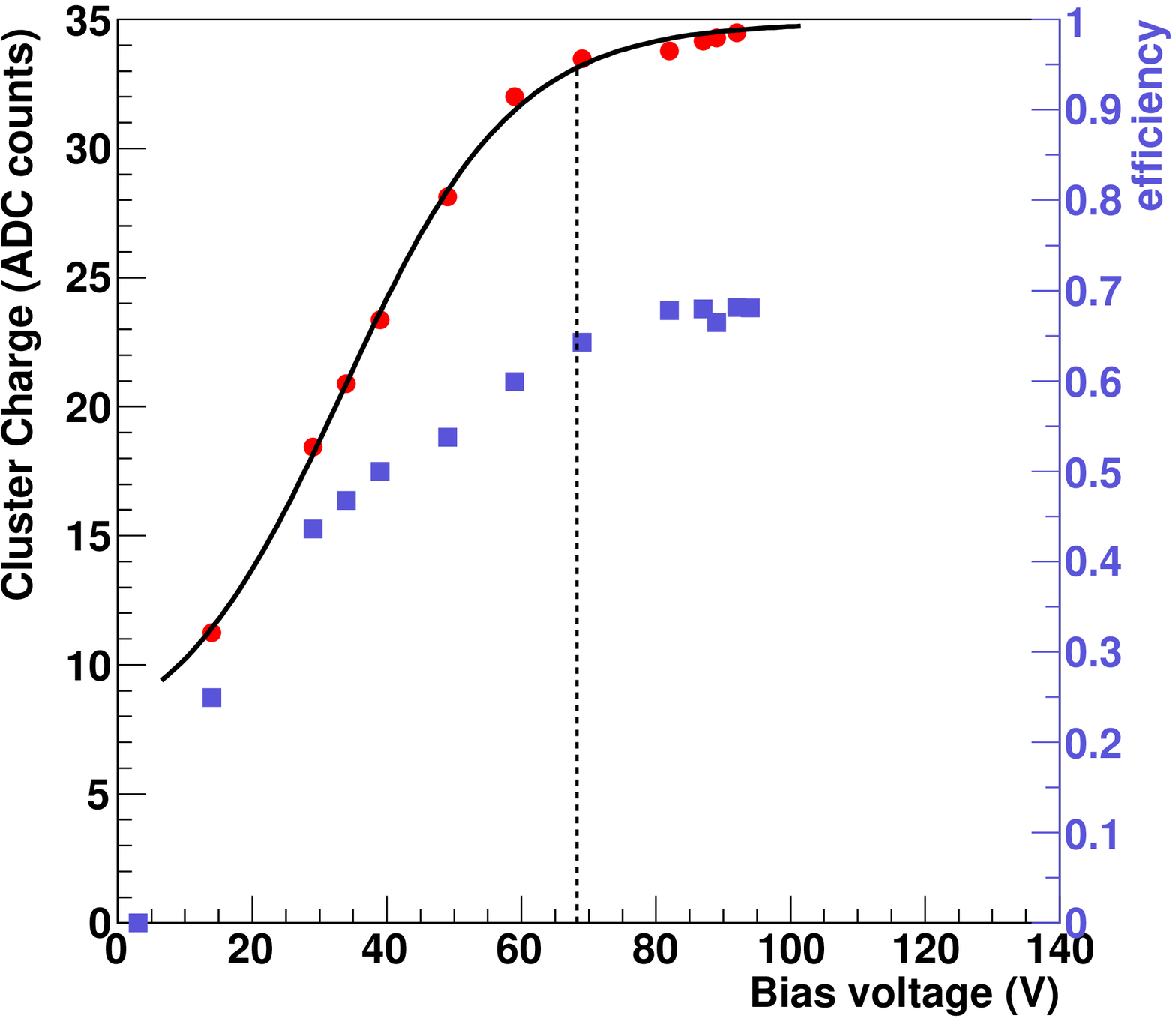}  &
\includegraphics[scale=0.3]{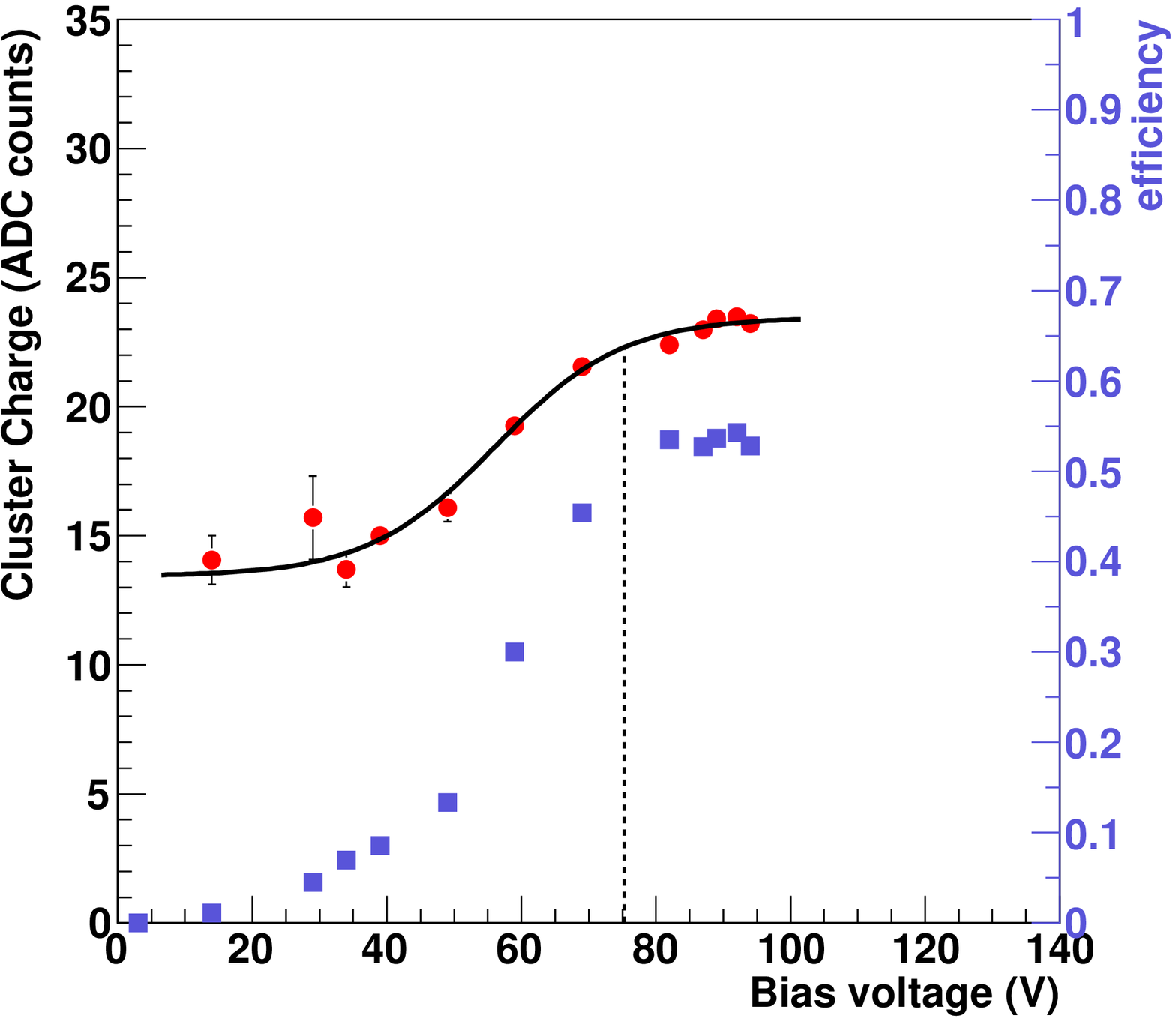} \\
\includegraphics[scale=0.3]{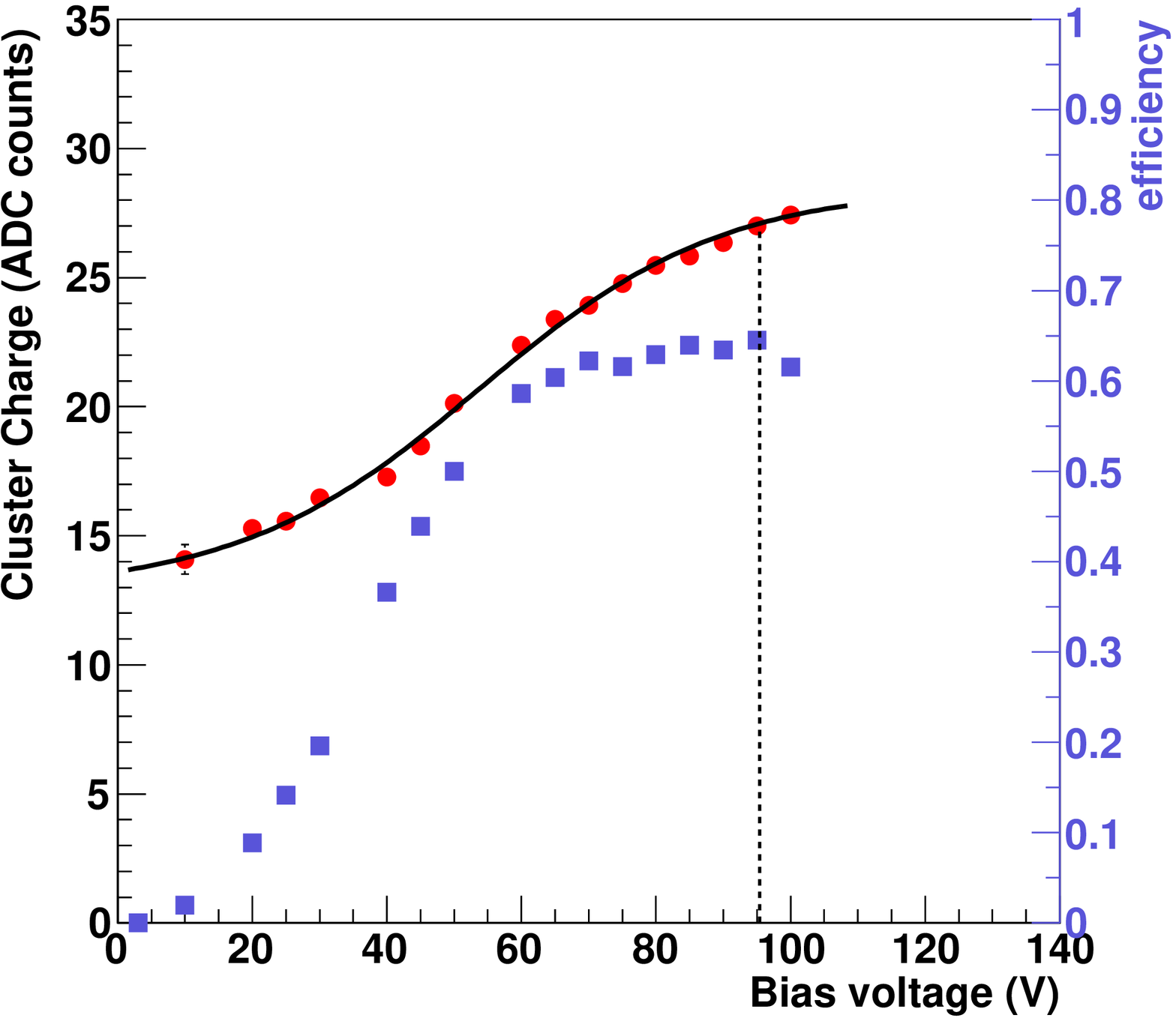}  &
\includegraphics[scale=0.3]{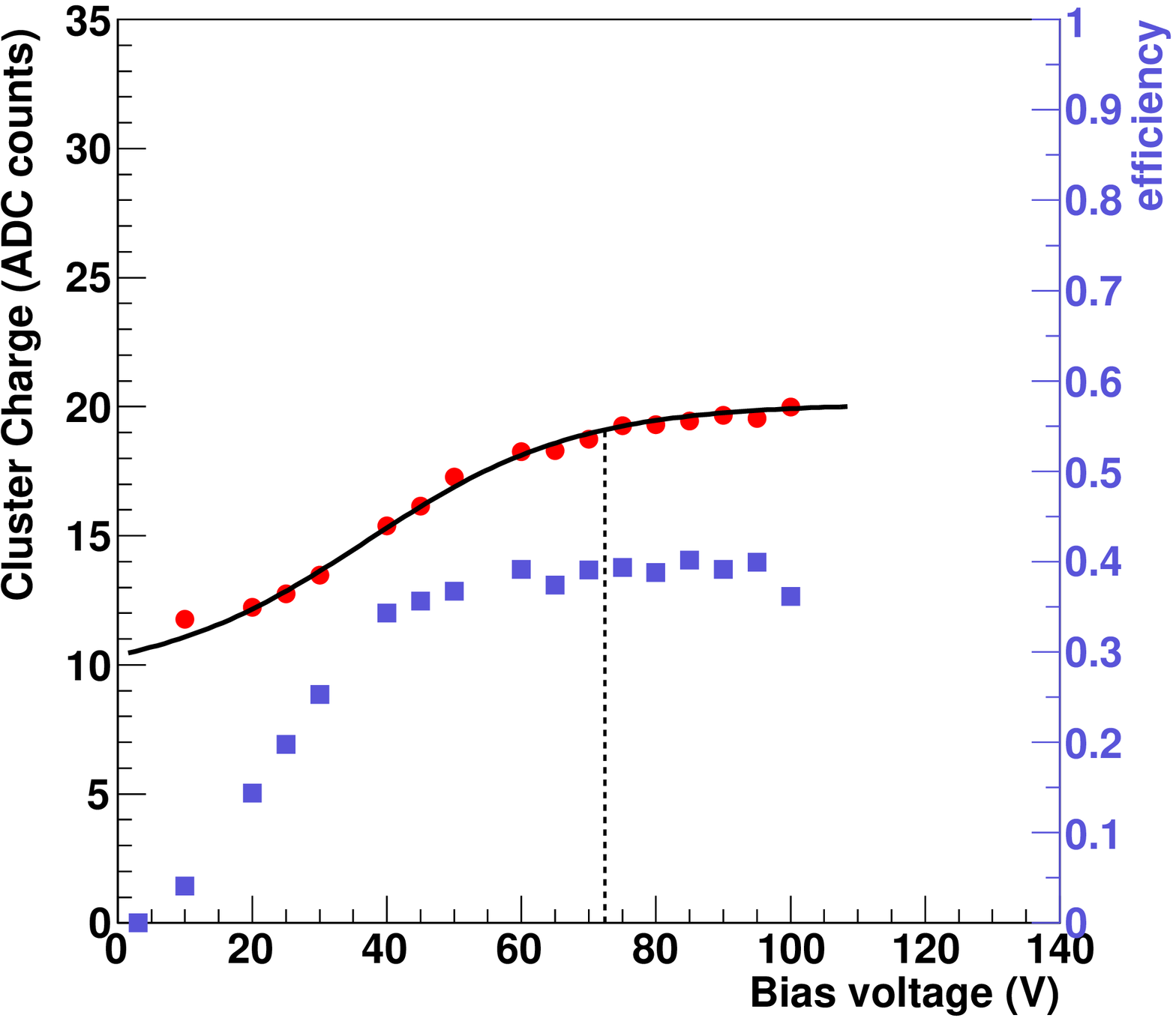} 
\end{tabular}
\end{center}
\caption{The cluster charge (circles) and efficiency (squares) as a function 
of bias voltage for the p-side (\rphi, left) and n-side (\rz, right) of a 
typical SVX-L0 sensor. The dashed line indicates the depletion voltage 
extracted from the measurement.
The upper plots contain data taken after 0.3~fb$^{-1}$ of delivered luminosity, 
the lower plots data taken after 6.9~fb$^{-1}$
for a sensor that inverted around 1.5~fb$^{-1}$.}
\label{dj}
\end{figure}

The data from the signal scans taken after significant radiation exposure are 
consistent with an electric field peaking at both faces of the sensor, and 
clearly inconsistent with the naive expectation of a linearly decreasing field 
with a single maximum at the n-side.   Fig.~\ref{dj} shows the signal scan 
data delivered luminosities of 0.3~fb$^{-1}$ (above) and 6.9~fb$^{-1}$ (below) 
for a typical sensor in SVX-L0. Because the signal was induced primarily by 
the motion of charge carriers in the depleted region adjacent to the electrode,
the measured cluster charge for a particular voltage was a measure of the size 
of a possible depleted region adjacent to the electrode. Because of the readout 
thresholds for the SVX sensors, clusters below 10-15~ADC counts are not 
detected.  However, the fraction of tracks with clusters above this threshold, 
shown as blue squares in Fig.~\ref{dj}, increased as the average charge 
collected increased. The upper measurement in Fig.~\ref{dj} was done when the 
sensor was only slightly irradiated.  The charge collection began at smaller
voltages on the p-side than the n-side, compatible with
a depleted region that began at the pn junction and grew toward the 
n-side electrode as the bias voltage increases.   The lower measurement
was done after 6.9~fb$^{-1}$ of luminosity, and the charge collection began
at similar bias voltages for each side and increased similarly with increasing 
voltage.  This latter behavior was compatible with an electric field that had 
two maximums, one at either face of the sensor, creating two depleted regions 
that started at either face and grew toward the center of the sensor as the 
bias voltage increased. Similarly, for the p-side of a single sided L00 sensor 
after 4~fb$^{-1}$ of luminosity (post-inversion), Fig.~\ref{DVmeas}(c)
shows an early onset and gradual increase of charge collection,
again consistent with a doubly peaked electric field.

The unexpected electric field behavior had an important consequence for the 
longevity of the CDF silicon detector. The SVX-L0 ladders could be operated in 
a slightly underdepleted state with only a small loss in charge collection and 
no loss in hit efficiency for a short time after the depletion voltage 
exceeded the maximum safe operating voltage. Because the $b$-tagging efficiency 
(Section~\ref{sec:b_tag}) of the detector
was insensitive to a small loss in charge collection in a fraction of 
the SVX-L0 sensors, it was decided to operate these 
sensors slightly underdepleted instead of risking damage at higher 
bias voltages.

\subsection{Signal-to-noise ratio}
\label{sec:s2n}
During Run II, the signal-to-noise ratio (S/N) of L00 and SVX-II sensors were 
monitored using well-measured tracks from events selected by the low momentum 
dimuon trigger. The signal $S$ was defined as the summed charge of a cluster 
of strips associated with a track and corrected for path length. 
The noise for individual strips was measured during special calibration runs 
performed bi-weekly with beam, as described in Section~\ref{sec:calib}.  
The noise of a cluster $N$ was defined as the average noise of the individual 
strips belonging to the cluster.

\begin{figure}
\centering
\includegraphics[scale =0.30]{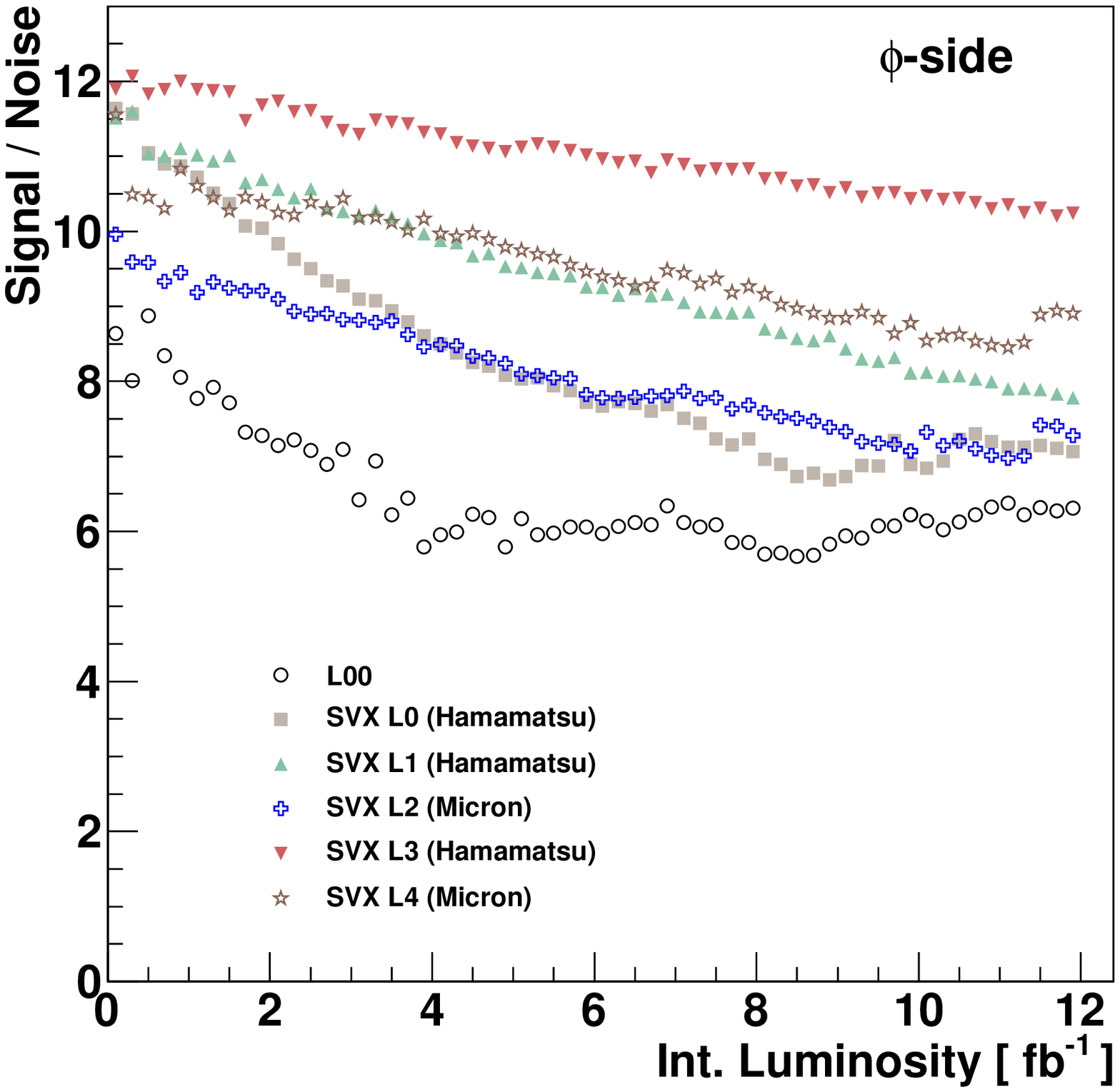} 
\includegraphics[scale =0.30]{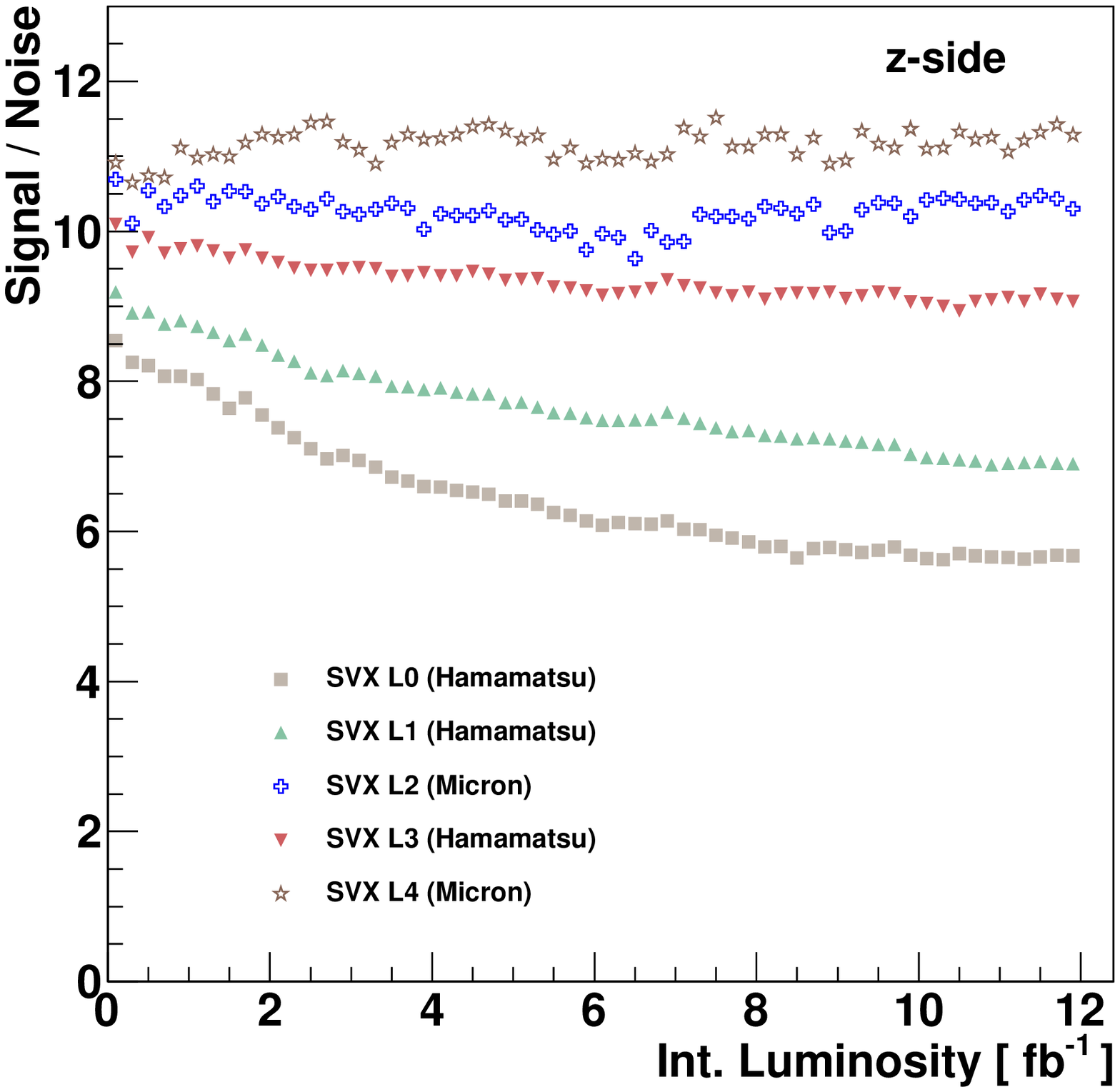}
\caption{Measured signal-to-noise ratio for L00 (left), the \rphi\ side of 
SVX-II (left) and the \rz\ side of SVX-II (right).}
\label{fig:snPlot}
\end{figure}

Fig.~\ref{fig:snPlot} shows the average measured $S/N$ ratio for L00 and
SVX-II, separately for the \rphi\ and \rz\ sides. All ladders that operated 
consistently well throughout
Run II are included,  corresponding to roughly 75\% of all ladders.  As
expected, the $S/N$ ratio decreased more quickly for L00 and SVX-L0
since they were closer to the interaction point and suffered from more
radiation damage.  The dip in $S/N$ values for L00 and SVX-L0 near 
$8 \mathrm{fb}^{-1}$ corresponded to a period of slight underdepletion.

\begin{figure}
\centering
\includegraphics[scale =0.30]{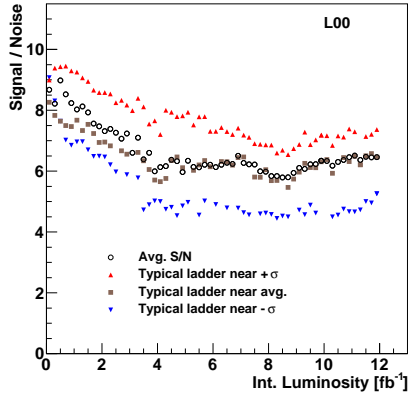}
\caption{Measured signal-to-noise ratio for the L00 average, and for
  three ladders that represent typical ladders close to the average,
  and close to the $\pm 1 $ RMS signal-to-noise values with respect to
  the average.  The RMS represents the spread of signal-to-noise
  values for the various ladders included in the average.}
\label{fig:snPlot_typ}
\end{figure}

We observed plateaus in the ratios in the L00 and
SVX-L0 $S/N$ ratios beginning at $5\ \mathrm{fb}^{-1}$ and $8\
\mathrm{fb}^{-1}$, respectively.  To verify that the leveling-off was
not an artifact of the averaging, the $S/N$ curves for the individual
ladders were investigated.  Fig.~\ref{fig:snPlot_typ} shows the
$S/N$ trends for the overall L00 average and for three typical
ladders whose $S/N$ values are close to (a) the overall L00 average,
(b) the plus-one RMS variation with respect to the overall average,
and (c) the minus-one RMS variation with respect to the overall
average, where the RMS was defined as the spread of $S/N$ values for
the ladders used in the average.  As we observed a plateau-like feature
for each ladder, we concluded that it was not due to an averaging
artifact.  The source of the plateau remains under investigation.

%
%
\section{Physics performance of the silicon detector}
\label{sec:perf}
Good performance of the silicon detectors was vital to the success of 
CDF's physics program. 
In this section we present some of the performance quantities which directly
impact the results from analyses requiring silicon tracks or
displaced secondary vertices found by SVT. Reported here are studies on the
impact parameter resolution with and without L00, the b-tagging efficiency and
the SVT efficiency.

\subsection{Impact parameter resolution}
The impact parameter, $d_{0}$ is defined as the shortest distance in the \rphi\ 
plane between the beam line and the trajectory of the particle obtained from 
the track fit. The impact parameter resolution $\sigma_{d_0}$ is a key 
performance indicator of the CDF silicon detector. This resolution affects 
identification of long-lived hadrons as well as the ability to study 
time-dependent phenomena, such as the mixing of $B_s$~\cite{cdfbsmixing} and 
charm mesons~\cite{cdfcharmmixing}. The detector provided good impact parameter 
resolution. L00 improved the performance particularly for  particles with 
low momentum or that passed through large amount of passive 
material~\cite{bbraunim}.

The resolution is parameterized as a function of the particle transverse 
momentum $p_T$ as
\begin{equation}
\sigma_{d_0}=\sqrt{A^2+\left(B/p_{\rm T}\right)^2 + r_{\rm{beam}}^2},
\label{eq:resolution}
\end{equation}
where $A$ is the asymptotic resolution parameter, and $B$ is the 
multiple-scattering component. The finite beam size $r_{\rm{beam}}=32~\mu$m 
accounts for the uncertainty in the location of the primary interaction.  

\begin{figure}[hbt]
\centering
\includegraphics[width=0.45\columnwidth]{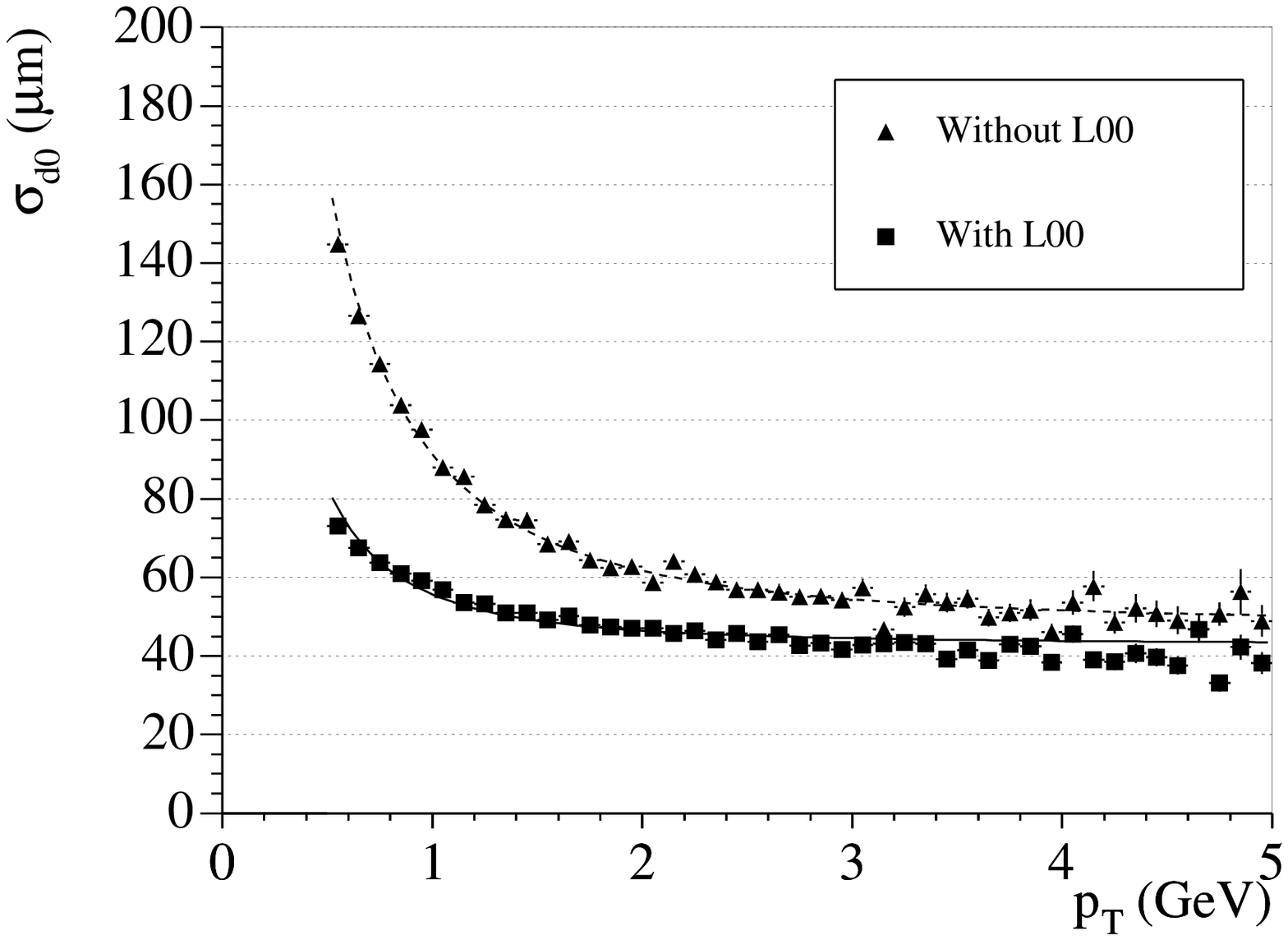}\hglue 0.2cm
\includegraphics[width=0.45\columnwidth]{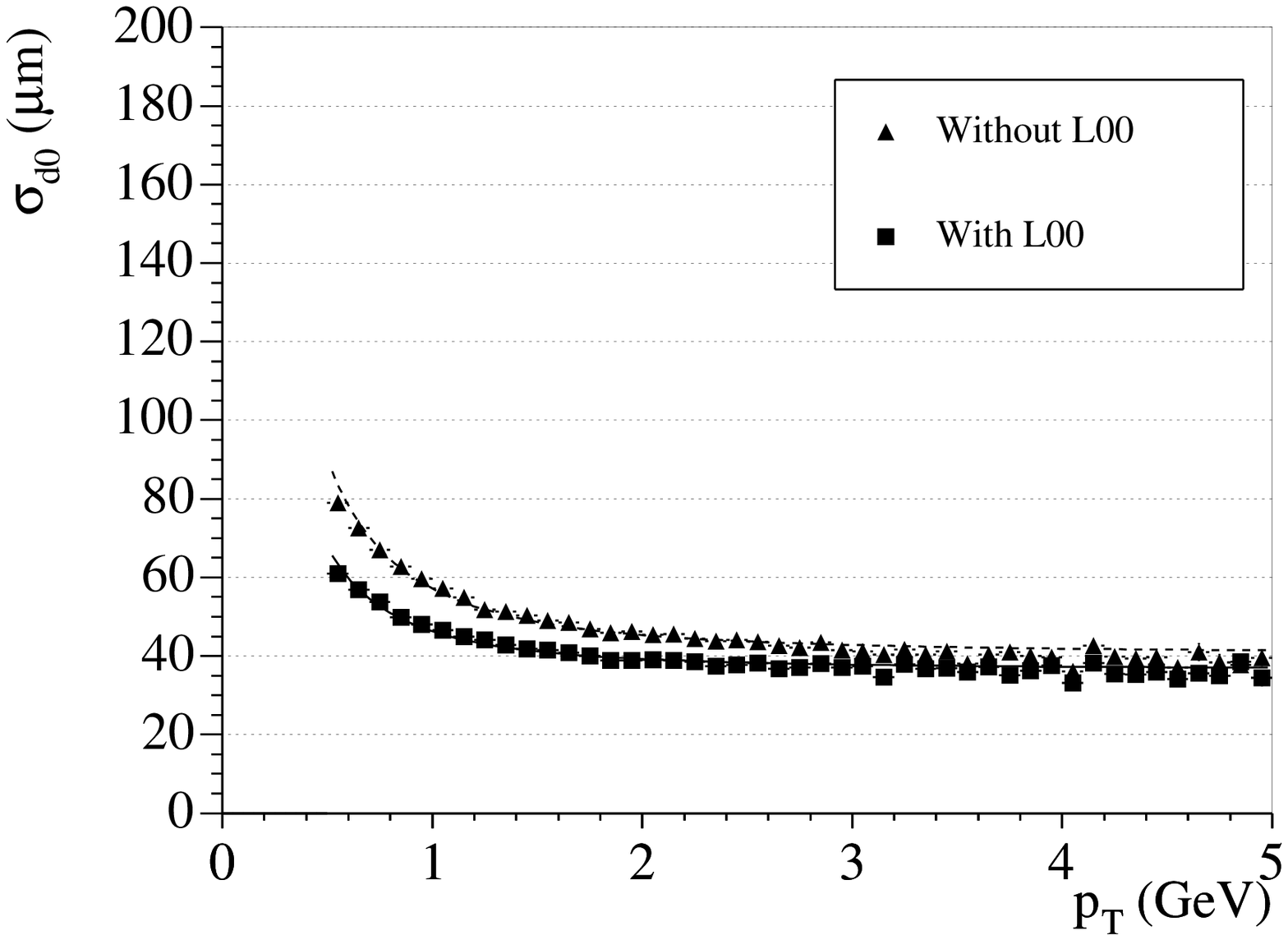}
\caption{\label{fig:resolution}
Impact parameter resolutions for tracks as a function of track $p_{\rm T}$.
Both plots show the resolutions for tracks before and after the addition of
L00 hits.  The plot on the left shows the resolutions for tracks that pass 
through readout hybrids which are mounted on some of the sensors, and the
plot on the right shows the performance for tracks that do not pass through 
hybrids. These plots include the $r_{\rm{beam}}^2$ term in 
Eq.~\ref{eq:resolution}.
}
\end{figure}

Fig.~\ref{fig:resolution} shows the fitted widths of the cores of the 
impact parameter distributions for tracks with and without L00 hits as a 
function of the track $p_{\rm T}$. 
%
The inclusion of L00 enhances the impact parameter resolution at low momentum. 
For tracks that pass through the electrical readout hybrids, the impact 
parameter resolution is somewhat degraded.

Table~\ref{tab:resolution} lists the fit parameters for the tracks with and 
without L00 hits that do not pass through the hybrids, shown in the right plot 
of Fig.~\ref{fig:resolution}. Tracks used in these fits originate dominantly 
from interactions in the central region of the detector (in $z$), where using 
a constant resolution term for the beam envelope is a reasonable approximation. 
The addition of L00 hits to tracks vastly improves the impact parameter
resolution for low momentum tracks and provides a relatively modest improvement
of the asymptotic resolution at high momentum. Furthermore, knowledge of the 
impact parameter is limited by the beam size unless a primary interacting 
vertex is reconstructed with many high-momentum tracks.

\begin{table}
\centering
\caption[Fits for impact parameter resolutions.]{\label{tab:resolution}
Fit parameters for the resolutions, shown in the right plot of 
Fig.~\protect{\ref{fig:resolution}}, for tracks that do not pass through 
the hybrids. The definitions of the fit parameters are given in 
Eq.~\protect{\ref{eq:resolution}}.  A fixed beam size, 
$r_{\rm{beam}}=32$~$\mu$m, is assumed.}
\begin{tabular}{|c|c|c|}\hline
Track Category & $A$~($\mu$m) & $B$~($\mu$m) \\ \hline
No Hybrid, No L00 & $28.6\pm 0.3$ & $35.5\pm 0.3$ \\
No Hybrid, L00    & $17.8\pm 0.2$ & $28.4\pm 0.1$ \\ \hline
\end{tabular}
\end{table}

\subsection{b-Tagging efficiency}~\label{sec:b_tag}
Many of the physics goals of the CDF experiment rely on the identification of 
weakly decaying bottom hadrons.  The mean lifetime of these hadrons is 
approximately 1.5~ps, and the mean decay length is order of a few millimeters.  
The fact that the weakly decaying hadrons have large boosts means that the 
particles from the decay travel in approximately the same direction as the 
parent hadron, with their kinematic distributions depending on the mass of the 
parent hadron.  Precise measurement of the track positions allows tracks 
originating from displaced vertices to be distinguished from tracks that 
originate at the primary vertex.  Most jets of hadrons produced in $p{\bar{p}}$ 
collisions do not contain bottom or charm hadrons, and very strong rejection
of falsely tagged light-flavor jets is another important figure of merit for 
the tracker.

CDF uses a displaced-vertex algorithm, SECVTX~\cite{secvtx}, to identify --- 
or \emph{b-tag} --- secondary vertices that are significantly displaced from 
the beamline.  It has two configurations, referred to as ``loose'' and 
``tight'', which refer to the track and vertex requirements used to form the 
displaced-vertex candidates.  With the loose requirements, more displaced 
vertices from heavy hadron decay are identified than the tight requirements, 
but with a higher rate of falsely tagged light-flavor jets.

The assignment of silicon hits to tracks has a large impact on the efficiency 
of the algorithm to identify the decays of heavy hadrons, as at least two 
well-measured tracks are required to form a displaced vertex, and the presence 
of hits in multiple silicon layers in the inner tracking volume improves the 
chances of finding that the vertex is significantly displaced from the primary. 
The tails in the impact parameter resolution --- due to hard nuclear collisions 
with detector material, multiple scattering, and mis-assigned hits in the 
silicon detector and COT --- determine the false tag rate.

\begin{figure}[hbt]
\begin{center}
\includegraphics[width=0.45\columnwidth]{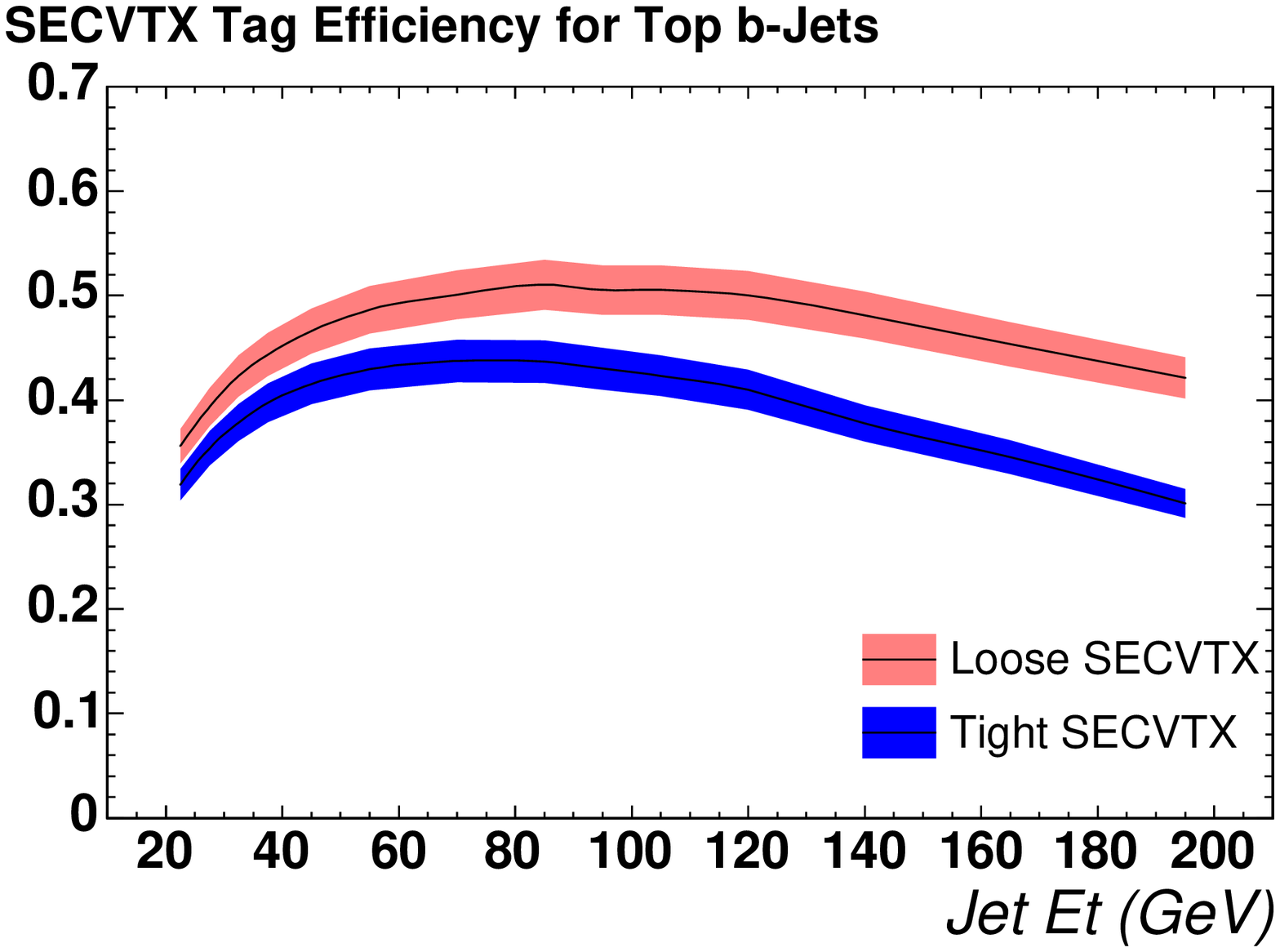}
\includegraphics[width=0.45\columnwidth]{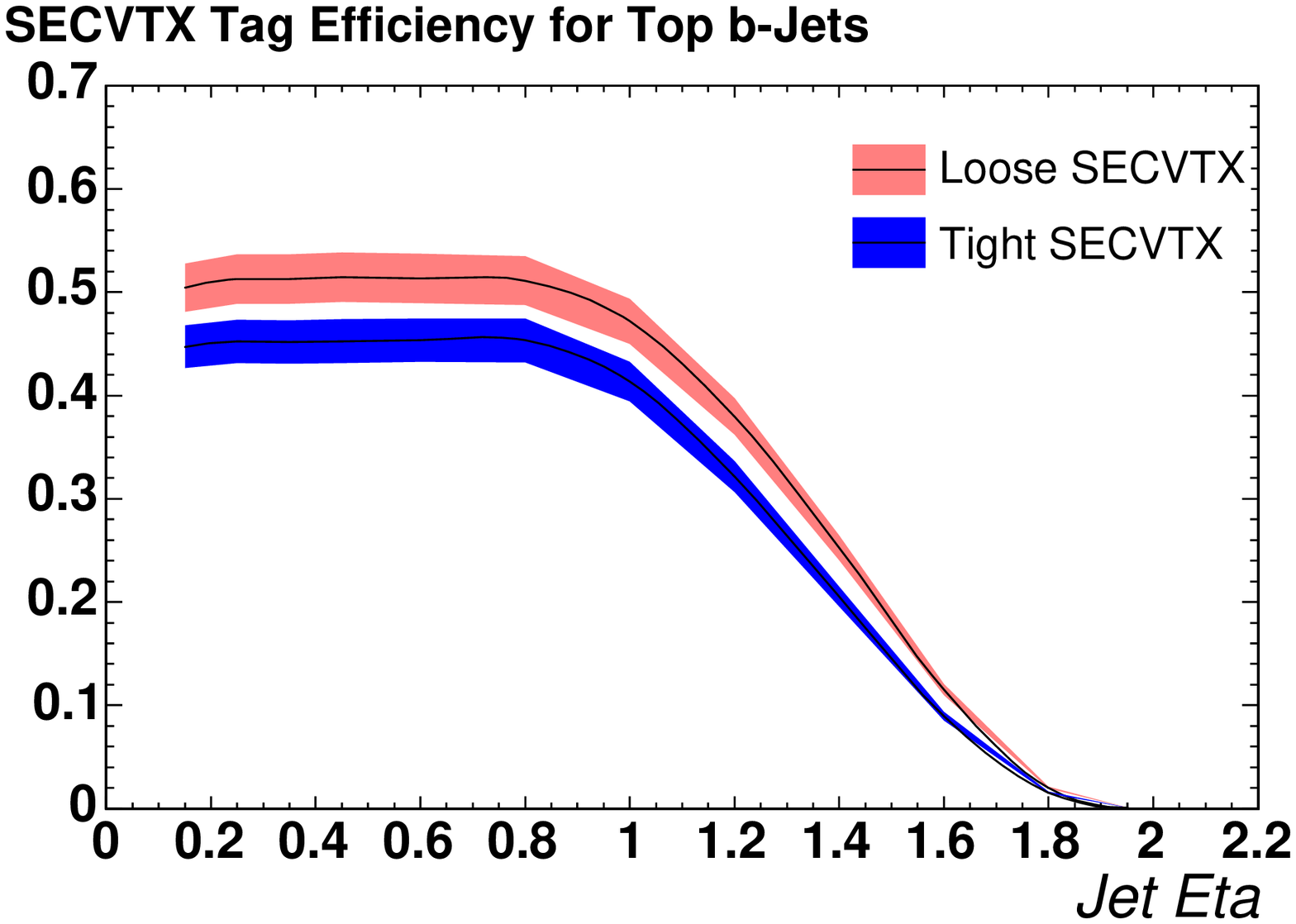}
\end{center}
\caption{
Efficiency of the displaced vertex $b$-tagger, as functions of jet $E_{\rm T}$ 
and jet pseudorapidity, for two configurations of the $b$-tagger.
The efficiency is obtained from tagging jets which have been
matched to $b$ quarks in Monte Carlo top quark decays,
multiplied by data/MC scale factors.
}
\label{fig:perf_secvtxeff}
\end{figure}

\begin{figure}[hbt]
\begin{center}
\includegraphics[width=0.45\columnwidth]{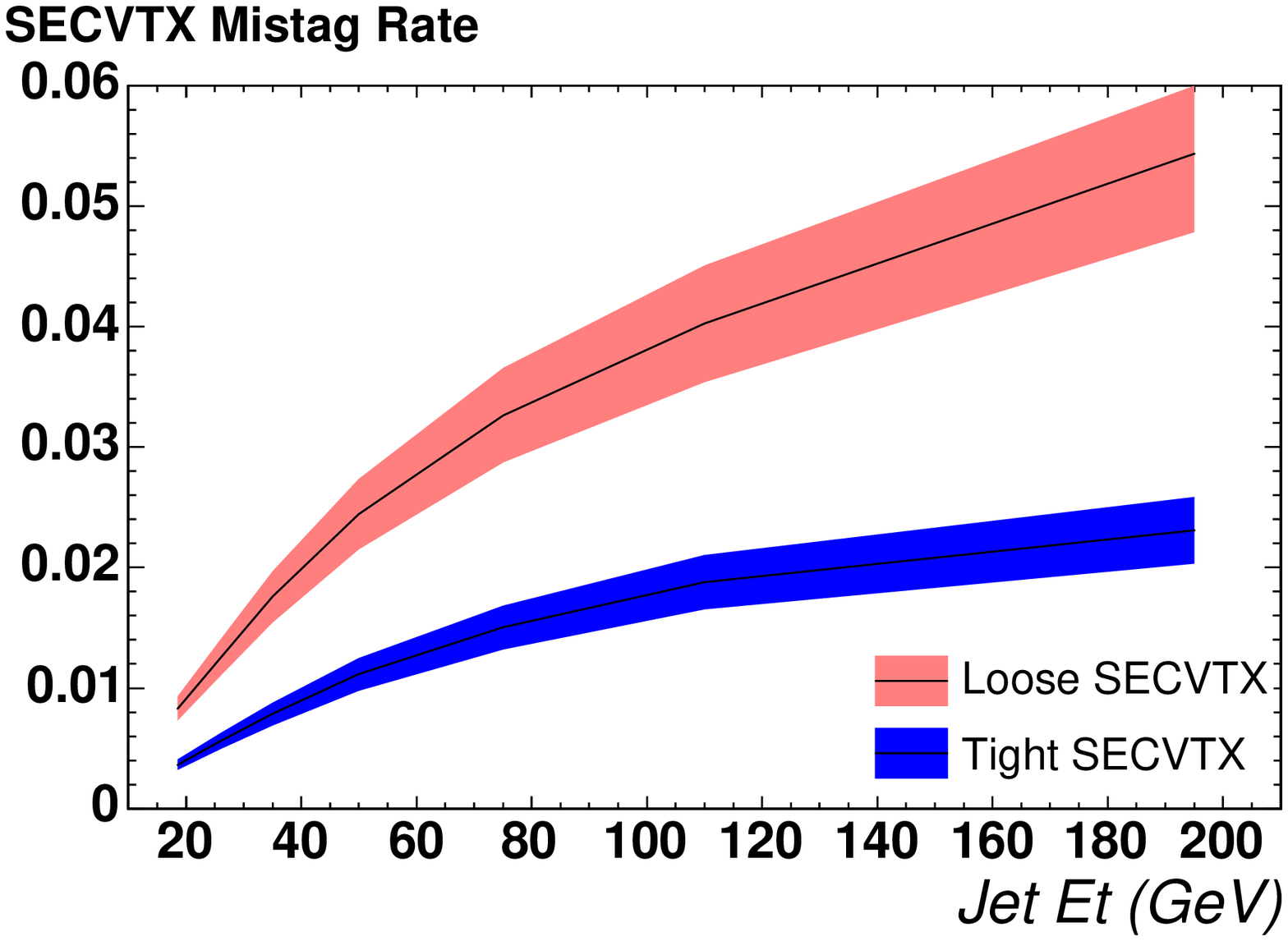}
\includegraphics[width=0.45\columnwidth]{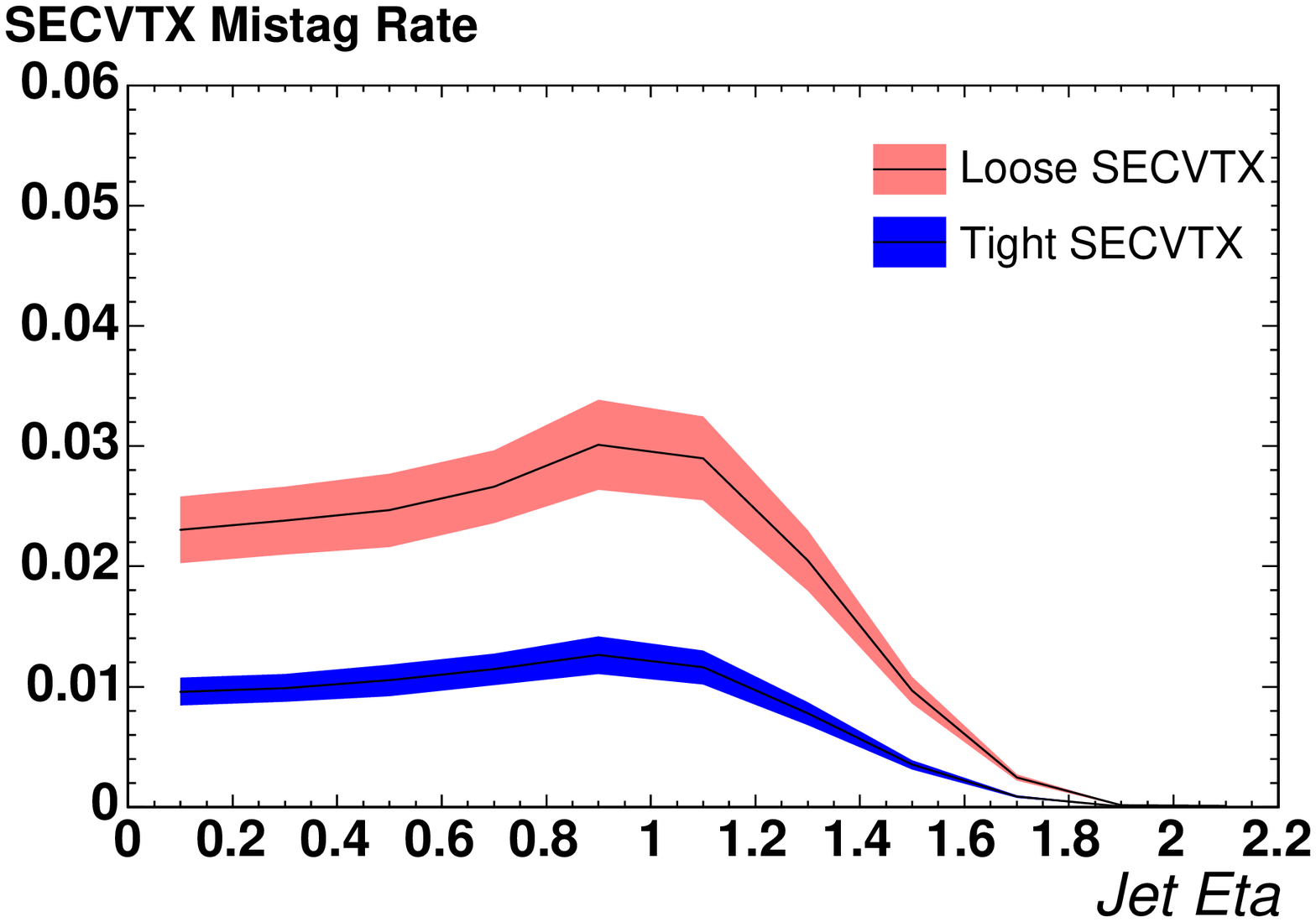}
\end{center}
\caption{
Probability of non-$b$-jets to be $b$-tagged for the 
displaced vertex $b$-tagger,
as functions of jet $E_{\rm T}$ and jet pseudorapidity, for two
configurations of the $b$-tagger.
The probabilities have been measured from inclusive jet data.
}
\label{fig:perf_secvtxmistag}
\end{figure}

The efficiency of the algorithm to identify heavy hadrons increases with the 
hadron momentum.  The efficiency is not measured directly since not all hadron 
decay products are reconstructed; instead it is parameterized as a function 
of the transverse energy $E_{\rm T}$ of the jet and shown in 
Fig.~\ref{fig:perf_secvtxeff}. The efficiency rises with momentum because 
the tracks that result from the decay are better
measured at higher energy, and the decay flight distance is also longer. At 
very high $E_{\rm T}$, the tagging efficiency
drops, as tracks begin to share hits, with the jets becoming more collimated.  
Fig.~\ref{fig:perf_secvtxeff} also shows the $b$-tag efficiency as a 
function of the pseudorapidity, $|\eta|$, of the jet.  Jets at higher
pseudorapidity pass through more material and have fewer COT hits, reducing the
tag efficiency and raising the false tag rate as shown in 
Fig.~\ref{fig:perf_secvtxmistag}. At very high pseudorapidity, the loss of 
tracking efficiency in the COT also reduces the false tag rate.

\begin{figure}[hbt]
\begin{center}
\includegraphics[width=0.45\columnwidth]{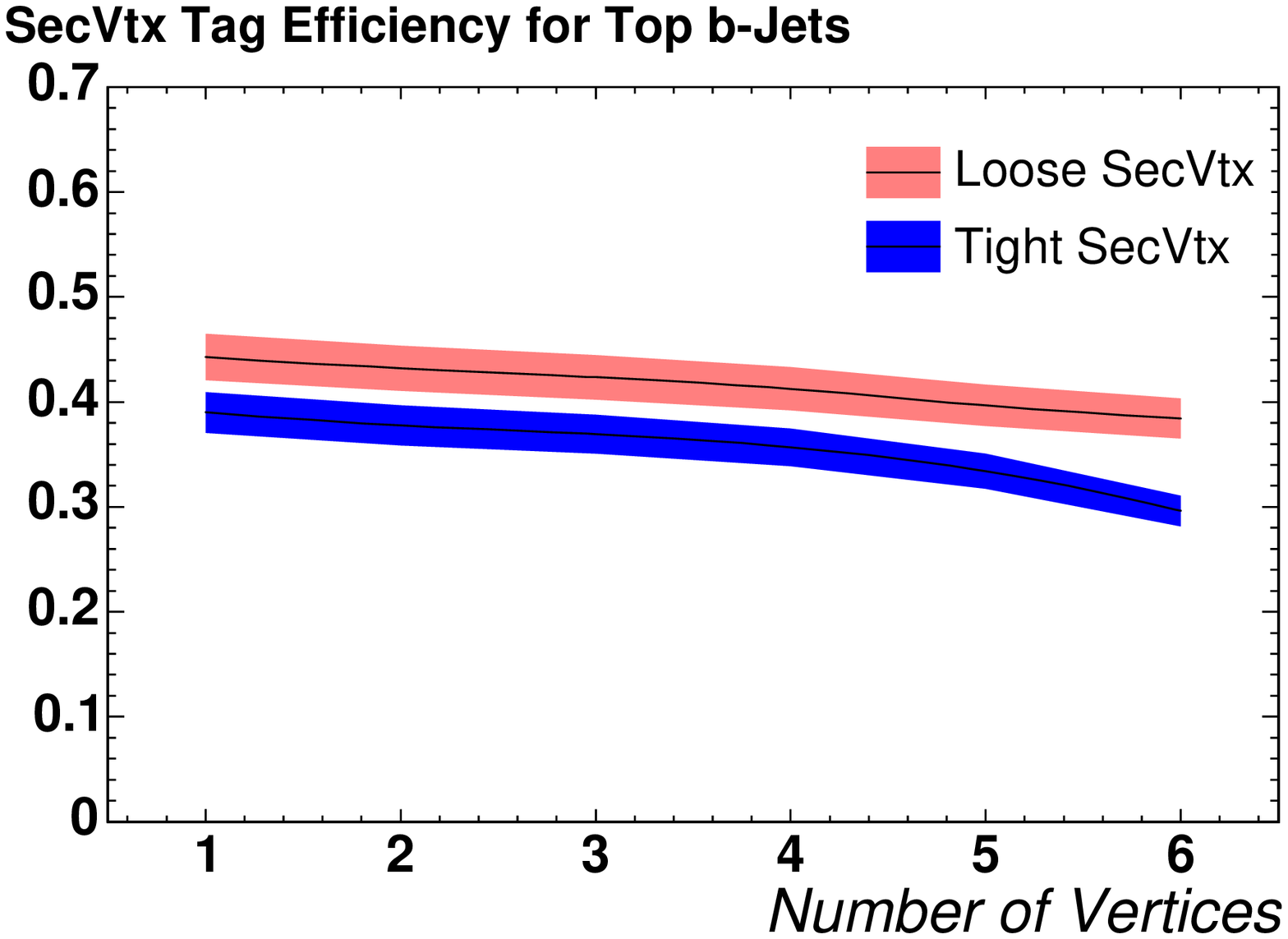}
\end{center}
\caption{
Efficiency of the displaced vertex $b$-tagger, as a function of the number of
reconstructed $p{\bar{p}}$ collision vertices per beam crossing, for two
configurations of the $b$-tagger.
The efficiency is obtained from tagging jets which have been
matched to $b$ quarks in Monte Carlo top quark decays,
multiplied by data/MC scale factors.
}
\label{fig:perf_secvtxeff_zvert}
\end{figure}

At high instantaneous luminosities, the average occupancy of the silicon 
detector and the inner layers of the COT rises. This leads to increased chance 
of assigning noise hits to the tracks as well as missed COT hits due to 
ambiguity in resolving many overlapping tracks. As a result, at higher 
instantaneous luminosities, the $b$-tag efficiency drops slightly and the 
fake tag rate rises. The $b$-tag efficiency is shown in 
Fig.~\ref{fig:perf_secvtxeff_zvert} as a function of the number of 
reconstructed primary vertices per beam crossing, which is more directly 
related to the tracking occupancy than the instantaneous luminosity.
The $b$-tagging performance is remarkably robust at high luminosities due to 
the high granularity of the silicon detector covering the low radii tracking 
region.

\subsection{SVT efficiency study}

\begin{figure}
\centering
\includegraphics[width=0.55\columnwidth]{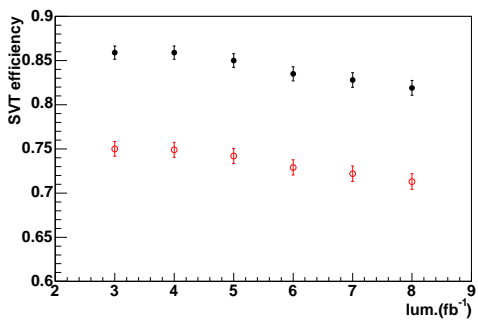}
\caption{
Prediction of the track finding efficiency of SVT as a function of total 
integrated luminosity. The efficiency is measured in di-muon events from 
$J/\psi$ decays. Closed black circles show the efficiency when only 
four layers of SVX-II are used by the SVT algorithm; empty red circles show 
the efficiency including all five layers.
}
\label{fig:svt_efficiency_study}
\end{figure}

Aging of and radiation damage to the silicon detector resulted in increased 
noise and reduction of the level of the signal (see Section~\ref{sec:s2n}).
These factors degraded the performance and efficiency of the SVT 
(see Section~\ref{sec:svt}).
A study was performed during the Tevatron run to measure the SVT efficiency 
as a function of total integrated luminosity and estimate any potential 
impact on physics analyses.
Starting from the measured level of signal and noise in the silicon detector 
for a reference data sample at 3 fb$^{-1}$, simulated samples were produced by 
applying extrapolations of the signal and noise up to 8 fb$^{-1}$. Each of the 
samples was used as input to SVT simulation software to estimate track finding 
efficiency in $J/\psi$ events at the corresponding integrated luminosity.
Fig.~\ref{fig:svt_efficiency_study} shows the prediction of the SVT 
efficiency as a function of total integrated luminosity for four (open 
circles) and all five (closed circles) layers of SVX-II.

A decrease of SVT efficiency of about 4\% was predicted between 3-8 
fb$^{-1}$, while the impact of losing an SVX-II layer was about 13\%.


%
%
\section{Summary}
\label{sec:sum}
The CDF silicon detector, consisting of the SVX-II, ISL, and L00
components, was designed to withstand only 2-3~fb$^{-1}$ of 
integrated $p{\bar{p}}$ collision luminosity and was expected to be
replaced in 2004 by an upgrade. It ran successfully for over 10 years,
through 2011, and was exposed to about 12~fb$^{-1}$ of integrated
luminosity.  About 90\% of its ladders took data with high efficiency
until the end of Tevatron Run II.  This was an unprecedented feat 
compared to any silicon detector in the same category prior to
it. It was also the first silicon detector system to be incorporated
into a hardware trigger to identify tracks from secondary vertices.
It provided precise measurements of the trajectories of charged
particles which were important to identify and measure heavy-flavored
hadrons, which in turn were crucial to CDF's physics program, including
top quark, $b$~hadron, and Higgs boson physics.

The detector consisted of about 722,000 readout channels,
with approximately 500 independent ladders, which required
voltages to bias the sensors as well as run the data acquisition
electronics mounted on the detector.  Elaborate data acquisition,
trigger, cooling, and monitoring systems were required to collect
the data used for analysis.  The detector itself was located inside
the drift chamber in a volume heavily congested with cabling, cooling
pipes, and the beam pipe, that rendered it largely inaccessible for repair.
Due to its inherent complexity, the detector operations involved
detailed procedures and required specially trained experts to execute
them. Keeping up with the loss of experts and training new ones to
replace them was one of the major challenges faced by the silicon
operations team. In addition, inaccessibility of the detector required
that every detector access be planned elaborately and well in advance
to take advantage of the Tevatron shutdowns.  Future
experiments must take extra care in designing the accessibility
aspects of their subsystems to avoid these situations.

The silicon detector system had its share of failure modes, most of
which were addressed during commissioning and the rest 
were mitigated during
the operation of the detector.  The most serious failure modes, those
that required immediate response, such as power supply failures,
cooling system failures, and unsafe beam conditions, had dedicated
hardware and software systems designed to protect the silicon detector
from damage.

Unanticipated failure modes and exposures to detector damage became
evident as Run~II of the Tevatron progressed.  These had been
addressed with hardware modifications, monitoring, review, and
improved operational procedures to reduce the chance of damage to the
detector.  For example,
the Lorentz force on bond wires perpendicular to the magnetic field,
connecting one side of a ladder to another, on rare instances caused
the wires to vibrate at their mechanical resonant frequency and
eventually break (Section~\ref{DAQ:Wirebond}).  A hardware device was
devised and installed to stop data acquisition when the trigger system
requested readouts at high frequencies and at regular intervals.
Another issue involved spontaneous energizing of the kicker magnets or
separators sparking at unexpected times, which steer the beams near
the silicon detector and caused damage to it due to the acute
radiation dose (Section~\ref{sec:beam:kickerPrefire}).  The addition
of a collimator near CDF and improvement to the high voltage
conditioning of the separators had minimized the effect of such
incidents on the silicon detector. Careful monitoring of beam
conditions, automatic ramp-down of the bias voltage in case of bad
beam conditions and beam abort requests from the diamond sensors had
also protected the silicon detector from damage due to beam incidents.

Certain electronic components necessary for the functioning of the
silicon detector were located in the collision hall.  Radiation caused
temporary and permanent failures of data acquisition electronics,
particularly FPGAs (Section~\ref{DAQ:SEU}).  The power supplies were
also susceptible to radiation-induced failures
(Section~\ref{subsec:psradvulnerable}).  Many of the temporary
failures could be addressed simply by resetting and re-initializing
the affected components, and these procedures were highly automated so
that the reset and recovery to the nominal data-taking configuration
resulted in a minimum of downtime.  Permanent failures of collision
hall electronics typically required short accesses to replace the
affected components.  A sufficiently large pool of spares was kept on
hand to maintain a high availability of the detector.  Many of the
components were repaired at Fermilab, and others required sending
equipment back to the manufacturer.

The cooling system for SVX-II and L00 worked remarkably well, while
that of ISL experienced higher failure rates
(Section~\ref{sec:cool}).  The initial epoxy blockages were cleared
with laser light guided inside the small cooling tubes with fiber
optics.  A later incident stemming from acidified ISL coolant caused
leaks in the piping cooling the portcards. These leaks were sealed
with epoxy from inside the pipes as the outside was inaccessible.  The
ISL coolant was returned to distilled water, as in early running phase, 
and care was taken to
monitor and respond rapidly to changes in the cooling
system. Nonetheless, as the system aged, leaks in the piping became
larger. These did not impact the operation as the coolant ran below
atmospheric pressure.

The unexpectedly high longevity of the silicon detector, which came as
a welcome surprise, is in part due to the slow aging of the sensors as
radiation dose was accumulated (Section~\ref{sec:age}).  The inner
layers of the detector type-inverted as expected. The depletion voltage, 
signal response and noise behaved as expected also after type inversion.  
There was evidence that the electric field was
not a linear function of the position within the sensor, but instead
strengthened near the p+ and n+ implants.  This created two
depletion regions, one on either side of an underdepleted bulk.
Sensors that were not fully depleted at the end of Run-II still
provided usable data on both sides of the sensor, 
with slightly reduced charge collection on the p-side.

As a result of  to its size and complexity, the CDF silicon detector
required a dedicated team of
experts to operate and maintain it, ensuring the 
continuous harvest of high quality data.  Despite the challenges
from a prolonged run and the gradual reduction of spares and experts
toward the end of Run II, the data taking was concluded successfully due to
extensive efforts on procedure automation and diligent monitoring of
every subsystem.  Sufficient experience with detector aging,
operational failure modes and their mitigation gained along the way
ensured good performance during the final years of Tevatron running at
high luminosity.  Many profound successes of the CDF physics program
were the direct result of the high quality data provided by the silicon
detector and the dedicated displaced vertex triggers based on it.

\section{Acknowledgements}
The authors would like to thank Dr. D. Christian and Dr. T. Zimmerman
of Fermilab for useful discussions on radiation damage in silicon detectors 
and SVX3D chip functionalities, respectively. This work would
not have been possible without a strong support by the CDF operations
management and the spokespersons.
We also thank the Fermilab staff and the technical staffs of the
participating institutions for their vital contributions. This work
was supported by the U.S. Department of Energy and National Science
Foundation; the Italian Istituto Nazionale di Fisica Nucleare; the
Ministry of Education, Culture, Sports, Science and Technology of
Japan; the Natural Sciences and Engineering Research Council of
Canada; the National Science Council of the Republic of China; the
Swiss National Science Foundation; the A.P. Sloan Foundation; the
Bundesministerium f\"ur Bildung und Forschung, Germany; the Korean
World Class University Program, the National Research Foundation of
Korea; the Science and Technology Facilities Council and the Royal
Society, UK; the Russian Foundation for Basic Research; the Ministerio
de Ciencia e Innovaci\'{o}n, and Programa Consolider-Ingenio 2010,
Spain; the Slovak R\&D Agency; the Academy of Finland; and the
Australian Research Council (ARC).


%
%


\begin{thebibliography}{10}
\expandafter\ifx\csname url\endcsname\relax
  \def\url#1{\texttt{#1}}\fi
\expandafter\ifx\csname urlprefix\endcsname\relax\def\urlprefix{URL }\fi

\bibitem{cdftdr}
{CDF Collaboration}, {The CDF II Detector Technical Design Report},
  {FERMILAB-Pub-96/390-E} (1996).

\bibitem{RunIIPhys}
{K. Anikeev {\it et al.}}, {B Physics at the Tevatron: Run II and Beyond},
  {hep-ph/0201071v2} (2002).

\bibitem{RunIIPhys2}
{TEVNPH (Tevatron New Phenomena and Higgs Working Group), CDF Collaboration, D0
  Collaboration}, {Combined CDF and D0 search for standard model Higgs Boson
  production with up to 10.0 $fb^{-1}$ of Data}, {FERMILAB-CONF-12-065-E,
  CDF-NOTE-10806, D0-NOTE-6303} (2012).

\bibitem{RunIIPhys3}
{D. A. Toback},
 AIP Conference Proceedings {\bf 753} (2005) 373--382.

\bibitem{RunIIPhys4}
{P. Azzi},
 AIP Conference Proceedings {\bf 794} (2005) 66--69.

\bibitem{Akimoto2006459}
{T. Akimoto {\it et al.}},
 Nuclear Instruments and Methods in Physics Research Section A {\bf 556} (2006) 459 -- 481.

\bibitem{Jung2012ii}
{A. W. Jung {\it et al.}},
 Physics Procedia {\bf 37} (2012) 1003--1008.

\bibitem{Weber2008zz}
{M. Weber}, {Operational experience with the D0 silicon tracker}, PoS
  VERTEX2008 (2008) 001.

\bibitem{Desai2009zza}
{S. Desai}, {Radiation Damage Study of the D0 Silicon Microstrip Tracker}, PoS
  VERTEX2008 (2008) 013.

\bibitem{Merkel20031}
P.~Merkel,
 Nuclear Instruments and Methods in Physics Research Section A {\bf 501}~(1) (2003) 1 -- 6,
 {Proceedings of the 10th International Workshop on Vertex Detectors}.

\bibitem{l00_tdr2}
{C. S. Hill},
  Nuclear Instruments and Methods in Physics Research Section A {\bf 511}~(1-2) (2003) 118 -- 120,
  {Proceedings of the 11th International Workshop on Vertex Detectors}.

\bibitem{Hou2003166}
{S. Hou},
  Nuclear Instruments and Methods in Physics Research Section A {\bf 511}~(1-2) (2003) 166 -- 170,
  {Proceedings of the 11th International Workshop on Vertex Detectors}.

\bibitem{Bolla2004277}
{G. Bolla {\it et al.}},
 Nuclear Instruments and Methods in Physics Research Section A {\bf 518}~(1-2) (2004) 277 -- 280,
 {Frontier Detectors for Frontier Physics: Proceedings}.

\bibitem{Miller2004281}
L.~Miller,
 Nuclear Instruments and Methods in Physics Research Section A {\bf 518}~(1-2) (2004) 281 -- 285,
 {Frontier Detectors for Frontier Physics: Proceedings}.

\bibitem{Hill20041}
C.~S. Hill,
 Nuclear Instruments and Methods in Physics Research Section A {\bf 530}~(1-2) (2004) 1
  -- 6, {Proceedings of the 6th International Conference on Large Scale
  Applications and Radiation Hardness of Semiconductor Detectors}.

\bibitem{l00_tdr1}
T.~Nelson,
 International Journal of Modern Physics A {\bf 16S1C} (2001) 1091--1093.

\bibitem{svxii_tdr1}
{CDF Collaboration}, {SVXII} simulation and upgrade proposal,
  {CDF/DOC/SEC\_VTX/CDFR/} $\!\!$1922 (1992).

\bibitem{svxii_tdr2}
{CDF Collaboration}, {The CDF upgrade}, {CDF/DOC/CDF/PUBLIC/} $\!\!$3171
  (1995).

\bibitem{isl_tdr}
{A. Affolder {\it et al.}},
 Nuclear Instruments and Methods in Physics Research Section A {\bf
  461}~(1-3) (2001) 216 -- 218, {8th Pisa Meeting on Advanced Detectors}.

\bibitem{Rohacell}
{Evonik Industries}, {Rohacell}.
\newline\urlprefix\url{http://www.rohacell.com}

\bibitem{Holmes:2011ey}
S.~Holmes, R.~S. Moore, V.~Shiltsev,
 Journal of Instrumentation {\bf 6} (2011) T08001.

\bibitem{XFT}
{E.~J.~Thomson {\it et al.}},
 IEEE Transactions on Nuclear Science {\bf 49}~(3) (2002) 1063 -- 1070.

\bibitem{svt}
{S.~Belforte {\it et al.}}, {Silicon Vertex Tracker Technical Design Report},
  {CDF/DOC/TRIGGER/PUBLIC/} $\!\!$3108 (April 1995).

\bibitem{SVT2}
{B.~Ashmanskas {\it et al.}},
 Nuclear Instruments and Methods in Physics Research Section A {\bf 518}~(1-2) (2004) 532 -- 536,
 {Frontier Detectors for Frontier Physics: Proceeding}.

\bibitem{svtupg}
{J. Adelman {\it et al.}},
 Nuclear Instruments and Methods in Physics Research Section A {\bf 572}~(1) (2007) 361 -- 364,
  {Frontier Detectors for Frontier Physics, Proceedings of the 10th Pisa
  Meeting on Advanced Detectors}.

\bibitem{svx_3D_chip_1}
{T. Zimmerman {\it et al.}},
 Nuclear Instruments and Methods in Physics Research Section A {\bf 409}~(1-3)
  (1998) 369 -- 374.

\bibitem{svx_3D_chip_2}
{M. Garcia-Sciveres {\it et al.}},
 Nuclear Instruments and Methods in Physics Research Section A {\bf 435}~(1-2) (1999) 58 -- 64.

\bibitem{svx3d_rad}
{D. Sjoegren {\it et al.}}, {Radiation effects on the SVX3 chip},
  {CDF/DOC/SEC\_VTX/PUBLIC/} $\!\!$4461 (January 1998).

\bibitem{portcard}
{J. Andresen {\it et al.}}, {Radiation hardness of the compact port card for
  the CDF silicon tracking detector upgrade}, {CDF/PUB/PRODUCTION/PUBLIC/}
  $\!\!$5535 (January 2001).

\bibitem{chu}
{M.L. Chu {\it et al.}},
 Nuclear Instruments and Methods in Physics Research Section A {\bf 541}~(1-2) (2005) 208 -- 212,
 {Development and
  Application of Semiconductor Tracking Detectors: Proceedings of the 5th
  International Symposium on Development and Application of Semiconductor
  Tracking Detectors (STD 5)}.

\bibitem{VMEStandard}
{VME International Trade Association}, {VME Standard}.
\newline\urlprefix\url{http://www.vita.com/}

\bibitem{glink}
{Chu-Sun Yen {\it et al.}},
 Hewlett Packard Journal {\bf 43}~(5) (1992) 103--115.

\bibitem{daq:gb}
S.~Nahn, M.~Stanitzki, T.~Maruyama, {Silicon resonance detection using the
  Ghostbuster board}, {CDF/DOC/CDF/PUBLIC/} $\!\!$7749 (August 2005).

\bibitem{cdfbsmixing}
{A. Abulencia {\it et al.}},
 Physical Review Letters {\bf 97} (2006) 062003.

\bibitem{ps:seb}
{R.J. Tesarek {\it et al.}}, Radiation effects in {CDF} switching power
  supplies, {CDF/DOC/CDF/PUBLIC/} $\!\!$5903 (March 2002).

\bibitem{Siemans575PLC}
{Siemens AG}, {Siemens 575 PLC}.
\newline\urlprefix\url{http://support.automation.siemens.com/}

\bibitem{SiemansQuadlogPLC}
{Siemens AG}, {Siemens Quadlog PLC}.
\newline\urlprefix\url{https://eb.automation.siemens.com/}

\bibitem{aluminumhandbook}
G.~E. Totten, D.~S. Mackenzie (Eds.), {Alloy Production and Materials
  Manufacturing}, Vol.~2 of Handbook of Aluminium, CRC Press, Boca Raton,
  Florida, 2003.

\bibitem{ScotchDP190Expoxy}
3M, {Scotch DP190 Epoxy Adhesive}.
\newline\urlprefix\url{http://www.3m.com/}

\bibitem{DiamondPaper}
{P. Dong {\it et al.}},
 IEEE Transactions on Nuclear Science {\bf 55}~(1) (2008) 328 --332.

\bibitem{ge_fanuc}
{GE Fanuc}, {GE Intelligent Platforms}.
\newline\urlprefix\url{http://www.ge-ip.com/products/3311/}

\bibitem{TLD}
{R.J. Tesarek {\it et al.}},
 Nuclear Instruments and Methods in Physics Research Section A {\bf
  514}~(1-3) (2003) 188 -- 193, {Proceedings of the 4th International
  Conference on Radiation Effects on Semiconductor Materials, Detectors and
  Devices}.

\bibitem{Moll:1999kv}
{M. Moll}, Radiation damage in silicon particle detectors: microscopic
  defects and macroscopic properties, DESY-THESIS-1999-040 (1999).

\bibitem{Eremin:2002wq}
V.~Eremin, E.~Verbitskaya, Z.~Li,
 Nuclear Instruments and Methods in Physics Research Section A {\bf 476} (2002) 556 -- 564.

\bibitem{swartz}
{M. Swartz {\it et al.}},
 Nuclear Instruments and Methods in Physics Research Section A {\bf 565}~(1) (2006) 212 -- 220, {PIXEL
  2005: International Workshop on Semiconductor Pixel Detectors for Particles
  and Imaging}.

\bibitem{cdfcharmmixing}
{T. Aaltonen {\it et al.}},
 Physical Review Letters {\bf 100} (2008) 121802.

\bibitem{bbraunim}
B.~Brau,
 Nuclear Instruments and Methods in Physics Research Section A {\bf 541}~(1-2) (2005) 73 -- 77,
  {Proceedings of the 5th International Symposium on Development and
  Application of Semiconductor Tracking Detectors (STD 5)}.

\bibitem{secvtx}
{D. Acosta {\it et al.}},
 Physical Review D {\bf 71} (2005) 052003.

\end{thebibliography}
\end{document}